\documentclass{elsarticle}
\usepackage[letterpaper,top=1in,bottom=1in,left=1in,right=1in]{geometry}
\usepackage{lineno}
\usepackage[parfill]{parskip}
\usepackage{amstext,amssymb,bm,amsthm,mathtools}
\usepackage[per-mode=symbol,separate-uncertainty = true]{siunitx}
\usepackage[dvipsnames]{xcolor} 
\definecolor{light-gray}{gray}{0.95}
\usepackage{graphicx}
\usepackage{enumitem}
\usepackage{booktabs,tabu,tabularx}
\usepackage[ruled,vlined]{algorithm2e}
\usepackage{algpseudocode}
\usepackage{multirow}

\usepackage{float}
\usepackage{natbib}
\usepackage[bookmarks=true,colorlinks=true,linkcolor=blue,citecolor=red]{hyperref}
\makeatletter
\AtBeginDocument{\def\@citecolor{red}}
\makeatother
\biboptions{sort&compress}
\journal{Journal of Computational Physics}

\usepackage{subcaption}
\begin{document}


\begin{frontmatter}


\title{A Neural Surrogate Approach for Simulating Natural Convection Problems} 

\author[cs,isis]{Nurshat Menglik}
\ead{nulixiati.mangnike@vanderbilt.edu}
\author[cs,isis]{Alex Shao}
\ead{zhenxuan.shao@vanderbilt.edu}
\author[cs,isis]{David Hyde\corref{cor1}}
\cortext[cor1]{Corresponding author}
\ead{david.hyde.1@vanderbilt.edu}

\address[cs]{Department of Computer Science, Vanderbilt University, 1400 18\textsuperscript{th} Avenue South, Nashville, TN 37212-2846, USA}
\address[isis]{Institute for Software Integrated Systems, Vanderbilt University, 1025 16\textsuperscript{th} Avenue South, Nashville, TN 37212-2328, USA}


\begin{abstract}

This paper presents a neural surrogate approach for improving the accuracy of natural convection problems simulated with a Boussinesq flow model (incompressible flow with heat transfer).
Our approach, based on Fourier neural operators, uses training data consisting of matched pairs of simulations run under the computationally cheaper yet less accurate Boussinesq flow model and a more computationally expensive and more accurate compressible flow model.
In both cases, we implement our parallelized simulation codes based on an implicit monolithic mixed finite element method (FEM) approach using the open-source FEniCSx framework.
Our implementations are validated against a commercial software package, COMSOL, as well as standard test problems from the literature.
We include a careful discussion and analysis of data set generation and present learning results in two and three spatial dimensions.
Using compressible flow results as high-fidelity reference solutions, our learning approach, with a \textit{single} model evaluation per simulation, substantially improves the per-channel accuracy of Boussinesq predictions, with structural similarity (SSIM) close to unity across all flow variables and test distributions and corresponding mean-squared error reductions of one to nearly three orders of magnitude. 
All code and data is released as open-source.
\end{abstract}




\end{frontmatter}

\section{Introduction}
Practitioners select different models of fluid dynamics for problems depending on the physical characteristics of the problem and the application area \citep{TU20181}, as seen across aerospace \citep{Spalart2016OnTR,FUJII2005455}, automotive \citep{CHINTALA2013709}, biomedical \citep{YE2024112639,ESMAILYMOGHADAM201363,Reid2021}, chemical \citep{babanezhad2020high}, civil \citep{zawawi2018review}, and environmental engineering \citep{KIM1999145,TOMINAGA2024105741}.
Natural convection, a critical phenomenon in many of these domains (ranging from electronic cooling to atmospheric dynamics), has traditionally relied on the Boussinesq approximation \citep{boussinesq1903theorie,lappa2022incompressible} for computational efficiency.
This simplified framework assumes constant fluid properties except for density in the buoyancy term, which is modeled as linearly proportional to temperature differences.
The Boussinesq approximation fails for large temperature differences, where density variations depend non-linearly on thermodynamics \citep{GRAY1976545, mangnike2024toward}, necessitating compressible flow models.
However, the transition from an incompressible (or Boussinesq) formulation to a compressible flow formulation introduces significant numerical challenges, especially when flows are at a low Mach number ($ Ma \ll 0.1 $).
In low-Mach regimes, where the fluid velocity is much smaller than the speed of sound, the large disparity between acoustic wave speeds and convective flow speeds leads to system stiffness, making explicit solvers inefficient and standard density-based methods inaccurate.
Addressing these challenges, numerical methods for solving the compressible Navier--Stokes equations coupled with heat transfer in the low-Mach regime can be broadly categorized based on their formulation strategy: pressure-based methods, density-based methods, and monolithic and stabilized methods (see Section \ref{sec:rw-compressible-flow}).
We note that these approaches, while effective, typically add significant implementation complexity compared to Boussinesq flow and often require complex operator splitting and problem-dependent tuning.
Thus, improved algorithms for both Boussinesq and compressible flow remain of interest.

In parallel with these and other computational fluid dynamics (CFD) algorithms developed in recent years, there has been increasing interest in scalable open-source CFD frameworks, particularly in the context of machine learning and data-driven modeling.
With the continuous growth of the application of deep learning in computational simulations \cite{willard2020integrating,sharpdeep}, open-source CFD frameworks have become crucial for machine learning (ML)-based CFD applications, where large amounts of high-fidelity simulation data may be needed to train models \cite{mangnike2024toward}, or where ML models are used to augment the solver.
They also facilitate coupling with machine learning components---for instance, one can embed neural network models into an open-source CFD code or use Python interfaces to stream simulation data to an ML training pipeline in real-time.
Although commercial tools like COMSOL provide robust multiphysics interfaces, the authors have found them expensive and ill-suited for large-scale data generation. 



Motivated by joint considerations of algorithm and implementation, this work first (Sections \ref{sec:math_formulations}--\ref{sec:results}) develops and validates an open-source, fully implicit monolithic finite element framework for both Boussinesq and compressible natural convection problems in two and three dimensions, built on FEniCS/FEniCSx \cite{BarattaEtal2023,logg2012automated}. 
Our proposed solver is based on a globally coupled mixed finite element formulation with a single variational residual for the compressible Navier--Stokes and total-energy equations.
The paper then conducts a study of dataset generation (Section \ref{sec:data}) that illustrates how one can construct a ``good'' set of fluid simulations for training a machine learning model.
Section \ref{sec:learning} subsequently uses such a dataset to train our proposed Fourier neural operator (FNO)-based approach for single-evaluation corrections of Boussinesq fluid simulations.
This approach improves on the preliminary conference work of the authors based on a less performant U-Net architecture \cite{mangnike2024toward}.
Finally, Section \ref{sec:conc} concludes the paper and discusses future directions.

Our contributions include:

\begin{itemize}

\item \textbf{Robust open-source solvers:} We develop solvers using a monolithic, fully implicit mixed-element discretization based on the Taylor-Hood mixed element method \citep{TAYLOR197373} that handles both nearly-\hspace{0pt}incompressible and fully compressible regimes without explicit preconditioning or stabilization in the natural-convection regimes considered here.
Implicit backward Euler time integration and tight coupling through a single global weak form yield robust behavior over the flow regimes considered in this work and permit stable simulations with comparatively large time steps.
The complete solver is developed using FEniCS/FEniCSx with configurable direct (MUMPS) and iterative (FGMRES + BoomerAMG) linear solvers. 

\item \textbf{Fourier- and Voronoi-based stochastic data generation:} Given our algorithm and implementation, we present strategies for generating large, diverse natural convection simulation datasets via randomized wall temperatures, constructed as truncated random Fourier series with spectral decay in two dimensions and as random Voronoi tessellations with i.i.d.\ Gaussian amplitudes in three dimensions.
In both cases the perturbation fields are centered, normalized, and clamped to ensure smooth (or piecewise-constant), reproducible, and statistically independent boundary realizations.
This enables efficient large-scale dataset creation for machine learning without modifying the underlying fluid dynamics.
We demonstrate how this approach improves over na\"ive dataset generation, as used in works such as \citet{mangnike2024toward}.

\item \textbf{Improving Boussinesq simulations via FNO surrogate:} By training an FNO-based model on matched pairs of Boussinesq and compressible flow simulation data, we are able to run a Boussinesq flow simulation, query the trained model \textit{once}, and obtain a corrected fluid simulation result that much more closely resembles the output of the same fluid simulation if a full compressible flow model had been used.
We claim that this is a particularly appropriate application for learning in computational physics: since practitioners are already making a potentially significantly erroneous approximation by choosing the Boussinesq model, incorporating an imperfect learning component into one's simulation pipeline is less controversial than replacing a high-fidelity compressible flow solver with a physics-informed neural network (PINN), for instance
\cite{11447375}.

\end{itemize}

All code and data is released as open-source: \href{https://github.com/Nurshat317/NeuralConvection}{https://github.com/Nurshat317/NeuralConvection}. 


\section{Related Work}
\label{sec:rw}

\subsection{Boussinesq flows}

The development of algorithms for and involving Boussinesq flows remains an active topic in computational physics; as with any fluid dynamics model, a variety of discretization choices have been explored.
\citet{UECKERMANN2016390} present a hybridized discontinuous Galerkin (DG) discretization \cite{cockburn2009unified,nguyen2009implicit} with implicit-explicit Runge-Kutta time integration.
\citet{SCHROEDER2017760} propose a stabilized DG-FEM approach for Boussinesq flows, which requires stabilization parameters to be chosen, although their examples show robust behavior with a fixed choice of parameters.
A fictitious domain method is developed in \citet{YU2006424}, where the authors use the Boussinesq approximation in the context of particulate flow.
Prior to these works, \citet{liu2003fourth} gave a fourth-order scheme for the Boussinesq equations, albeit only in two spatial dimensions.

Beyond DG formulations, monolithic finite element strategies for thermally coupled incompressible flow have also been studied extensively.
For example, \citet{DAMANIK20093869} develop a monolithic FEM--multigrid solver for non-isothermal incompressible flow, with emphasis on multigrid efficiency and block-structured linear algebra.
At larger scales and in three dimensions, \citet{Smethurst2013UnstructuredFE} similarly emphasize efficient solver technology for unstructured finite element discretizations of 3D Boussinesq flow.
Three-dimensional Boussinesq dynamics in differentially heated enclosures has also been the subject of detailed numerical studies, including the emergence of oscillatory regimes at low Prandtl number~\citep{wakitani2001numerical}.
From the standpoint of mixed discretizations, \citet{MILLER2022150} propose a versatile mixed method for non-isothermal incompressible flow designed to remain robust under large buoyancy forces.
Related to these developments, several other recent works have sought to improve the efficiency of Boussinesq solvers or apply them in settings such as control and learning.
For instance, \citet{ELMAN20113900} provide fast solver and preconditioning techniques in the case of a Boussinesq solver using a stabilized trapezoid rule as the basis for their time integration scheme.
A numerical algorithm for optimal control of Boussinesq flows is presented in \citet{SONG2023112458}.
Recently, \citet{mangnike2024toward} presented a deep learning technique for improving Boussinesq flow simulations in regimes outside the traditional plausible regime of the Boussinesq approximation, based on a U-Net architecture.

Finally, a related body of work concerns non-isothermal incompressible flow coupled with additional physics such as phase change.
\citet{rakotondrandisa2020finite} present a FreeFem++ finite element toolbox for solid--liquid phase-change systems with natural convection, and \citet{SADAKA2020107492} develop parallel three-dimensional simulations for the same class of problems using a domain decomposition strategy.
While these tools target incompressible/Boussinesq regimes rather than fully compressible flow, they are directly relevant to the broader theme of open-source, scalable solvers for coupled thermal-fluid problems.
For further discussion of the Boussinesq approximation, we refer the interested reader to \citet{mayeli2021buoyancy,zeytounian2003joseph,mangnike2024toward,lappa2022incompressible}, and the references therein.
Consistent with our treatment of compressible flow, in the present work, we pursue a mixed finite element approach for simulating Boussinesq flow.

\subsection{Compressible flow strategies}
\label{sec:rw-compressible-flow}

When solving the compressible Navier--Stokes equations, especially (though not exclusively) in the context of FEM and in the low-Mach regime, we identify three common categories of approaches, each with their advantages and drawbacks.

\textbf{Pressure-Based Methods.} Pressure-based approaches derive a pressure equation (typically Poisson- or Helmholtz-type) from the coupled mass and momentum equations and then update density through an equation of state.
The SIMPLE algorithm \citep{PATANKAR19721787} and subsequent variants such as SIMPLEC \citep{Doormaal01041984} and PISO \citep{ISSA198640} were originally developed for incompressible flows and later extended to compressible regimes \citep{rhie1989pressure,karki1989pressure}. 
However, for low-speed compressible flows, the presence of fast acoustic waves introduces computational stiffness. 
The low Mach number approximation (LMNA), introduced by \citet{rehm1978equations}, addresses this by decomposing pressure into thermodynamic, hydrostatic, and dynamic components, replacing the pressure in the equation of state with only the thermodynamic pressure. 
Alternatively, \citet{WALL2002545} propose converting the pressure Poisson equation into a Helmholtz equation to avoid Courant number restrictions imposed by acoustic waves. 
This decouples density from dynamic pressure variations and filters acoustic waves, permitting large time steps for combustion and thermal convection problems. 
\citet{le1992chebyshev} demonstrate LMNA for non-Boussinesq convection, and more recently, \citet{HENNINK2021109877} developed a pressure-based discontinuous Galerkin method for low-Mach flows.
Inspired by \citet{WALL2002545}, \citet{yanaoka2025numerical} propose a simultaneous relaxation method that solves all variables (density, velocity, pressure, and internal energy) through iterative correction steps, converting the pressure Poisson equation into a Helmholtz equation to avoid acoustic time-step restrictions while emphasizing discrete conservation of momentum, energy, and entropy without requiring explicit stabilization or preconditioning.
While LMNA enables efficient low-Mach computation, it is restricted to truly low-Mach regimes (Ma $\ll$ 1) and requires careful treatment to maintain thermodynamic consistency.

\textbf{Density-Based Methods.} Density-based approaches advance density directly from mass conservation and obtain pressure from the equation of state, making them natural for compressible flows but often ill-conditioned at low Mach numbers due to acoustic stiffness. 
\citet{TURKEL1987277,turkel1999preconditioning} introduces local time-derivative preconditioning that modifies the system eigenvalues to reduce this disparity, enabling computation from Ma$\sim$0.01 to supersonic regimes. 
\citet{CHOI1993207} extend preconditioning to viscous flows with heat transfer, and \citet{weiss1995preconditioning} develop a formulation for both variable and constant density flows. 
For flux-based methods, \citet{guillard1999behaviour} analyze the failure of upwind schemes in the low-Mach limit, showing that the standard Roe flux produces $O(1/\text{Ma})$ pressure scaling errors, and proposed modifications to restore accuracy. 
While preconditioning enables existing compressible solvers to handle low-Mach flows, these methods require problem-dependent parameter tuning and may compromise accuracy if artificial dissipation is not properly scaled.

\textbf{Monolithic and Stabilized Methods.} 
Monolithic methods solve the momentum, continuity, and energy equations simultaneously via a single large nonlinear system. 
This approach often offers superior robustness and convergence rates but requires specialized stabilization techniques to ensure stability, particularly in convection-dominated or incompressibility-like regimes (low Mach number).
Stabilized finite element methods are a prime example of this strategy.
\citet{brooks1982streamline} pioneered this field by developing SUPG (Streamline Upwind Petrov–Galerkin), and \citet{hughes1986new} introduce PSPG (Pressure-Stabilizing Petrov-Galerkin), enabling equal-order velocity-pressure interpolation by adding streamline diffusion and pressure stabilization terms. 
These formulations require problem-dependent stabilization parameters that depend on local flow characteristics, element geometry, and material properties. 
Alternatively, \citet{LUO2012133} develop implicit discontinuous Galerkin methods for compressible Navier--Stokes, achieving high-order accuracy with local conservation but requiring numerical flux definitions at element interfaces.

In the present paper, we base our method on a monolithic system solve, and we pursue an implicit time discretization that allows us to take large time steps.
For the regimes of natural convection problems we test, we find that this discretization allows us to omit specialized stabilization terms (and the associated problem-specific parameter tuning).

\subsection{FEniCS and FEniCSx}

Due to the numerical instability and complexity of compressible flows, the number of high-fidelity open-source codes and CFD tools capable of performing three-dimensional, time-dependent, and large-scale compressible flow simulations remains limited. 
In the context of monolithic FEM, these challenges can be even more pronounced due to the fully coupled nature of the discretized system. 
Modern finite element frameworks, notably FEniCS/FEniCSx \citep{BarattaEtal2023}, provide powerful automation and flexibility for formulating and solving partial differential equations (PDEs).
FEniCS is designed to be capable of automatically transforming a variational formulation into a high-performance finite element code \citep{logg2012automated}. 
Coupled with linear algebra backends, such as PETSc \citep{petsc-efficient}, FEniCS enables efficient parallel computing on large-scale systems. 
Previous work has demonstrated the capabilities of FEniCS for incompressible flows (e.g., \citep{mortensen2011fenics,vynnytska2013benchmarking}) and coupled multiphysics problems (e.g., \citep{zhang2016mixed}), but full compressible flow solvers with parallel implementations and data-generation pipelines for ML applications remain largely absent from the literature.
FEniCSx is the latest version of the FEniCS platform and is recommended for new applications.
Our solver implementations are based on FEniCSx.

\subsection{Hybridizing Simulation and Learning}

An extensive body of literature explores hybridizing traditional simulation methods with ML \cite{sharpdeep,willard2020integrating,karniadakis2021physics}.
PINNs \cite{11447375,raissi2019physics,712178,dissanayake1994neural} embed PDEs and physical constraints in a neural network's loss function and architecture, generally replacing a traditional numerical solver.
Other methods use learning to accelerate specific aspects of classical solvers, such as preconditioning \cite{dgcm} and adaptive mesh refinement \cite{yang2023reinforcement}.
More broadly, neural surrogate models are often used to replace high-fidelity models (e.g., of material properties or subscale physics) with more efficient proxies of comparable quality \cite{viquerat2020supervised,SpatiallyLocalSurrogateModelingofSubgridScaleEffectsinIdealizedAtmosphericFlowsADeepLearnedApproachUsingHighResolutionSimulationData,eiximeno2025deep,bhattacharjee2026machine,stuckner2021optimal,khorrami2023artificial}.
The present work can be viewed as a neural surrogate for the compressible flow physics missing from the Boussinesq model; or, alternatively, one may view the present work as using a neural model to correct model-form error in the governing fluid dynamics.
Model-form correction for PDEs has been explored in recent works such as \citet{zhou2026latent}, which develops a latent-space model correction framework based on encoders, decoders, and Gaussian processes.
Additive discrepancy correction is another approach used for model-form correction wherein governing equations receive an additional correction term that is learned from data \cite{vagnoli2025local,duraisamy2019turbulence,farchi2021using}.
The surrogate in the present work directly outputs corrected field data, as opposed to predicting a correction to be added to the Boussinesq equations.


\section{Governing Equations}
\label{sec:math_formulations}
To begin, we introduce the mathematical formulations of Boussinesq and compressible flow equations, their non-dimensionalized forms, and ultimately, weak forms. 
In both cases, we base our weak form derivations on the non-dimensionalized governing equations.

\subsection{Boussinesq approximation}
The Boussinesq approximation \cite{boussinesq1903theorie}, commonly used for buoyancy-driven flows such as natural convection, assumes variations in the density of a fluid are negligible except for the effects of buoyancy, which may be induced by variations in temperature. 
This approximation yields a model that resembles the incompressible Navier--Stokes equations combined with a heat solve (cf.\ \cite{mangnike2024toward}):
\begin{equation}
\begin{aligned}
    \nabla \cdot \boldsymbol{u} &= 0, \\
    \rho_0 \left( \frac{\partial \boldsymbol{u}}{\partial t} + \boldsymbol{u} \cdot \nabla \boldsymbol{u} \right) &= -\nabla P + \mu \nabla ^2 \boldsymbol{u} - \rho_0\,g\,\beta\,(T-T_0) \boldsymbol{\hat{e}_g},\\
    \frac{\partial T}{\partial t} + \boldsymbol{u} \cdot \nabla T &= \alpha \nabla^2 T ,
\end{aligned}
\label{Bo_strong}
\end{equation}
where $\rho_0$ and $T_0$ are a reference density and temperature, $\alpha$ is thermal diffusivity, $\mu$ is dynamic viscosity, and $P = p - \rho_0\,g(\boldsymbol{\hat{e}_g}\cdot\boldsymbol{x})$ is a shifted pressure. We define $\boldsymbol{\hat e}_{\boldsymbol{g}}$ as a unit vector in the direction of gravity (along the $y$-axis in both 2D and 3D).
The Boussinesq approximation assumes that viscous heat dissipation and pressure work terms are negligible.

\subsubsection{Non-dimensionalized form}
\label{subsec:boussinesq_nondim}
We define the characteristic reference scales for our basic non-dimensionalized variables below, where L is the characteristic length, U is the characteristic velocity (equal to $\alpha / L$), and $\rho_0$ and $T_0$ remain reference density and temperature: 

\begin{equation*}
    \begin{aligned}
    x^* &= \frac{x}{L}, \quad \quad\quad
    t^* = \frac{t\,U}{L}, \quad \quad \quad\quad
    u^* = \frac{u}{U},\\
    \rho^* &= \frac{\rho}{\rho_0}, \quad\quad
    T^* = \frac{T - T_0}{\Delta T}, \quad\quad
    P^* = \frac{P\,L^2}{\rho_0\,\alpha^2}.
    \end{aligned}
\end{equation*}

The non-dimensionalized Prandtl number $Pr$ and Rayleigh number $Ra$ are
\begin{align*}
    Pr = \frac{C_p \mu}{\kappa}, \quad Ra = \frac{g \beta \Delta T L^3}{\nu \alpha},
\end{align*}
where $C_p$ is the specific heat capacity of a material, $\beta$ is the thermal expansion coefficient of a material, and $\Delta T$ is the temperature difference across distance $L$ (in the case of a natural convection problem such as a differentially heated thermal cavity, this is simply taken as the difference in temperature between the hot and cold walls).
Combining the above expressions, the non-dimensionalized form of the Boussinesq approximation can be written as
\begin{equation}
\begin{aligned}
    \nabla \cdot \boldsymbol{u^*} &= 0, \\
     \frac{\partial \boldsymbol{u^*}}{\partial t} + \boldsymbol{u^*} \cdot \nabla \boldsymbol{u^*}  &= -\nabla P^* + Pr \nabla ^2 \boldsymbol{u^*} - PrRaT^* \boldsymbol{\hat{e}_g},\\
    \frac{\partial T^*}{\partial t} + \boldsymbol{u^*} \cdot \nabla T^* &=  \nabla^2 T^* .
\end{aligned}
\label{Bo_strong_nondim}
\end{equation}

\subsubsection{Weak form}
We employ a mixed finite element method with the Taylor--Hood element pair to discretize velocity, temperature, and pressure. 
Let $\Omega\subset\mathbb{R}^d$ ($d=2,3$) denote the domain, and let
\begin{align*}
    \boldsymbol{u}\in \boldsymbol{V}_h,\quad T\in \Theta_h,\quad P\in Q_h,
    \qquad\text{with}\qquad
    \boldsymbol{V}_h=[P_2]^d,\;\Theta_h=P_1,\;Q_h=P_1 .
\end{align*}
We use a Taylor--Hood $P_2$--$P_1$ pairing for velocity and pressure, along with a $P_1$ temperature space.
We pose the weak form of our problem in the mixed space $\mathcal{W}_h^\text{B}=\boldsymbol{V}_h\times \Theta_h\times Q_h$.
This yields a trial function \( w \) and a test function \( \psi \) of the form
\begin{align*}
    w = 
    \begin{bmatrix}
     \mathbf{u}\\
     T\\
     P\\
    \end{bmatrix}
    \in \mathcal{W}_h^\text{B},\quad
    \psi = 
    \begin{bmatrix}
     \mathbf{v}\\
     \theta\\
     q\\
    \end{bmatrix}
    \in \mathcal{W}_h^\text{B} .\\
\end{align*}
Thus, the variational problem to solve is: Find \( w \in \mathcal{W}_h^\text{B} \) such that
\begin{align*}
    \mathcal{F}(w ; \psi) = \int_{\Omega} \mathcal{F}(w) \psi \, dx = 0 \quad \forall \psi \in \mathcal{W}_h^\text{B} .
\end{align*}

Based on Equation \ref{Bo_strong_nondim}, and dropping asterisks for ease of exposition, the weak form for the conservation of mass can be written as: 
\begin{align*}
    \int_{\Omega} \left( \nabla \cdot \mathbf{u} \right) q \, dx = 0,
\end{align*}
and the weak form for conservation of momentum as
\begin{align*}
    \int_{\Omega} \left( \frac{\partial \mathbf{u}}{\partial t} + \mathbf{u} \cdot \nabla \mathbf{u} \right) \cdot \boldsymbol{v} \, dx - \int_{\Omega} P(\nabla \cdot \mathbf{v}) \, dx  + \int_{\Omega} Pr  \nabla \mathbf{u} : \nabla \mathbf{v} \, dx -  \int_\Omega \mathbf{F_B} \cdot \mathbf{v}\ dx= 0,
\end{align*}
where the body force term $\boldsymbol{F_B} = -PrRa T \boldsymbol{\hat{e}_g}$.
The weak form for the conservation of energy may be written as
\begin{align*}
    \int_{\Omega} \frac{\partial T}{\partial t} \theta \, dx  + \int_{\Omega} (\mathbf{u} \cdot \nabla T)\,\theta \, dx + \int_{\Omega} \nabla T \cdot \nabla{\theta} \, dx= 0 .
\end{align*}

Integrating \( \mathcal{F} \) by parts yields the final weak form:
\begin{equation}
\begin{aligned}
    \mathcal{F}_{\text{mass}} &= \int_{\Omega} (\nabla \cdot \boldsymbol{u}) q \, d\Omega = 0, \\
    \mathcal{F}_{\text{mom}} &= \int_{\Omega} \left( \frac{\partial \boldsymbol{u}}{\partial t} + \boldsymbol{u} \cdot \nabla \boldsymbol{u} \right) \cdot \boldsymbol{v} \, d\Omega 
    - \int_{\Omega} P (\nabla \cdot \boldsymbol{v}) \, d\Omega  \\
    &\quad + \int_{\Omega} Pr \nabla \boldsymbol{u} : \nabla \boldsymbol{v} \, d\Omega 
    + \int_{\Omega} Pr Ra T (\boldsymbol{\hat{e}}_g \cdot \boldsymbol{v}) \, d\Omega = 0, \\
    \mathcal{F}_{\text{energy}} &= \int_{\Omega} \frac{\partial T}{\partial t} \theta \, d\Omega 
    + \int_{\Omega} (\boldsymbol{u} \cdot \nabla T) \theta \, d\Omega 
    + \int_{\Omega} \nabla T \cdot \nabla \theta \, d\Omega = 0, \\
    \mathcal{F}(w ; \psi) &=   \mathcal{F}_{\text{mass}} + \mathcal{F}_{\text{mom}} + \mathcal{F}_{\text{energy}} .
\end{aligned}
\label{eq:bo-weak}
\end{equation}

\subsection{Compressible flow}
For flows with large temperature variations, the Boussinesq approximation fails, and compressible flow models are required \cite{GRAY1976545}.
One can write the compressible Navier--Stokes equation coupled with the energy equation as follows:
\begin{equation}
\begin{aligned}
    \frac{\partial \rho}{\partial t} + \nabla \cdot (\rho \mathbf{u}) =& 0,\\
    \frac{\partial (\rho \boldsymbol{u})}{\partial t} + \nabla \cdot (\rho \boldsymbol{u} \otimes \boldsymbol{u})  =& -\nabla p + \nabla \cdot \boldsymbol{\tau} + \rho  g \boldsymbol{\hat{e}_g},\\
    \frac{\partial (\rho e)}{\partial t} + \nabla \cdot (\rho e \boldsymbol{u}) =& \nabla \cdot (\kappa \nabla T) - p\nabla \cdot \mathbf{u} + \Phi,
\end{aligned}
\end{equation}
where $e$ is the internal energy per unit mass, $g = ||\boldsymbol{g}||$, $e=c_{v}T$ (for an ideal gas), $\rho_0$ is the reference density, $\kappa$ is the thermal conductivity, and $p\nabla \cdot \mathbf{u}$ is the pressure work term.
The viscous stress tensor $\boldsymbol{\tau}$ and the viscous dissipation term $\Phi$ are defined as:
\begin{align*}
    \boldsymbol{\tau} =& \mu \left( \nabla \mathbf{u} + (\nabla \mathbf{u})^T \right) - \frac{2}{3} \mu (\nabla \cdot \mathbf{u}) \mathbf{I},\\
    \Phi =& \boldsymbol{\tau} : \nabla \mathbf{u}.
\end{align*}

Throughout this work, the dynamic viscosity $\mu$ and thermal conductivity $\kappa$ are taken to be constants of the reference state; temperature-dependent property models such as Sutherland's law are not used.

An equation of state is needed to close the equations.
For an ideal gas, the equation of state is
\begin{align*}
    p = (\gamma - 1)\rho e,
\end{align*}
where $\gamma$ is the ratio of specific heat of a gas at a constant pressure to heat at a constant volume (1.4 for air).
The energy equations can also be written in total energy form:
\begin{align}
    \frac{\partial (\rho E)}{\partial t}
    + \nabla\cdot(\rho E\,\boldsymbol{u})
    =  \nabla\cdot(\kappa \nabla T) - \nabla \cdot (p\boldsymbol{u})+  \nabla\cdot(\boldsymbol{\tau}\cdot\boldsymbol{u})
      + \rho\,g\boldsymbol{\hat{e}_g}\cdot\boldsymbol{u} ,
      \label{eq:energy-total-energy}
\end{align}
where $E$ is total energy:
    \begin{align*}
        E=e+\frac{1}{2}\boldsymbol{u}\cdot\boldsymbol{u} .
    \end{align*}
Following standard practice for low-Mach formulations, the $E$ here is the sum of internal and kinetic energy only, the gravitational work enters the energy balance solely through the body-force power term $\rho g \hat{\mathbf{e}}_g \cdot \boldsymbol{u}$ on the right-hand side.
\subsubsection{Non-dimensionalized form}

Following \cite{yanaoka2025numerical}, the characteristic reference scales needed to express compressible flow in non-dimensionalized form are
\begin{align*}
    \boldsymbol{x^*} &= \frac{\boldsymbol{x}}{L}, \quad \quad
    t^* = \frac{t\,U}{L}, \quad \quad \quad
    \boldsymbol{u^*} = \frac{\boldsymbol{u}}{U}, \quad \quad\quad
    e^* = \frac{e}{e_0}\\
    \rho^* &= \frac{\rho}{\rho_0}, \quad\quad
    T^* = \frac{T}{T_0}, \quad\quad
    p^* = \frac{p-p_0}{\rho_0U^2},
\end{align*}
where a naught subscript represents reference values and where $L$ represent characteristic length. Here the reference velocity is the buoyancy (free-fall) velocity $U=\sqrt{g\beta\,\Delta T\,L}$, which differs from the thermal-diffusion velocity $U=\alpha/L$ used in the Boussinesq non-dimensionalization (Section~\ref{subsec:boussinesq_nondim}), the symbol $U$ thus denotes a different characteristic scale in each of the two formulations.
This choice is what produces the parameter relation $Re=\sqrt{Ra/Pr}$ used for the compressible runs, matching each case to its Boussinesq counterpart at the same Rayleigh number, with $U=\alpha/L$ one would instead obtain the constant $Re=1/Pr$.
Furthermore, the non-dimensionalized Mach number $Ma$, Reynolds number $Re$ and Froude number $Fr$ are defined as
\begin{align*}
    Ma = \frac{U}{c}, \quad
    Re = \frac{\rho_0 U L}{\mu}, \quad
    Fr = \frac{U}{\sqrt{gL}},
\end{align*}
where $c = \sqrt{\frac{\gamma p_0}{\rho_0}}$ is the speed of sound in a material.
Using the total energy form of the energy equation (Equation \ref{eq:energy-total-energy}), the non-dimensionalized forms of the compressible flow governing equations can be written as
\begin{equation}
\begin{aligned}
    \frac{\partial \rho^*}{\partial t^*} + \nabla \cdot (\rho^* \mathbf{u}^*) =& 0,\\
    \frac{\partial \rho^* \mathbf{u}^*}{\partial t^*} +  \nabla^* \cdot (\rho^* \boldsymbol{u}^* \otimes \boldsymbol{u^*})   
    =& -\nabla^* p^* + \frac{1}{Re} \nabla^* \cdot \boldsymbol{\tau^*} + \frac{\rho^*}{Fr^2} \boldsymbol{\hat{e_g}},\\
    \frac{\partial (\rho^* E^*)}{\partial t^*} +  \nabla^* \cdot(\rho^* E^* \boldsymbol{u^*})  =& \frac{\gamma}{RePr} {\nabla^*}^2 T^* - (\gamma - 1)\nabla \cdot ((\gamma Ma^2 p^* +1)\boldsymbol{u^*})\\
    &+ \frac{(\gamma -1)\gamma Ma^2}{Re}\nabla^*\cdot(\boldsymbol{\tau^*}\cdot\boldsymbol{u^*}) + \frac{(\gamma -1)\gamma Ma^2}{Fr^2} \rho^* \boldsymbol{\hat{e}_g}\cdot\boldsymbol{u^*},\\
    \rho^*e^* =& 1 + \gamma Ma^2\,p^*.
\end{aligned}
\end{equation}

\subsubsection{Weak form}
Similarly to our derivation for Boussinesq flow, we solve for $(\boldsymbol{u},e,\rho,p)$ in the mixed Taylor--Hood-type space
\begin{align*}
    \mathcal{W}_h^\text{C}=\boldsymbol{V}_h\times E_h\times R_h\times P_h
    =[P_2]^d\times P_1\times P_1\times P_1,
\end{align*}
where these solution variables are defined as components of the associated trial function.
The trial and test functions are defined as
\begin{align*}
    w=\begin{bmatrix}\boldsymbol{u}\\ e\\ \rho\\ p\end{bmatrix}\in \mathcal{W}_h^\text{C},\qquad
    \psi=\begin{bmatrix}\boldsymbol{v}\\ w_e\\ w_\rho\\ q\end{bmatrix}\in \mathcal{W}_h^\text{C}.
\end{align*}
Then, the weak form for the non-dimensionalized compressible equations (dropping asterisks, as before, for ease of exposition) can be written as:\\
    \textbf{Continuity Equation:}
    \begin{align}
    \mathcal{F}_{\rho}(\rho,\boldsymbol{u};q)=
    \int_\Omega \left (\frac{\partial \rho}{\partial t} + \nabla \cdot (\rho \boldsymbol{u}) \right ) q \, d\Omega = 0.
\end{align}

\textbf{Momentum Equation:}
\begin{align}
    \mathcal{F}_{\boldsymbol{u}}(\rho,\boldsymbol{u},p;\boldsymbol{v})=&\int_\Omega \left ( \frac{\partial ( \rho \boldsymbol{u})}{\partial t}\cdot\boldsymbol{v}
    +\nabla \cdot(\rho\boldsymbol{u}\otimes\boldsymbol{u})\cdot\boldsymbol{v}
    -p\nabla\cdot\boldsymbol{v}  
    + \frac{1}{Re}\boldsymbol{\tau}:\nabla\boldsymbol{v}
    -\frac{\rho}{Fr^2}(\boldsymbol{\hat{e_g}}\cdot\boldsymbol{v}) \right ) \;\mathrm d\Omega
= 0.
\end{align}

\textbf{Energy Equation:}

\begin{align}
    \mathcal{F}_{e}(\rho, e, \boldsymbol{u},p;w_e) &= \int_\Omega \left ( \frac{\partial(\rho E)}{\partial t}\, w_e +\nabla \cdot (\rho E \boldsymbol{u}) w_e
    +\frac{\gamma}{\mathrm{Re}\,\mathrm{Pr}} \nabla e \cdot \nabla w_e - (\gamma-1)\!\left(\gamma\,\mathrm{Ma}^2 p+1\right)\boldsymbol{u}\cdot\nabla w_e \right ) d\Omega\\
    &+\int_\Omega \left ( \frac{(\gamma-1)\gamma\,\mathrm{Ma}^2}{\mathrm{Re}} (\boldsymbol{\tau}\cdot \boldsymbol{u}) \cdot \nabla w_e - \frac{\gamma(\gamma-1)\mathrm{Ma}^2}{\mathrm{Fr}^2} \rho (\boldsymbol{\hat{e}}_g \cdot \boldsymbol{u})\, w_e \right )  d\Omega= 0.
\end{align}

\textbf{Equation of State}

\begin{align}
    \mathcal{F}_q(\rho, e, p;w_\rho)=\int_\Omega ((\gamma Ma^2p + 1) - \rho e)\,w_\rho \, d\Omega = 0.
\end{align}

Altogether,
\begin{align}
\mathcal{F}(w ; \psi) =   \mathcal{F}_{\rho} + \mathcal{F}_{\boldsymbol{u}} + \mathcal{F}_{e} + \mathcal{F}_q.
\label{eq:weak_form_compressible}
\end{align}

For pedagogical purposes, we include detailed derivations of the non-dimensionalized weak forms for both Boussinesq and compressible flows in \ref{sec:app-derivations}.

\section{Numerical methodology}
\label{sec:methodology}

\begin{figure}
    \centering
    \includegraphics[width = 0.4\textwidth]{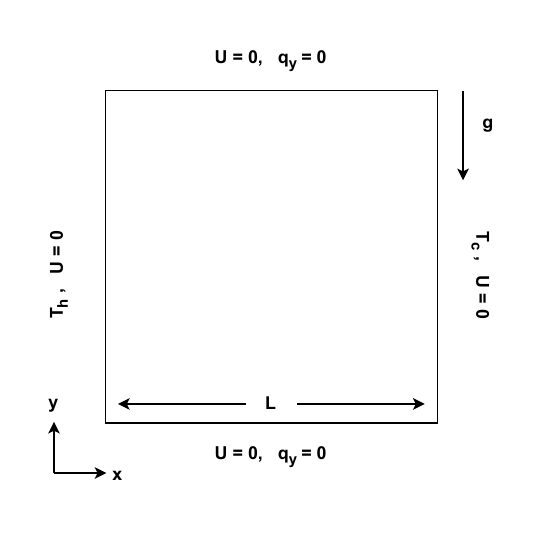}
    \caption{Domain geometry and boundary conditions for differentially heated thermal cavity simulations.  A temperature gradient is set across the domain with a hot left wall (temperature $T_\text{h}$) and a cooler right wall (temperature $T_\text{c}$).  Figure reproduced from \cite{mangnike2024toward}, licensed under CC BY-NC-SA 4.0.} 
    \label{fig:Domain}
\end{figure}

Our primary test case is the differentially heated square cavity problem \cite{de1983natural} in two or three spatial dimensions, the canonical benchmark for natural convection. 
The domain $\Omega = [0, L]^d$ ($d=2,3$) consists of adiabatic horizontal walls and vertical walls held at constant temperatures $T_h$ and $T_c$ (Figure \ref{fig:Domain}). 
The flow is governed by the non-dimensionalized Rayleigh number ($Ra$), Prandtl number ($Pr$), Mach number ($Ma$), and Froude number ($Fr$) (as appropriate, in the Boussinesq or compressible flow cases).
We employ a unified, monolithic finite element formulation implemented in the FEniCSx ecosystem (DOLFINx, FFCx, PETSc). 
A distinct feature of this approach is the use of a fully coupled implicit solver for all variables simultaneously in a mixed element space. 
This avoids the splitting errors inherent in partitioned pressure-correction schemes and allows for robust convergence at large time steps ($\Delta t$) without the need for ad-hoc stabilization (such as SUPG/PSPG) or turbulence modeling in the regimes considered.
The following subsections summarize the mesh and function spaces, temporal discretization, and the nonlinear/linear solvers employed. 

\subsection{Mesh, geometry, and function spaces}
Let the domain $\Omega\subset\mathbb{R}^d$ ($d=2,3$) be partitioned into a shape-regular mesh
$\mathcal{T}_h=\{K\}$ of triangles (2D) or hexahedra/brick cells (3D).
Geometry is represented isoparametrically: each cell $K\in\mathcal{T}_h$ is the image of a reference element $\hat K$ under a mapping
\[
\mathbf{x} \;=\; \Phi_K(\hat{\mathbf{x}}) \;=\;
\sum_{i} \mathbf{X}_{K,i}\,\hat{\phi}_i(\hat{\mathbf{x}}),
\]
where $\{\mathbf{X}_{K,i}\}$ are geometry nodes, $\{\hat{\phi}_i\}$ are reference Lagrange basis functions, and the Jacobian $J_K=\nabla_{\hat{\boldsymbol{x}}}\Phi_K$ and its determinant are provided to all quadrature loops via automatic differentiation in FFCx. 
FEniCSx stores topology and geometry (coordinate field) separately, enabling efficient mesh adaptation and parallel repartitioning.

For Boussinesq flow, we employ the mixed finite element space
\[
\mathcal{W}_h^\text{B}
\;=\;
\mathbf{V}_h \times \Theta_h \times P_h
\;=\;
[\mathbb{P}_2]^d \times \mathbb{P}_1 \times \mathbb{P}_1
\]
for velocity $\boldsymbol u$, temperature $T$, and pressure $p$, respectively.
Here $[\mathbb{P}_2]^d$ denotes vector-valued quadratic Lagrange elements for velocity (the Taylor–Hood pairing ensures inf-sup stability), while $\Theta_h$ and $P_h$ are scalar linear Lagrange spaces ($\mathbb{P}_1$) for temperature and pressure.
For compressible flow, we extend the mixed space to include density and internal energy:
\[
\mathcal{W}_h^\text{C}
\;=\;
\mathbf{V}_h \times E_h \times R_h \times P_h
\;=\;
[\mathbb{P}_2]^d \times \mathbb{P}_1 \times \mathbb{P}_1 \times \mathbb{P}_1
\]
for velocity $\boldsymbol u$, specific internal energy $e$ (proportional to temperature), density $\rho$, and total pressure $p$, respectively.
The vector-valued quadratic Lagrange elements for velocity $[\mathbb{P}_2]^d$ capture momentum and viscous effects accurately, and $E_h$, $R_h$ and $P_h$ are scalar linear Lagrange spaces ($\mathbb{P}_1$) for $e$, $\rho$ and $p$. 
Each subspace is $H^1$-conforming and assembled into a globally conforming mixed space in DOLFINx with consistent orientation and numbering.

\subsection{Time discretization}
Time integration is performed with a first-order implicit backward Euler scheme. 
For each time step, we pose a Galerkin weak problem: find the \emph{trial} field
\(w^{n+1}\in\mathcal W_h\)
such that
\[
  \mathcal{F}^{n+1}(w^{n+1};z)=0 \quad \text{for all \emph{test} functions } z\in\mathcal W_h .
\]
Here \(F^{n+1}(\,\cdot\,;\,\cdot\,)\) is the discrete residual, obtained by inserting the finite element unknowns into the weak form of the PDEs and testing against \(z\).
Temporal derivatives are approximated by first-order backward differences.
For the Boussinesq approximation, with $w^{n+1}=(\boldsymbol u^{n+1},T^{n+1},p^{n+1})$, we have
\begin{equation}
\frac{\partial \boldsymbol{u}}{\partial t}\bigg|^{n+1} \approx \frac{\boldsymbol{u}^{n+1} - \boldsymbol{u}^n}{\Delta t}, \qquad
\frac{\partial T}{\partial t}\bigg|^{n+1} \approx \frac{T^{n+1} - T^n}{\Delta t}.
\label{eq:time_bouss}
\end{equation}

For the compressible formulation, with $w^{n+1}=(\boldsymbol u^{n+1},e^{n+1},\rho^{n+1},p^{\,n+1})$, define the non-dimensionalized kinetic energy coefficient
\[
\alpha_{\mathrm{ke}} \;=\; \tfrac12\,\gamma(\gamma-1)\,\mathrm{Ma}^2.
\]
The specific \emph{total} energy is then
\[
E^{n+1} \;=\; e^{\,n+1} + \alpha_{\mathrm{ke}}\,\boldsymbol u^{\,n+1}\!\cdot\boldsymbol u^{\,n+1},
\qquad
E^{n} \;=\; e^{\,n} + \alpha_{\mathrm{ke}}\,\boldsymbol u^{\,n}\!\cdot\boldsymbol u^{\,n},
\]
and the conserved variable in the energy equation is the total energy density $\rho E$.
The temporal derivatives are
\begin{align}
\frac{\partial \boldsymbol{u}}{\partial t}\bigg|^{n+1} &\approx \frac{\boldsymbol{u}^{n+1} - \boldsymbol{u}^n}{\Delta t}, \qquad
\frac{\partial(\rho\boldsymbol{u})}{\partial t}\bigg|^{n+1} \approx \frac{\rho^{n+1}\boldsymbol{u}^{n+1} - \rho^n\boldsymbol{u}^n}{\Delta t},\\
\frac{\partial \rho}{\partial t}\bigg|^{n+1} &\approx \frac{\rho^{n+1} - \rho^n}{\Delta t},
\qquad
\frac{\partial(\rho E)}{\partial t}\bigg|^{n+1} \approx \frac{\rho^{n+1}E^{n+1} - \rho^n E^{n}}{\Delta t}.
\label{eq:time_derivatives}
\end{align}
Thus, the scheme is fully implicit in all conserved variables $(\rho, \rho\boldsymbol u, \rho E)$, while $e$ itself remains the thermodynamic variable used in heat conduction and the equation of state.

\subsection{Unified global weak formulation}

The Boussinesq system is discretized as a single variational residual acting on the mixed space $\mathcal{W}_h^\text{B}$. As before, $\boldsymbol{\hat e}_g$ denotes the unit vector in the direction of gravity. The residual is
\begin{equation}
\begin{aligned}
\mathcal{F}^\text{B,\,n+1}(w^{n+1}; z)
&= \int_{\Omega} \Big[
\frac{\boldsymbol{u}^{\,n+1}-\boldsymbol{u}^{\,n}}{\Delta t}\cdot\boldsymbol{v}
\;+\; \frac{T^{\,n+1}-T^{\,n}}{\Delta t}\,\theta
\Big]\,\mathrm{d}x
\\
&\quad + \mathcal{N}^\text{B}(w^{n+1}; z),
\end{aligned}
\label{eq:residual_bouss}
\end{equation}
where the spatial part is
\begin{equation}
\begin{aligned}
\mathcal{N}^\text{B}(w^{n+1}; z)
&= \int_{\Omega} \Big[
(\nabla\!\cdot\boldsymbol{u}^{\,n+1})\,q
\;+\; (\boldsymbol{u}^{\,n+1}\!\cdot\!\nabla\boldsymbol{u}^{\,n+1})\cdot\boldsymbol{v}
\;-\; p^{n+1}\,\nabla\!\cdot\boldsymbol{v}
\\
&\qquad +\; \mathrm{Pr}\,\nabla\boldsymbol{u}^{\,n+1}:\nabla\boldsymbol{v}
\;+\; \mathrm{Pr}\,\mathrm{Ra}\,T^{\,n+1}(\boldsymbol{\hat e}_g\cdot\boldsymbol{v})
\Big]\,\mathrm{d}x
\\
&\quad + \int_{\Omega} \Big[
(\boldsymbol{u}^{\,n+1}\!\cdot\nabla T^{\,n+1})\,\theta
\;+\; \nabla T^{\,n+1}\!\cdot\nabla\theta
\Big]\,\mathrm{d}x.
\end{aligned}
\label{eq:N_bouss}
\end{equation}

The fully coupled compressible system is discretized as a single variational residual acting on the mixed space $\mathcal{W}_h^\text{C}$. Let $\boldsymbol{\sigma}(\boldsymbol{u})=\nabla \boldsymbol{u}+\nabla \boldsymbol{u}^{\top}-\tfrac{2}{3}(\nabla\!\cdot \boldsymbol{u})\boldsymbol{I}$ denote the deviatoric stress. The residual is
\begin{equation}
\begin{aligned}
\mathcal{F}^\text{C,\,n+1}(w^{n+1}; z)
&= \int_{\Omega} \Big[
\frac{\rho^{\,n+1}-\rho^{\,n}}{\Delta t}\,q
\;+\; \frac{\rho^{\,n+1}\boldsymbol{u}^{\,n+1}-\rho^{\,n}\boldsymbol{u}^{\,n}}{\Delta t}\cdot\boldsymbol{v}
\;+\; \frac{\rho^{\,n+1}E^{\,n+1}-\rho^{\,n}E^{\,n}}{\Delta t}\,w_e
\Big]\,\mathrm{d}x
\\
&\quad + \mathcal{N}^\text{C}(w^{n+1}; z)
\;+\; \mathcal{B}(w^{n+1}; z),
\end{aligned}
\label{eq:residual_split_totalE}
\end{equation}
where $\mathcal{N}^\text{C}$ collects all spatial operators (mass, momentum, total–energy transport, equation of state) and $\mathcal{B}$ collects the body-force contributions (momentum buoyancy and the associated gravity power in the energy equation).
The spatial part is given by
\begin{equation}
\begin{aligned}
\mathcal{N}^\text{C}(w^{n+1}; z)
&= \int_{\Omega} \Big[
\nabla\!\cdot(\rho^{\,n+1}\boldsymbol{u}^{\,n+1})\,q
\;+\; \nabla\!\cdot\big(\rho^{\,n+1}\boldsymbol{u}^{\,n+1}\!\otimes\!\boldsymbol{u}^{\,n+1}\big)\cdot\boldsymbol{v}
\;-\; p^{n+1}\,\nabla\!\cdot\boldsymbol{v}
\\
&\qquad +\; \frac{1}{\mathrm{Re}}\,\boldsymbol{\sigma}(\boldsymbol{u}^{\,n+1}):\nabla\boldsymbol{v}
\Big]\,\mathrm{d}x
\\
&\quad + \int_{\Omega} \Big[
\nabla\!\cdot(\rho^{\,n+1}E^{\,n+1}\boldsymbol{u}^{\,n+1})\,w_e
\;+\; \frac{\gamma}{\mathrm{Re}\,\mathrm{Pr}}\,\nabla e^{\,n+1}\!\cdot\nabla w_e
\\
&\qquad -\; (\gamma-1)\big(\gamma \mathrm{Ma}^2 p^{n+1}+1\big)\,\boldsymbol{u}^{\,n+1}\!\cdot\nabla w_e
\;+\; \frac{\gamma(\gamma-1)\mathrm{Ma}^2}{\mathrm{Re}}\big(\boldsymbol{\sigma}(\boldsymbol{u}^{\,n+1})\cdot\boldsymbol{u}^{\,n+1}\big)\cdot\nabla w_e
\Big]\,\mathrm{d}x
\\
&\quad + \int_{\Omega}
\big(\gamma \mathrm{Ma}^2 p^{n+1}+1-\rho^{\,n+1}e^{\,n+1}\big)\,w_\rho\,\mathrm{d}x,
\end{aligned}
\label{eq:N_part_totalE}
\end{equation}
and the buoyancy term is
\begin{equation}
\mathcal{B}(w^{n+1}; z)
= - \int_{\Omega} \frac{1}{\mathrm{Fr}^2}\,\rho^{\,n+1}\,\big(\boldsymbol{\hat e}_g\cdot\boldsymbol{v}\big)\,\mathrm{d}x
\;-\; \int_{\Omega} \frac{(\gamma-1)\gamma\,\mathrm{Ma}^2}{\mathrm{Fr}^2}\,\rho^{\,n+1}\,\big(\boldsymbol{\hat e}_g\cdot\boldsymbol{u}^{\,n+1}\big)\,w_e\,\mathrm{d}x.
\end{equation}

In this formulation, the conserved scalar is the total energy density $\rho E$, but the thermodynamic scalar that appears in heat conduction and the equation of state is the internal energy $e$. 
The system is still treated as a single global mixed problem in the unknowns $(\boldsymbol u,e,\rho,p)$, and the remarks on robustness for large $\Delta t$, coarse meshes, and the absence of additional stabilization or special preconditioners remain unchanged.

\subsection{Monolithic nonlinear and linear solvers}
On each time step, we solve the discrete nonlinear system by a damped Newton method. 
Given an iterate \(w^{(k)}\) at \(t^{n+1}\), we locally linearize and assemble the residual vector and Jacobian:
\begin{equation}
\mathcal{R}^{(k)}_i \;=\; \mathcal{F}^{n+1}(w^{(k)};\psi_i),\qquad
\boldsymbol{J}^{(k)} \;=\; \frac{\partial \mathcal{F}^{n+1}}{\partial w}\Big|_{w^{(k)}}.
\end{equation}

A critical advantage of the FEniCSx framework is the use of the Unified Form Language (UFL) to derive the exact Jacobian matrix via automatic differentiation\footnote{We are aware that automatic differentiation does not necessarily simplify expressions, which in some cases can lead to substantial numerical errors, as studied in \citet{johnson2025software}.}. 
In code, this is expressed simply as \texttt{J = ufl.derivative(F, w)}, where \texttt{F} is the UFL representation of the weak residual $\mathcal{F}^{n+1}$ and \texttt{w} is the mixed trial function.
UFL applies symbolic differentiation rules (product rule, chain rule) to all terms in the residual—including nonlinear products ($\rho\boldsymbol{u}\otimes\boldsymbol{u}$, $\rho E\boldsymbol{u}$), viscous dissipation ($\boldsymbol{\sigma}:\nabla\boldsymbol{u}$), and the equation-of-state coupling—to produce the exact linearized bilinear form.

Because the Jacobian is differentiated from the same discrete residual that generates $\mathcal{R}^{(k)}$, it is algebraically consistent and includes all cross-coupling terms (e.g., $\partial \mathcal{F}_\rho/\partial p$, $\partial \mathcal{F}_e/\partial \rho$, $\partial \mathcal{F}_u/\partial p$) automatically.
This completeness supports rapid Newton convergence even in stiff regimes (e.g., low-Mach compressible flow) and, in the regimes tested by our numerical experiments, it eliminates the need for artificial stabilization schemes, achieving global convergence strictly through standard line-search methods.
In contrast, many existing solvers either use finite-difference Jacobian approximations (prone to truncation error), hand-derived simplified Jacobians (which often omit off-diagonal blocks to reduce implementation complexity), or operator-splitting schemes (which decouple the system and require SUPG/PSPG-type stabilization to compensate for splitting errors).

For the compressible formulation with $w=(\boldsymbol{u}, e, \rho, p)$, the Jacobian has the natural $4\times 4$ block structure
\begin{equation}
\boldsymbol{J}^{(k)}
\;=\;
\begin{bmatrix}
J_{uu} & J_{ue} & J_{u\rho} & J_{up} \\[0.5em]
J_{eu} & J_{ee} & J_{e\rho} & J_{ep} \\[0.5em]
J_{\rho u} & J_{\rho e} & J_{\rho\rho} & J_{\rho p} \\[0.5em]
J_{pu} & J_{pe} & J_{p\rho} & J_{pp}
\end{bmatrix},
\label{eq:jacobian_block}
\end{equation}
where each block captures the sensitivity of one residual component to variations in another variable.
The diagonal blocks ($J_{uu}$, $J_{ee}$, etc.) contain diffusion, time-derivative, and linearized convection terms.
The off-diagonal blocks encode key physical couplings: $J_{up}$ is the pressure-gradient term in momentum, $J_{u\rho}$ arises from inertia and buoyancy ($\rho\boldsymbol{u}\cdot\nabla\boldsymbol{u}$ and $\rho\boldsymbol{g}$), $J_{eu}$ captures kinetic energy and viscous dissipation feedback, $J_{ep}$ represents the pressure-work term, and $J_{p\rho}, J_{pe}, J_{pp}$ encode the equation-of-state linearization $p = (\gamma-1)\rho e$.
For Boussinesq flow, the structure simplifies to a $3\times 3$ block matrix for $(u, T, p)$.
All blocks are computed automatically by UFL and assembled element-wise via FFCx-generated kernels.

After assembling the Jacobian and residual, we then compute the standard Newton update $\delta w^{(k)}\in\mathcal W_h$ by solving the linearized problem
\begin{equation}
\boldsymbol{J}^{(k)}\,\delta w^{(k)} \;=\; -\,\mathcal{R}^{(k)}.
\label{eq:newton_linear}
\end{equation}
Rather than directly using $\delta w^{(k)}$ to update the solution iterate, we perform an Armijo line search \citep{armijo1966minimization} to find an optimal damped update
\begin{equation}
w^{(k+1)} \;=\; w^{(k)} + \alpha^{(k)}\,\delta w^{(k)},\qquad \alpha^{(k)}\in(0,1] .
\end{equation}
Our solution loop (see Algorithm \ref{alg:solver-loop} for pseudocode for the compressible case) iterates until convergence, which we define as either of the following two conditions being satisfied:
\begin{align}
\|\mathcal{R}^{(k)}\| &\le \text{rtol}\,\|\mathcal{R}^{(0)}\| + \text{atol}, &
\|\delta w^{(k)}\| &\le \text{rtol}_\delta\,\|w^{(k)}\| + \text{atol}_\delta .
\end{align}
In our implementation, the nonlinear residual tolerances are strictly enforced at $\text{rtol} = \text{atol} = 10^{-9}$. $ \text{rtol}_\delta$ and $ \text{atol}_\delta$ are maintained at the standard PETSc SNES defaults (step tolerance $= 10^{-8}$). The matrices and vectors in Equation \ref{eq:newton_linear} are assembled element–wise by FFCx–generated kernels using the isoparametric map, with essential boundary conditions enforced through consistent lifting in the residual and row modifications of the Jacobian on the constrained subspaces. 
We consider solving Equation \ref{eq:newton_linear} via direct and iterative solvers:

\paragraph{Direct solver (baseline)} We solve $\boldsymbol{J}^{(k)}\,\delta w^{(k)} = -\mathcal{R}^{(k)}$ using multifrontal sparse LU factorization via MUMPS (MUltifrontal Massively Parallel Sparse direct solver) \citep{amestoy2013mumps}:
\[
P\,\boldsymbol{J}^{(k)}\,Q = L\,U,
\]
with permutations $P,Q$ for sparsity and numerical stability. 
The Newton update is obtained by forward and backward triangular solves. 
This approach is robust, requires no preconditioning tuning, and serves as the reference baseline for smaller problems where direct factorization is feasible.

\paragraph{Iterative solver (scalability)} For larger systems, we solve the linearized problem in Equation \ref{eq:newton_linear} by  employing right-preconditioned FGMRES (Flexible GMRES) \citep{saad1993flexible} with Algebraic Multigrid (AMG) preconditioning from the Hypre library (BoomerAMG \citep{yang2002boomeramg}):
\begin{equation}
\boldsymbol{J}^{(k)}\,Z_k = V_{k+1}\,\underline{H}_k,\qquad Z_k = [M_1^{-1}v_1, \ldots, M_k^{-1}v_k],
\label{eq:fgmres}
\end{equation}
where $Z_k$ contains the preconditioned search directions, $V_{k+1}=[v_1,\ldots,v_{k+1}]$ is the orthonormal Krylov basis, $\underline{H}_k\in\mathbb{R}^{(k+1)\times k}$ is the upper Hessenberg matrix encoding the recurrence, and $M_j^{-1}$ denotes one BoomerAMG V-cycle applied to the $j$-th basis vector. 
The solution is found by minimizing the residual in the subspace spanned by $Z_k$. 
AMG constructs a hierarchy of coarse operators $A_{\ell-1}=R_\ell A_\ell P_\ell$ via Galerkin projections and employs damped Jacobi pre/post-smoothing, sparsity-preserving restriction/prolongation, and a sparse direct solve at the coarsest level. 
This reduces memory per rank and improves parallel scalability for very large matrices. 

The mixed structure of $\boldsymbol{J}^{(k)}$ is handled naturally by the single global assembly and direct/iterative linear algebra in PETSc; the interleaved mixed-space layout enables efficient sparse matrix operations without explicit field partitioning.
Parallelism is handled via PETSc's MPI-distributed vectors and matrices, with ghost value synchronization managed by the associated star forest object \cite{zhang2021petscsf}.

\medskip

\newcommand{\Div}{\nabla\!\cdot}
\newcommand{\Grad}{\nabla}
\newcommand{\Id}{\mathbf{I}}
\newcommand{\Rey}{\mathrm{Re}}
\newcommand{\Ma}{\mathrm{Ma}}
\newcommand{\Prr}{\mathrm{Pr}}
\newcommand{\Fr}{\mathrm{Fr}}
\newcommand{\dx}{\,\mathrm{d}x}
\newcommand{\sig}{\sigma(u)} 

\SetKwInput{KwData}{IC/BCs}
\begin{algorithm}[!t]

\caption{Fully Compressible Implicit Monolithic Solver}
\label{alg:solver-loop}
\KwIn{Mesh $\mathcal{T}_h$, mixed function space $\mathcal{W}=[\mathbb{P}_2]^d
\oplus\mathbb{P}_1\oplus\mathbb{P}_1\oplus\mathbb{P}_1$ for $(\boldsymbol{u},e,\rho,p)$, time step $\Delta t$, final time $T_{\max}$, tolerances $(\text{rtol}, \text{atol}, \text{rtol}_\delta, \text{atol}_\delta)$.}
\KwData{Constants $\gamma,\Prr,\Fr,\Ma,\Rey$; initial values $(\boldsymbol{u}^0,e^0,\rho^0,p^0)$; impose $\boldsymbol{u}|_{\partial\Omega}=0$ and $e|_{\partial\Omega}=T_{\mathrm{wall}}$ (Dirichlet).}

Set $t^0\gets 0$, $n\gets 0$\;

\While{$t^n < T_{\max}$}{
    \textbf{Newton initial guess:} $w^{(0)} \gets (\boldsymbol{u}^n,e^n,\rho^n,p^n)$\;
    
    \For{$k=0,1,2,\dots$}{
        \textbf{Time discretization:} update time derivatives (Eq.~\ref{eq:time_derivatives})\;

        Assemble residual vector $\mathcal{R}^{(k)}_i = \mathcal{F}^{n+1}(w^{(k)};\psi_i)$ with
        $\mathcal{F} = \mathcal{F}_{\rho} + \mathcal{F}_{\boldsymbol{u}} + \mathcal{F}_{e} + \mathcal{F}_q$
        (Eq.~\ref{eq:weak_form_compressible})\;

        Assemble exact Jacobian $\boldsymbol{J}^{(k)} = \partial\mathcal{F}^{n+1}/\partial w \big|_{w^{(k)}}$ via automatic differentiation (UFL)\;

        Solve linear system $\boldsymbol{J}^{(k)}\,\delta w^{(k)} = -\mathcal{R}^{(k)}$ with PETSc (MUMPS or FGMRES+AMG)\;

        Find optimal damping factor $\alpha^{(k)} \in (0,1]$ (Armijo line search)\;
        $w^{(k+1)} \gets w^{(k)} + \alpha^{(k)}\,\delta w^{(k)}$\;

        \If{$\| \mathcal{R}^{(k)} \| \le \mathrm{rtol}\, \| \mathcal{R}^{(0)} \| + \mathrm{atol}$ \textbf{or} $ \| \delta w^{(k)} \| \le \mathrm{rtol}_\delta\, \| w^{(k)} \| + \mathrm{atol}_\delta$}{
            \textbf{break}\;
        }
    }

    \textbf{Accept step:} $(\boldsymbol{u}^{n+1},e^{n+1},\rho^{n+1},p^{n+1})
    \gets w^{(k+1)}$\;

    $t^{n+1}\gets t^n+\Delta t$, $n\gets n+1$\;
}
\end{algorithm}

\section{Solver Experiments}
\label{sec:results}
\subsection{Convergence studies}

To establish the accuracy and reliability of our monolithic finite element solver, we performed systematic studies of both spatial (mesh) and temporal convergence.

\subsubsection{Mesh convergence}

We conducted mesh self-convergence studies for both the Boussinesq and fully compressible formulations using a sequence of systematically refined triangular meshes with resolutions of $16 \times 16$, $32 \times 32$, $64 \times 64$, $128 \times 128$, $256 \times 256$, and $512 \times 512$ elements. 
The finest mesh ($512 \times 512$) was used as the reference solution, and the relative $L^2$ errors were computed for all primary field variables at each coarser resolution. 
The mesh size $h$ represents the characteristic element spacing, defined as $h = L/N$, where $L$ is the domain size and $N$ is the number of elements in each direction.

\begin{figure}[!t]
    \centering                                                                                                
    \begin{subfigure}[t]{0.48\textwidth}
        \centering
        \includegraphics[width=\textwidth]{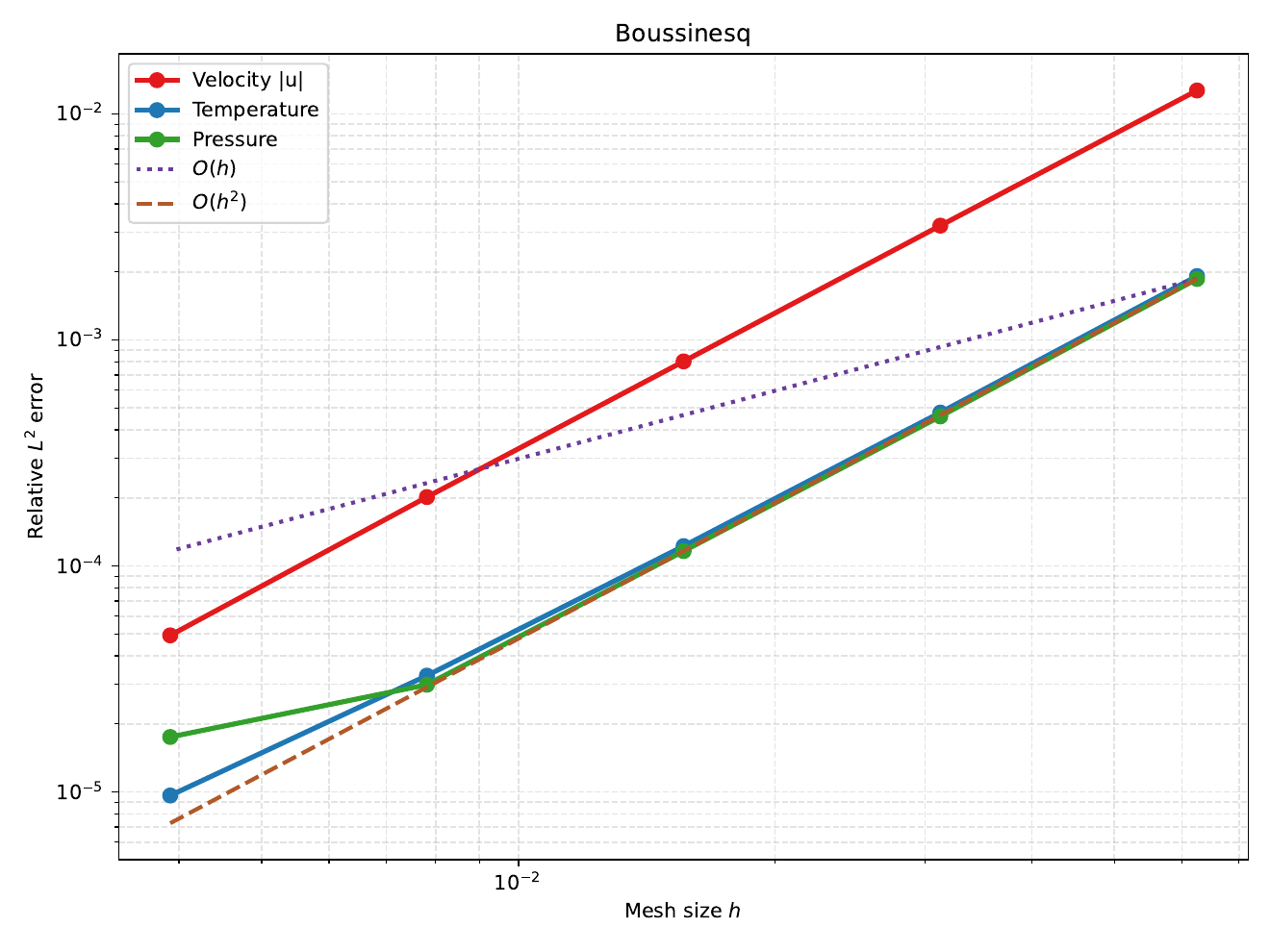}
        \caption{Boussinesq}
        \label{fig:B_conv}
    \end{subfigure}
    \hfill
    \begin{subfigure}[t]{0.48\textwidth}
        \centering
        \includegraphics[width=\textwidth]{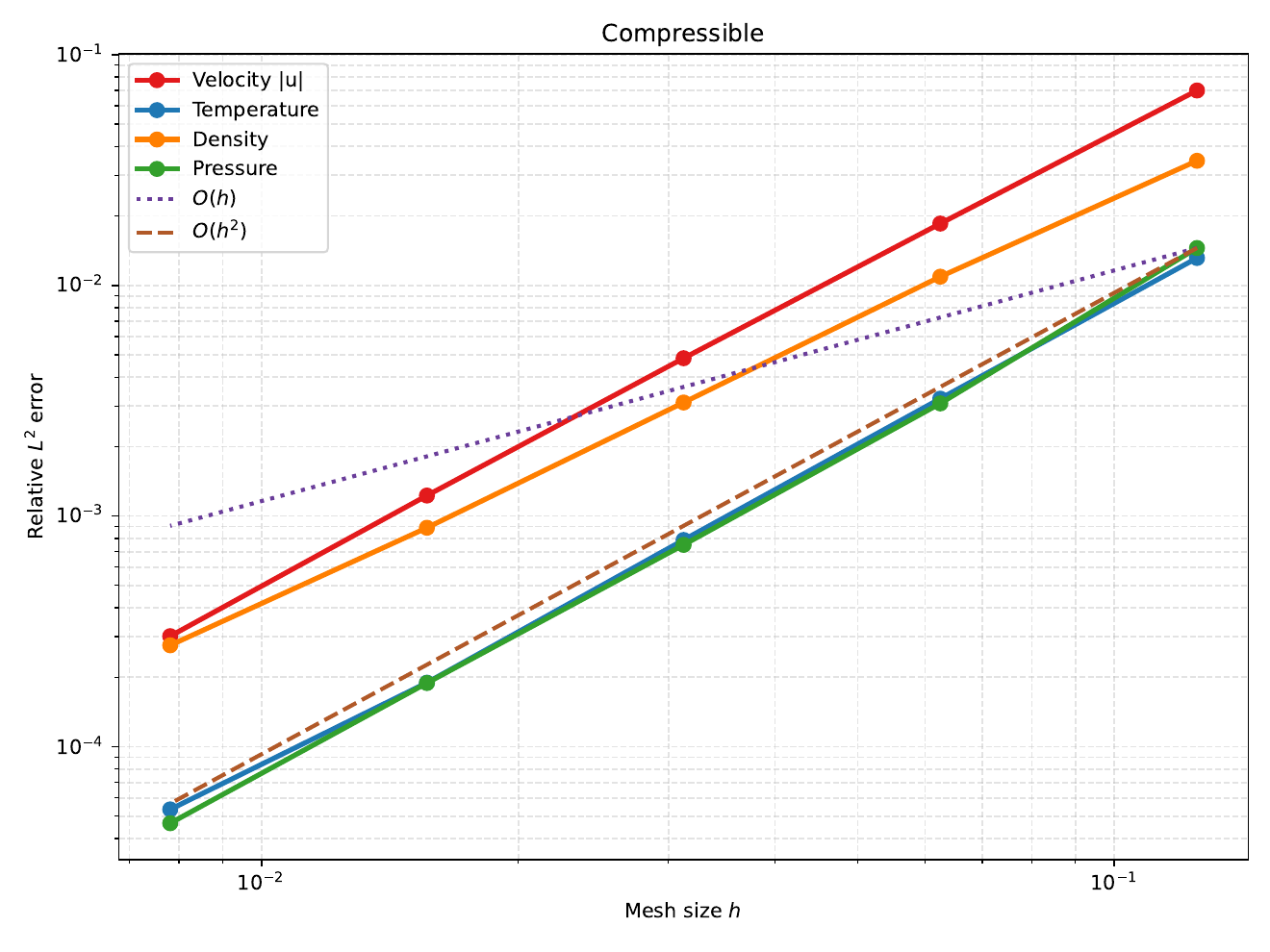}
        \caption{Compressible}
        \label{fig:C_conv}
    \end{subfigure}
    \caption{Self-convergence with mesh size for Boussinesq and fully compressible formulations. Relative $L^2$ error versus mesh size $h$ for all primary field variables, compared against the reference solution obtained with $512 \times 512$ mesh resolution.}
    \label{fig:mesh_conv_combined}
\end{figure}

Figure~\ref{fig:mesh_conv_combined} presents the mesh convergence results for both formulations. 
In Figure~\ref{fig:B_conv}, the Boussinesq approximation shows that the temperature and pressure fields converge at approximately second-order accuracy, closely following the $O(h^2)$ reference line. 
This behavior is expected given the $P_2$--$P_1$ Taylor-Hood finite element discretization, where velocity is approximated with quadratic elements and temperature/pressure with linear elements.
Figure~\ref{fig:C_conv} shows the convergence behavior of the compressible formulation, which includes an additional density field. 
All four fields—velocity magnitude, temperature, density, and pressure—demonstrate consistent second-order convergence rates across the mesh refinement sequence. The temperature and pressure errors closely track the $O(h^2)$ reference line, while the velocity field shows a slightly higher error magnitude, consistent with the Boussinesq results. 
The density field exhibits convergence comparable to temperature and pressure, validating the coupling between the equation of state and the energy equation in the compressible solver.

Comparing the two formulations, the absolute errors are slightly higher in the compressible case due to the increased complexity of the coupled variables and the absence of a low-Mach simplification.
Nevertheless, the consistent $\mathcal{O}(h^2)$ rate across both formulations confirms the accuracy and correct implementation of the chosen mixed finite element spatial discretization.

\subsubsection{Temporal convergence}

To assess the temporal discretization accuracy, we performed time-step convergence studies using both first-order (backward Euler) and second-order (BDF2) implicit time integration schemes based on the fully compressible formulation. 
Two reference cases were selected with different time step sizes: $\Delta t = 0.0005$ for the first-order scheme and $\Delta t = 0.1$ for the second-order scheme. 
Coarser time steps were systematically increased, and the relative $L^2$ errors were computed by comparing the solution at a fixed final time against the reference solution obtained with the smallest time step.

\begin{figure}[!t]
    \centering
    \begin{subfigure}[t]{0.48\textwidth}
        \centering
        \includegraphics[width=\textwidth]{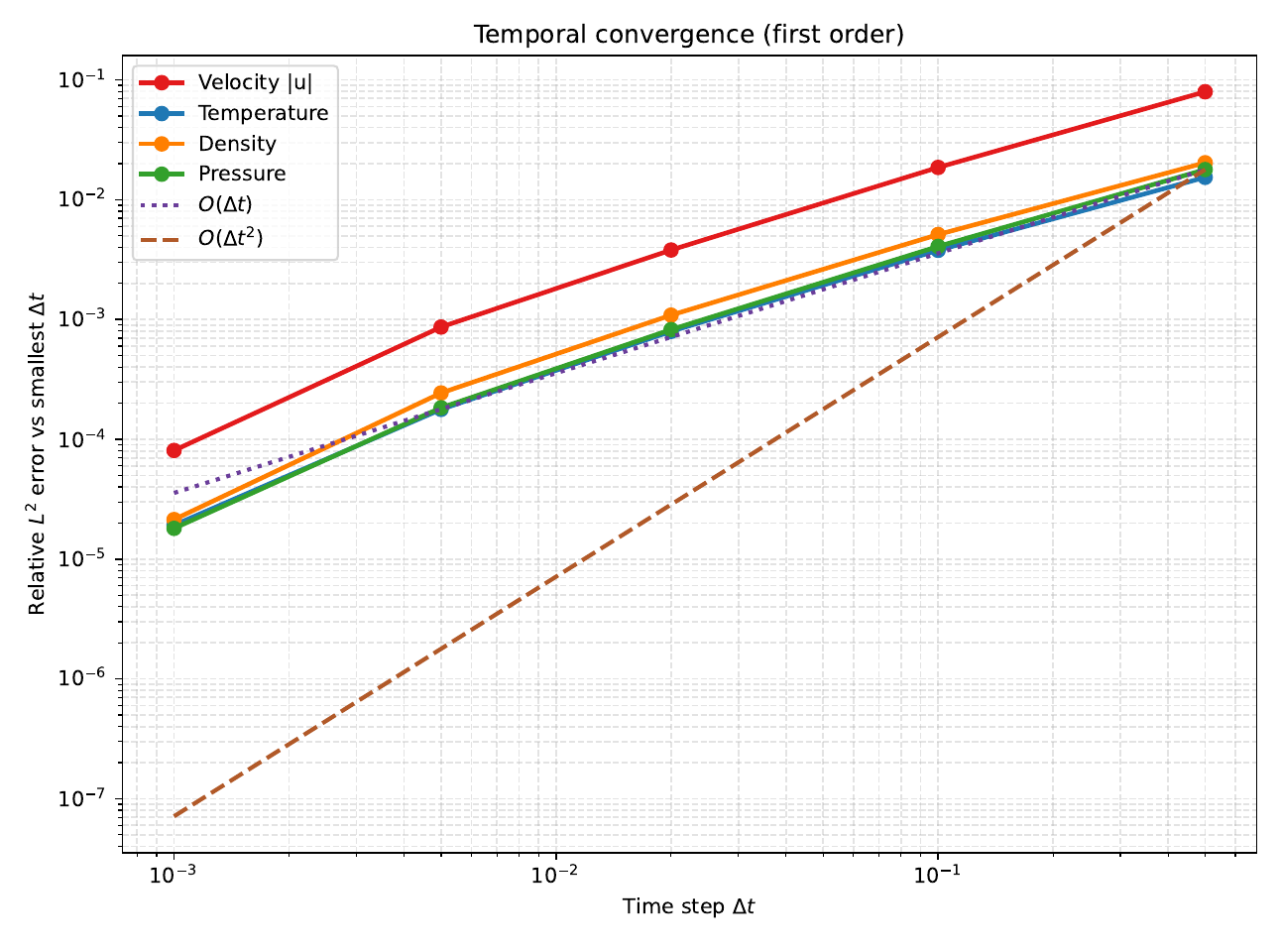}
        \caption{First-order convergence (Backward Euler)}
        \label{fig:BE}
    \end{subfigure}
    \hfill
    \begin{subfigure}[t]{0.48\textwidth}
        \centering
        \includegraphics[width=\textwidth]{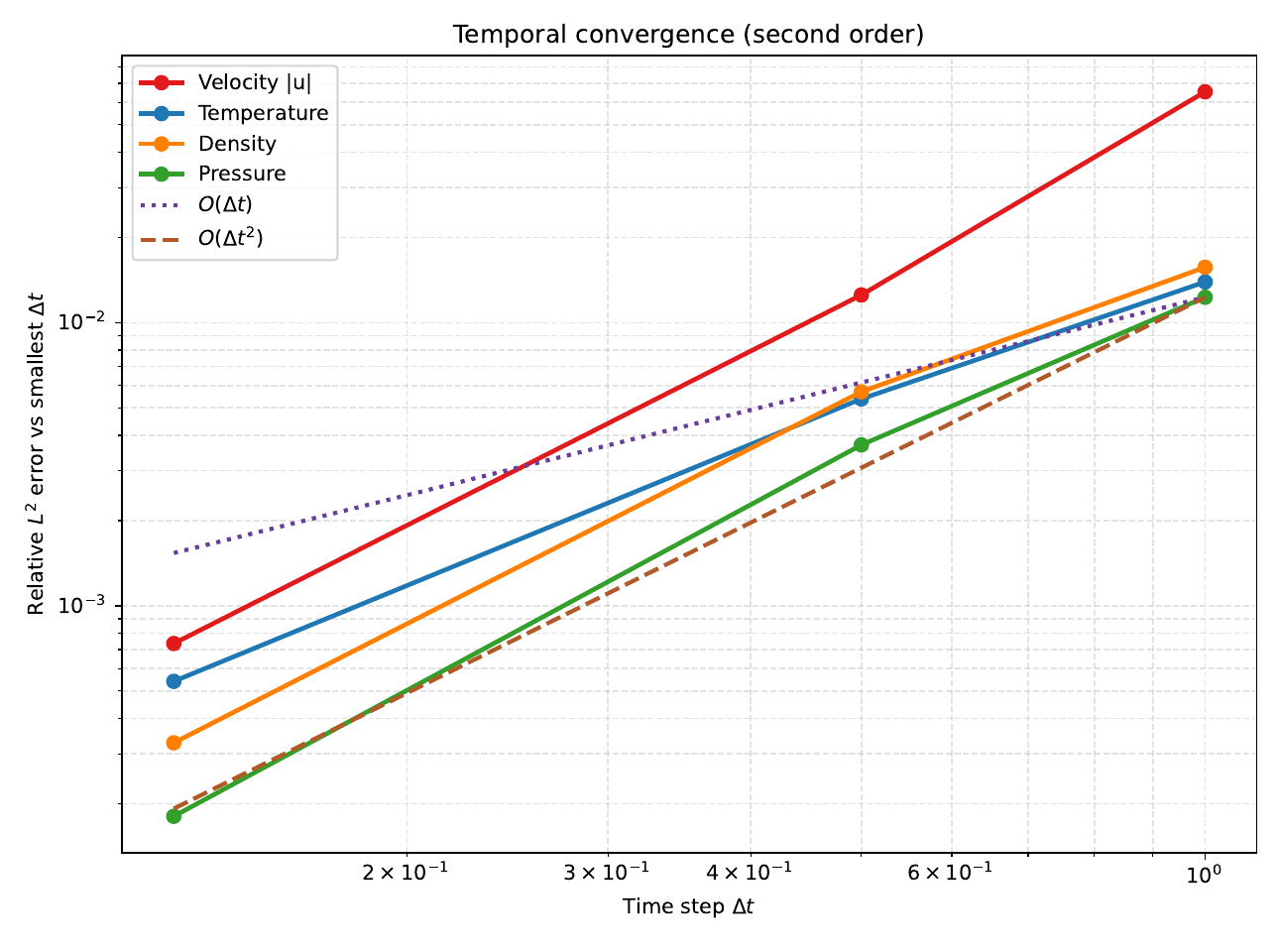}
        \caption{Second-order convergence (BDF2)}
        \label{fig:BDF2}
    \end{subfigure}
    \caption{Time-step convergence using first-order and second-order implicit time discretization schemes for compressible formulation. Relative $L^2$ error versus time step $\Delta t$ for all primary field variables.}
    \label{fig:time_conv_combined}
\end{figure}

Figure~\ref{fig:time_conv_combined} illustrates the temporal convergence behavior for both schemes. 
Figure~\ref{fig:BE} shows the first-order backward Euler scheme, where all field variables—velocity magnitude, temperature, density, and pressure—converge at the expected first-order rate, closely following the $O(\Delta t)$ reference line. 
The velocity magnitude exhibits the largest error accumulation with increasing $\Delta t$, highlighting the strong influence of the nonlinear momentum terms. 
Figure~\ref{fig:BDF2} presents the results for the second-order time integration scheme. 
All fields demonstrate clear second-order convergence, closely tracking the $O(\Delta t^2)$ reference line over the range of time steps tested. 
Thus, our solvers can achieve first- or second-order accuracy depending on the needs of the user.
Our remaining numerical results use backward Euler for sake of computational efficiency.


\subsection{Numerical validation}
We validate our monolithic mixed finite element solver against established benchmarks and commercial software (COMSOL Multiphysics 6.1) \citep{multiphysics1998introduction} based on the classic differentially heated square (or cubic) cavity, as illustrated in Figure \ref{fig:Domain}. The physical domain is defined as a unit square for 2D cases ($\Omega = [0, 1] \times [0, 1]\text{ m}$) and a unit cube for 3D cases ($\Omega = [0, 1] \times [0, 1] \times [0, 1]\text{ m}$).
The validation spans Boussinesq and fully compressible formulations in two and three dimensions, covering both field-level solutions and integrated heat transfer metrics (Nusselt numbers).
All 2D simulations employ a $64\times64$ mesh, while 3D simulations use a $24\times24\times24$ mesh.

\begin{figure}[!t]
    \centering
    \begin{subfigure}[t]{0.48\textwidth}
        \centering
        \includegraphics[width=\textwidth]{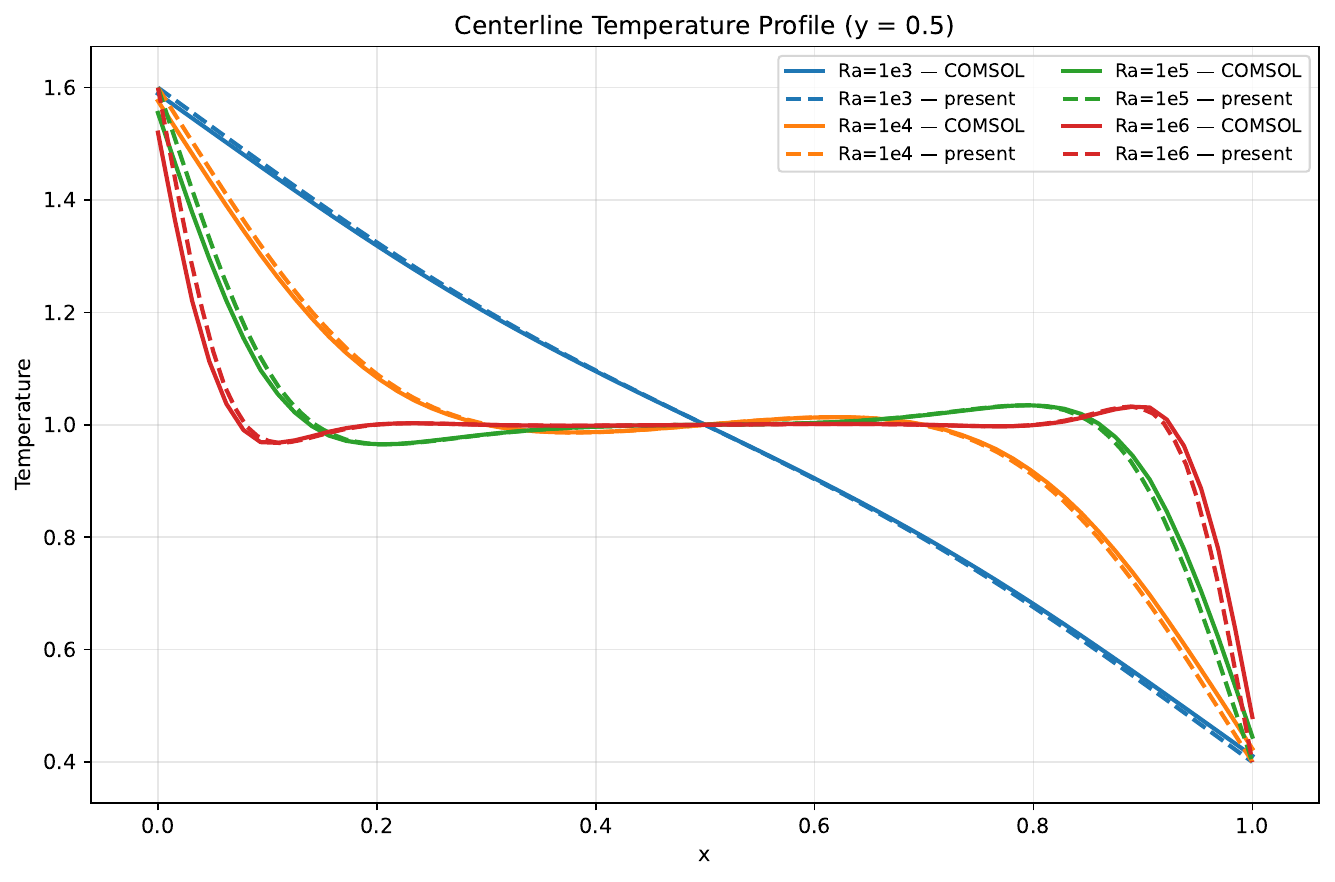}
        \caption{Boussinesq}
        \label{fig:B_COM}
    \end{subfigure}
    \hfill
    \begin{subfigure}[t]{0.48\textwidth}
        \centering
        \includegraphics[width=\textwidth]{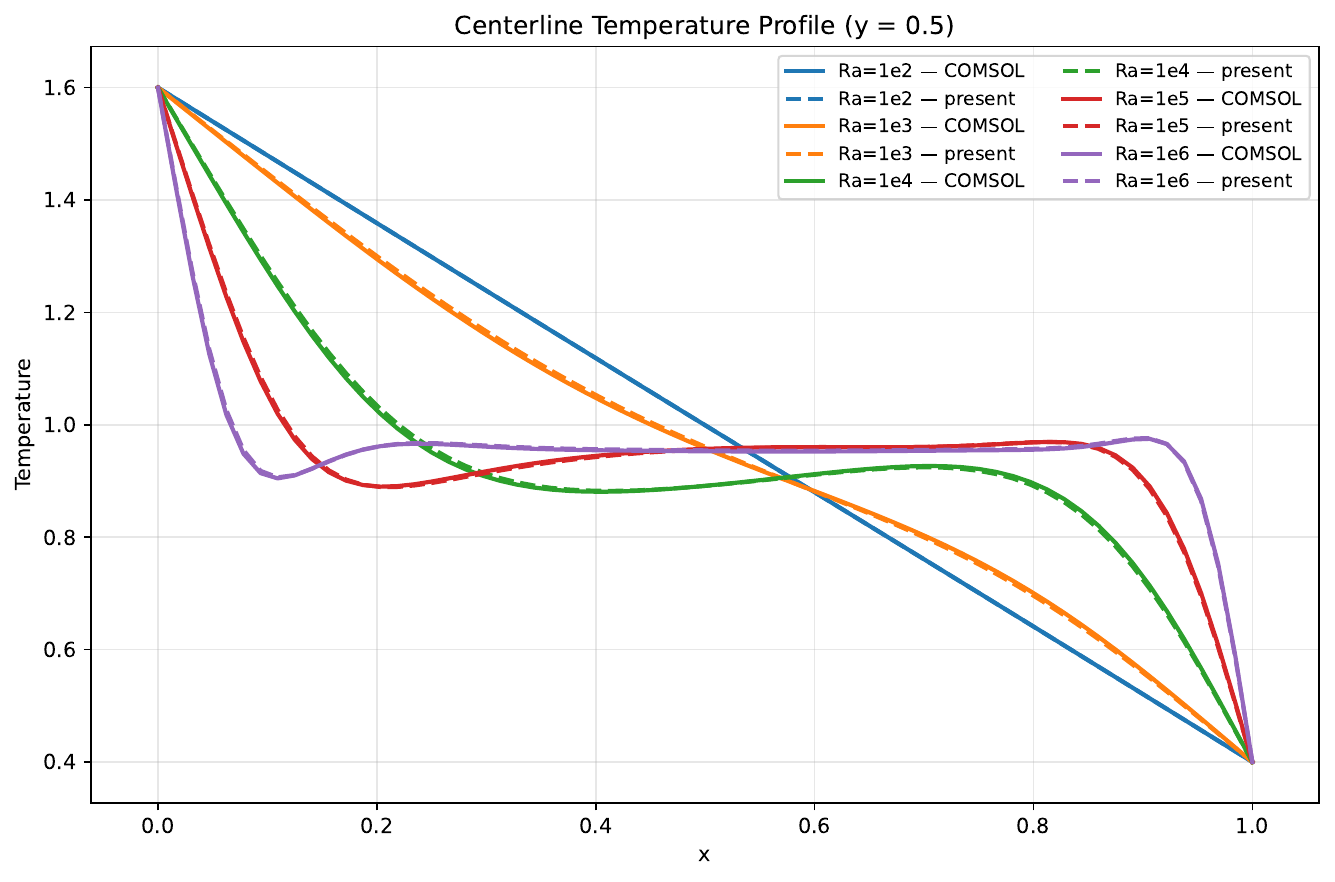}
        \caption{Compressible}
        \label{fig:C_COM}
    \end{subfigure}
    \caption{Temperature profile validation with COMSOL at $y=0.5$.}
    \label{fig:vali_centerline}
\end{figure}

\subsubsection{Reference field solutions}
\paragraph{Boussinesq natural convection}

\begin{figure}
    \centering
    \includegraphics[width = 0.9\textwidth]{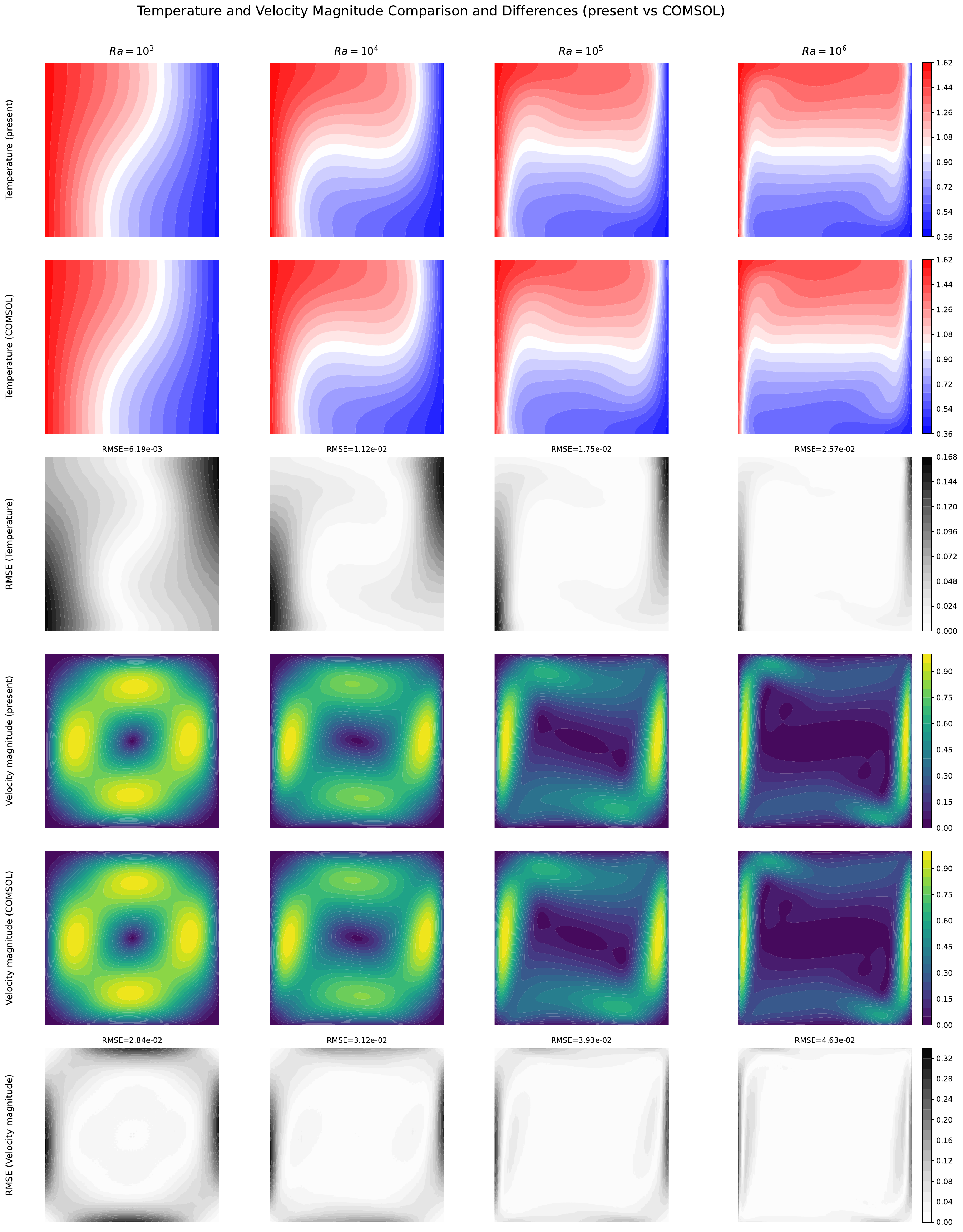}
    \caption{Validation with COMSOL (Boussinesq, 2D).  Despite using different discretizations and solvers, the two sets of solutions are visually and numerically quite similar, suggesting the validity of our implementation.}
    \label{fig:B_vali}
\end{figure}

The Boussinesq approximation is validated by comparing detailed temperature, velocity, and pressure fields computed by our solver against COMSOL Multiphysics at identical parameters and mesh resolution. 
Figure~\ref{fig:B_vali} shows a comprehensive side-by-side comparison for Rayleigh numbers $Ra \in \{10^3, 10^4, 10^5, 10^6\}$, displaying temperature contours, velocity magnitude fields, streamlines, and RMS error maps. 
The RMS error fields (Figure~\ref{fig:B_vali}, difference panels) show that discrepancies are confined to small-amplitude secondary features and boundary layer regions, with peak relative errors typically of order $10^{-3}$ to $10^{-2}$, unsurprising given the distinct discretizations and solvers used in COMSOL and the present work. 
A more stringent point-wise validation is provided by comparing temperature profiles along the horizontal centerline ($y=0.5$). 
As shown in Figure~\ref{fig:B_COM}, our solutions (solid lines) superimpose closely on the COMSOL results (dashed lines) across all Rayleigh numbers, confirming the high point-wise accuracy and regularity of the computed solutions.

\paragraph{Compressible natural convection}

\begin{figure}
    \centering
    \includegraphics[width = 0.9\textwidth]{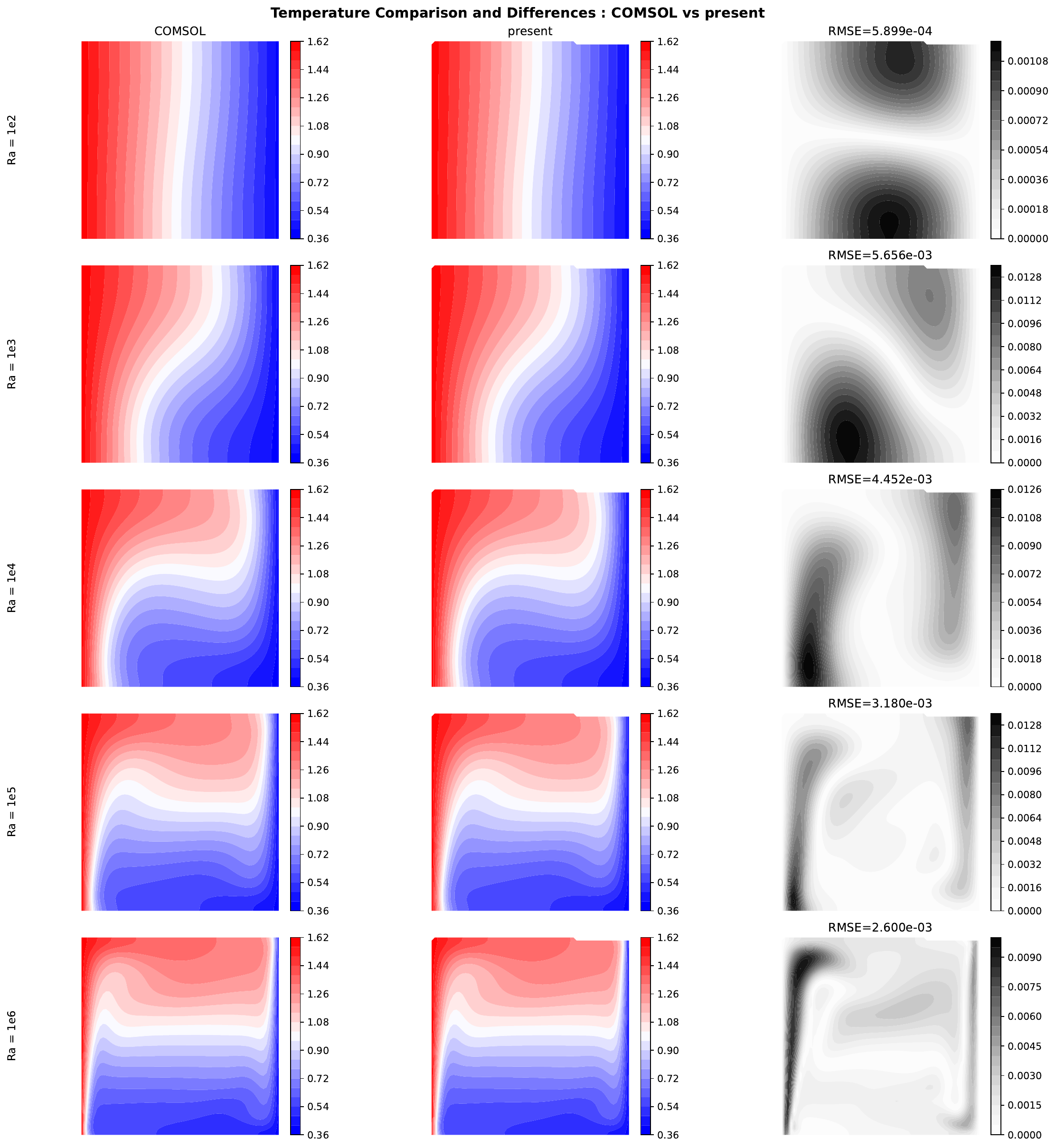}
    \caption{Validation with COMSOL (Compressible, 2D). Despite using different discretizations and solvers, the two sets of solutions are visually and numerically quite similar, suggesting the validity of our implementation.} 
    \label{fig:C_vali}
\end{figure}

\begin{figure}[!t]
    \centering
    \includegraphics[width = 1.0\textwidth]{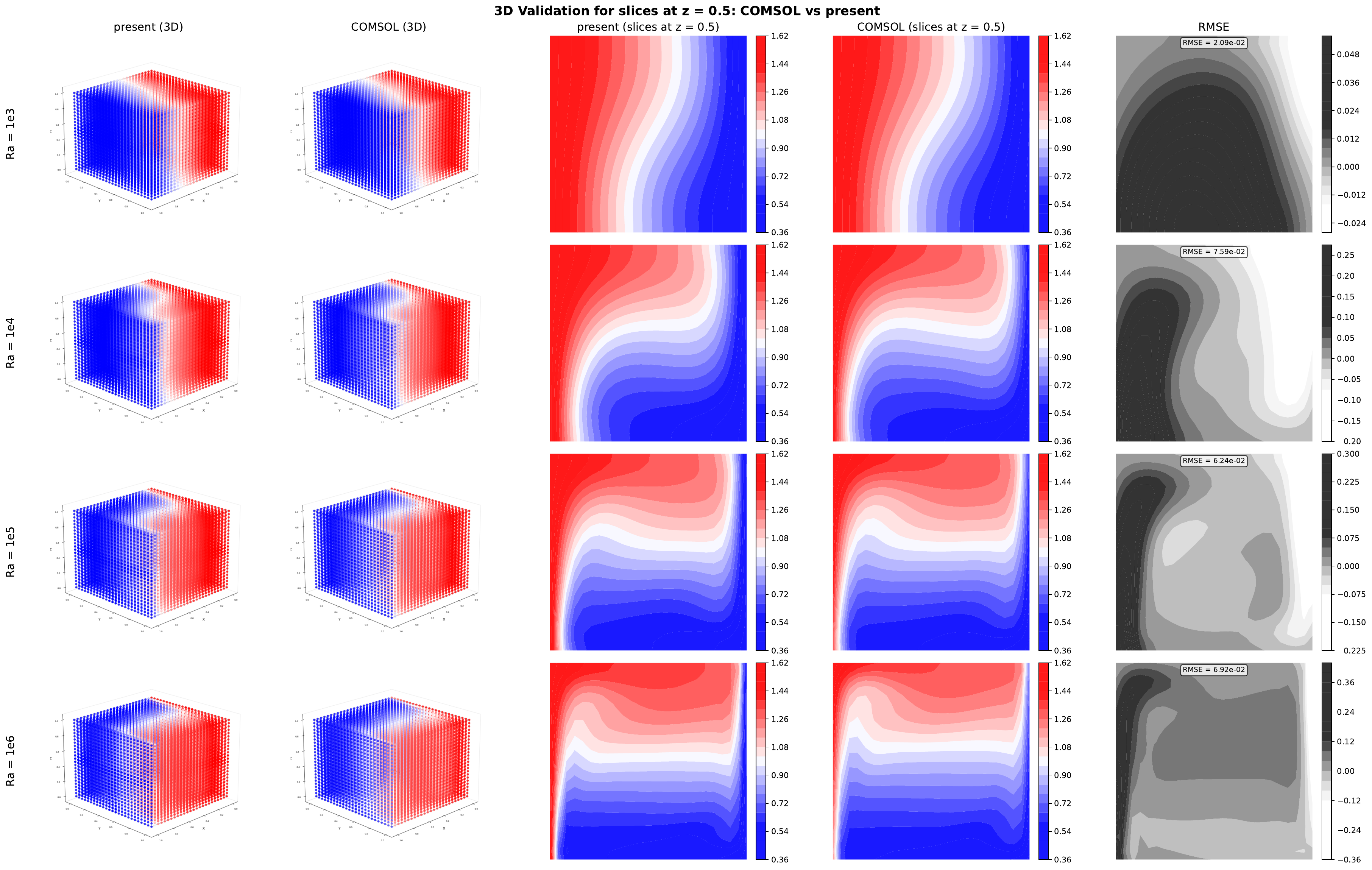}
    \caption{Validation with COMSOL (Compressible, 3D). Despite using different discretizations and solvers, the two sets of solutions are visually and numerically quite similar, suggesting the validity of our implementation.} 
    \label{fig:C_vali_3D}
\end{figure}

The fully compressible formulation is validated against COMSOL for Rayleigh numbers $Ra \in \{10^2, 10^3, 10^4, 10^5, 10^6\}$ and a corresponding Mach number $Ma \in\ \{3.84\times 10^{-4}, 5.63\times 10^{-4}, 8.27\times 10^{-4}, 1.21\times 10^{-3}, 1.78\times 10^{-3} \}$ (low-Mach regime \citep{yanaoka2025numerical}). 
Figure~\ref{fig:C_vali} presents 2D side-by-side comparisons showing temperature fields, streamlines, and RMS error maps. 
The proposed and COMSOL solutions exhibit excellent agreement, confirming that the monolithic treatment of density, momentum, and total energy correctly handles the compressible flow physics without the need for numerical stabilization (SUPG/PSPG) or specialized preconditioning.
Figure~\ref{fig:C_COM} displays the centerline temperature profiles for the compressible formulation, showing tight agreement between our monolithic compressible solver and COMSOL. 
We also validated our compressible monolithic solver for 3D cases against COMSOL.
Figure~\ref{fig:C_vali_3D} shows a mid-plane slice ($z=0.5$) comparison on the $24\times24\times24$ mesh.
The temperature fields are in close agreement across all $Ra$, with slice RMS errors of order $10^{-2}$ (rising to $\approx 7.6\times10^{-2}$ at $Ra=10^{4}$). 
These larger 3D discrepancies relative to the 2D case reflect the coarser $24^3$ mesh, which under-resolves the thermal boundary layers at higher Rayleigh numbers.
The corresponding Nusselt numbers in Table~\ref{tab:compressible_nusselt_combined} deviate from the reference by a few percent at $Ra=10^{6}$ for the same reason. 
As in 2D, we also expect modest differences from COMSOL due to the distinct discretizations and solvers.

\subsubsection{Nusselt number validation}

The Nusselt number, a dimensionless measure of convective heat transfer, is defined as
\[
\overline{Nu} = \frac{1}{L}\int_0^L \left|\frac{\partial T}{\partial x}\right| \, dy,
\]
where $L$ is the cavity height and the integral is evaluated at a vertical wall (hot or cold).
This metric integrates local temperature gradients and is sensitive to the overall flow intensity and heat transfer mechanism, making it an excellent benchmark for solver validation.
Table~\ref{tab:compressible_nusselt_combined} presents spatially-averaged Nusselt numbers on the hot and cold walls ($\overline{Nu}_{\text{hot}}$ and $\overline{Nu}_{\text{cold}}$) for the compressible formulation across a wide range of Rayleigh and Mach numbers. 
The 2D results ($64\times64$ mesh) span $Ra \in \{10^2, 10^3, 10^4, 10^5, 10^6\}$ with corresponding Mach numbers $Ma \in \{3.84\times 10^{-4}, 5.63\times 10^{-4}, 8.27\times 10^{-4}, 1.21\times 10^{-3}, 1.78\times 10^{-3}\}$, while the 3D results ($24\times24\times24$ mesh) cover $Ra \ge 10^3$ at the same Ma values.
We compare our results against those reported by \citet{Vierendeels2003BenchmarkSF} and \citet{FUSEGI19911543} for the same test cases.

In both 2D and 3D, we observe tight agreement between hot and cold wall Nusselt numbers across all Rayleigh numbers, confirming energy balance and the symmetry of the discrete problem. 
The computed values capture the expected scaling behavior of natural convection heat transfer, with $\overline{Nu}$ increasing monotonically from approximately $1.0$ at $Ra=10^2$ to $8.7$ at $Ra=10^6$ in 2D.
Our proposed solver demonstrates excellent agreement with reference values from the literature across all tested Rayleigh numbers. 
This validates the compressible formulation's ability to accurately capture heat transfer across the entire low-Mach regime from very low ($Ra=10^2$) to moderately high ($Ra=10^6$) Rayleigh numbers.

\begin{table}[htbp]
\centering
\begin{tabular}{c|c|ccc|ccc}
\toprule
& & \multicolumn{3}{c|}{\textbf{2D ($64\times64$)}} & \multicolumn{3}{c}{\textbf{3D ($24\times24\times24$)}} \\
$\textbf{Ra}$ & $\textbf{Ma}$ & $\overline{Nu}_{\mathrm{hot}}$ & $\overline{Nu}_{\mathrm{cold}}$ & $\overline{Nu}_{\mathrm{ref}}$ & $\overline{Nu}_{\mathrm{hot}}$ & $\overline{Nu}_{\mathrm{cold}}$ & $\overline{Nu}_{\mathrm{ref}}$ \\
\midrule
$10^{2}$ & $3.84\times 10^{-4}$ & 1.0010 & 1.0006 & 0.9787 & — & — & — \\
$10^{3}$ & $5.63\times 10^{-4}$ & 1.1190 & 1.1164 & 1.1077 & 1.105 & 1.103 & 1.105 \\
$10^{4}$ & $8.27\times 10^{-4}$ & 2.2282 & 2.2273 & 2.2180 & 2.211 & 2.238 & 2.302 \\
$10^{5}$ & $1.21\times 10^{-3}$ & 4.4945 & 4.5029 & 4.4800 & 4.59 & 4.50 & 4.646 \\
$10^{6}$ & $1.78\times 10^{-3}$ & 8.6833 & 8.6935 & 8.6870 & 8.497 & 8.271 & 8.012 \\
\bottomrule
\end{tabular}
\caption{Comparison of spatially-averaged Nusselt numbers with reference values \citep{Vierendeels2003BenchmarkSF,FUSEGI19911543} for compressible flows, the 3D reference values \citep{FUSEGI19911543} are from the Boussinesq cubical-cavity benchmark. Results are reported for both 2D ($64\times64$ mesh) and 3D ($24\times24\times24$ mesh) simulations across Rayleigh numbers $Ra \in \{10^2, 10^3, 10^4, 10^5, 10^6\}$ with corresponding Mach numbers $Ma \in \{3.84\times 10^{-4}, 5.63\times 10^{-4}, 8.27\times 10^{-4}, 1.21\times 10^{-3}, 1.78\times 10^{-3}\}$ (3D results available for $Ra \ge 10^3$).}
\label{tab:compressible_nusselt_combined}
\end{table}





\subsubsection{Conservation analysis}
We evaluate the discrete conservation properties of the monolithic solver by monitoring global mass, energy, and entropy balance over time. 
These tests were conducted for the 2D fully compressible cavity case at $Ra=10^{4}$ and $Ma=8.27\times 10^{-4}$ with $\gamma=1.4$, $Pr=0.71$, and $\Delta t=0.02$, using no-slip walls and fixed temperature on the vertical boundaries ($x=0$ hot, $x=1$ cold) with the mesh resolution of $64 \times 64$.

\begin{figure}[!h]
    \centering
    \includegraphics[width=0.90\textwidth]{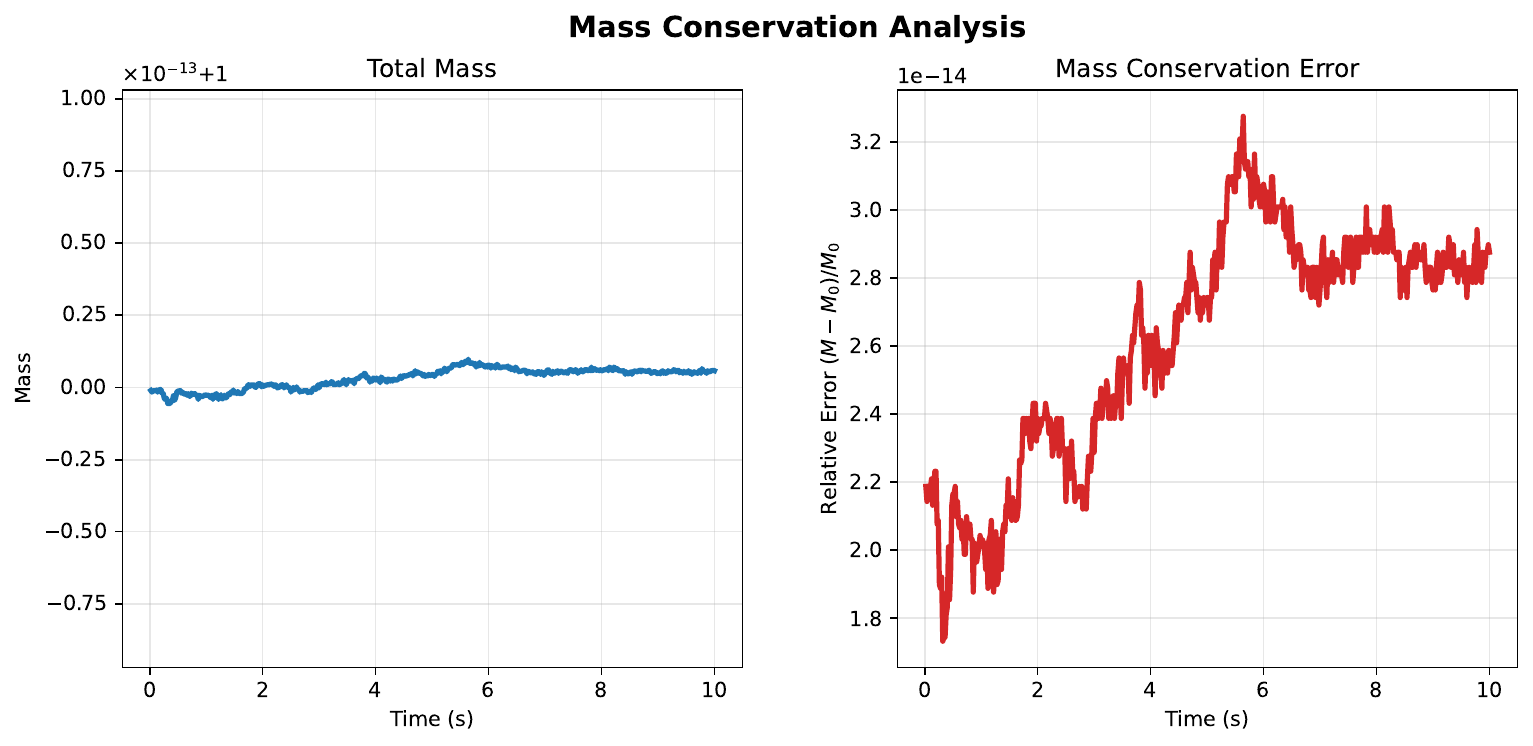}
    \caption{Mass conservation diagnostics for the 2D compressible cavity case (Mesh $64 \times 64$, $Ra=10^{4}$, $Ma=8.27\times 10^{-4}$, $\Delta t=0.02$). Left: total mass $M(t)=\int_{\Omega}\rho\,d\Omega$. Right: relative drift $(M(t)-M_0)/M_0$, showing machine-precision level conservation ($O(10^{-14})$).}
    \label{fig:mass_conservation}
\end{figure}

We compute global quantities by integrating the finite element fields over the domain $\Omega$. Because the cavity is closed (no mass flux through $\partial\Omega$), the total mass
\begin{equation}
M(t) = \int_{\Omega} \rho \, d\Omega
\end{equation}
should remain constant up to nonlinear/linear solver tolerances. 
Figure~\ref{fig:mass_conservation} reports $M(t)$ and the relative error $(M(t)-M_0)/M_0$, showing variation at approximately the machine precision-level throughout the run. 
These results confirm that the monolithic coupling and Newton solver do an excellent job of enforcing the discrete continuity equation. 

The total energy $E_{\mathrm{tot}}$ and total entropy $S$ are defined as:
\begin{align}
    E_{\mathrm{tot}} &= \int_\Omega \rho\left(e + \tfrac{1}{2}\gamma(\gamma-1)Ma^2|\mathbf{u}|^2\right)dx, \\
    S &= \int_\Omega \rho\left(c_v \ln e - (\gamma-1)\,c_v \ln \rho\right) d\Omega
\end{align}
With no-slip walls, mechanical boundary work vanishes, and the domain-integrated total-energy balance reduces to a conductive heat-flux term and a gravity power term,
\begin{equation}
\frac{dE_{\mathrm{tot}}}{dt} + Q_{\mathrm{out}}(t) - P_{g}(t) = 0,
\qquad
Q_{\mathrm{out}}(t)=\int_{\partial\Omega_{T}} \mathbf{q}\cdot\mathbf{n}\, dS,
\qquad
\mathbf{q}=-\kappa \nabla e,
\end{equation}
where $\partial\Omega_{T}$ denotes the temperature-prescribed walls and $P_{g}(t)= \int_\Omega \rho \mathbf{g}\cdot\mathbf{u}\,dx$ is the volumetric gravity power contribution induced by buoyancy in the non-dimensionalized energy equation. 
To verify the First and Second Laws of Thermodynamics, we monitor the following residuals:
\begin{align}
    \mathcal{R}_{\mathrm{energy}} &= \left| \frac{dE_{\mathrm{tot}}}{dt} + Q_{\mathrm{out}}(t) - P_{g}(t) \right|, \\
    \mathcal{R}_{\mathrm{entropy}} &= \left| \frac{dS}{dt} + \Phi_{S} - \Pi_{\mathrm{prod}} \right|.
\end{align}
Here, $\Phi_S$ is the boundary entropy flux, and $\Pi_{\mathrm{prod}}$ represents entropy production due to viscous and thermal dissipation.

\begin{figure}[!h]
    \centering
    \includegraphics[width=0.98\textwidth]{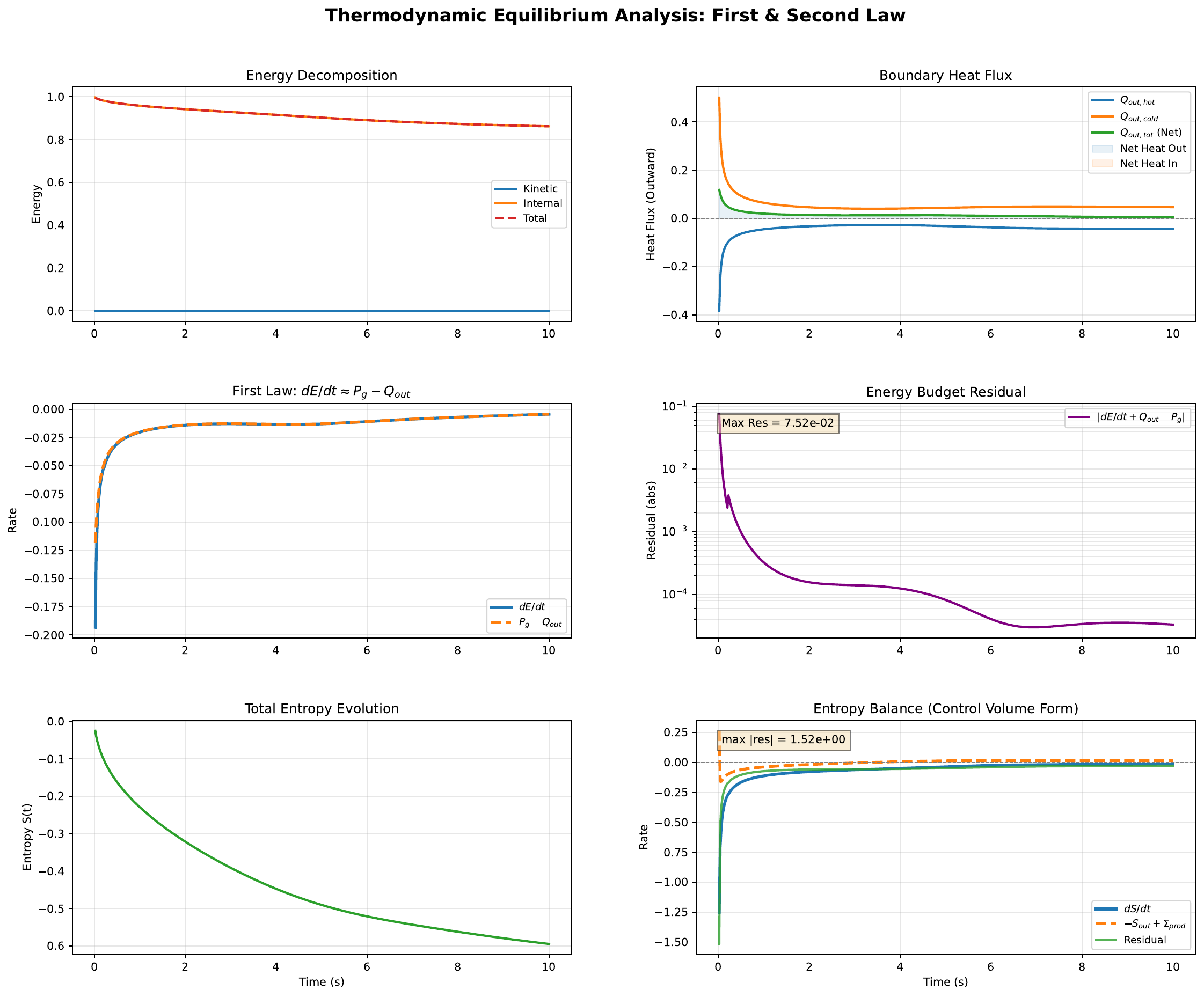}
    \caption{Analysis for First \& Second Law of Thermodynamics. Top: Energy components and boundary heat fluxes. Middle: Verification of the First Law ($dE/dt = P_g - Q_{\mathrm{net}}$) and associated residual. Bottom: Evolution of total entropy and verification of the Second Law balance.}
    \label{fig:thermodynamic_analysis}
\end{figure}

Figure~\ref{fig:thermodynamic_analysis} presents the energy and entropy evolution. 
The top row shows the system relaxing from its initial state, where the decrease in total energy is primarily driven by the internal energy component (orange), as kinetic energy (blue) remains negligible throughout the low-Mach simulation. 
This internal energy loss corresponds directly to the net boundary heat flux (top-right), which is initially large and negative (outward) but approaches zero as the system equilibrates to a quasi-steady state. 
The middle row verifies the First Law by comparing the time rate of change of total energy, $dE/dt$, against the net physical forcing, defined as the gravitational work minus the boundary heat flux ($P_{g}(t) - Q_{\mathrm{out}}(t)$). 
The substantial overlap of these curves and the decay of the residual to $O(10^{-5})$ confirm that the discrete formulation---albeit not to machine precision---correctly balances energy storage with external work and heat transfer. 
Finally, the bottom row presents the entropy balance, comparing the entropy rate $dS/dt$ with its forcing term $-\Phi_S + \Pi_{\mathrm{prod}}$ (net entropy flux plus internal production). 
After the initial transient ($t \gtrsim 0.5$~s) the entropy residual decays and remains bounded; the order-unity spike at startup, visible in the bottom-right panel, is consistent with the implicit backward-Euler discretization smoothing the discontinuous boundary heating into the interior over the first few time steps. 
We interpret the reported $\max|\mathrm{res}|$ as dominated by this transient artifact rather than as a converged measurement of entropy production, and read the entropy balance as a diagnostic of the post-transient quasi-steady regime.

\subsection{Scalability analysis}

\begin{figure}
    \centering
    \includegraphics[width = 0.75\textwidth]{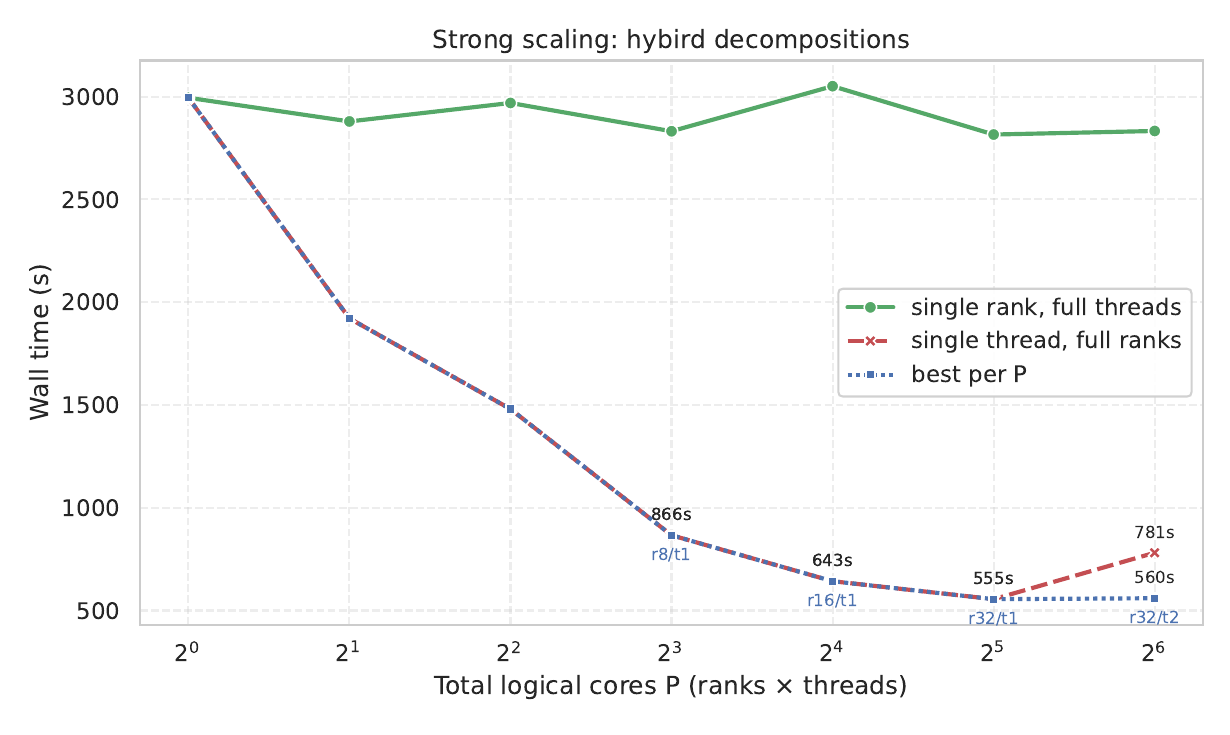}
    \caption{Strong scaling analysis for the 2D compressible flow simulation ($Ra=10^4$, $Ma=8.27\times 10^{-4}$). The plot compares wall time against total logical cores $P$ for three decomposition strategies: pure multithreading (green), pure MPI (red), and the optimal hybrid MPI+OpenMP configuration (blue). Annotated text indicates the specific rank ($r$) and thread ($t$) counts for key data points.}
    \label{fig:C_scaling}
\end{figure}
To evaluate the computational efficiency and parallel performance of the monolithic solver, we performed a strong scaling study for the 2D compressible flow simulation on a single multi-core node.
The problem size was fixed with a $2048\times2048$ mesh (approximately $4.6\times10^7$ DOFs with mixed P$_2$--P$_1$ elements), while the computational resources were systematically increased from $P=1$ to $P=64$ logical cores.
The physical parameters were selected to represent a low-Mach regime ($Ra=10^4$, $Ma=8.27 \times 10^{-4}$), and the simulation was integrated for 10 time steps with $\Delta t=0.01$.
The linear system was solved using the Flexible GMRES (FGMRES) method preconditioned with BoomerAMG. 
We investigated three parallel decomposition strategies to identify the optimal balance between distributed-memory parallelism (MPI ranks) and shared-memory parallelism (OpenMP threads).
Figure~\ref{fig:C_scaling} compares three strategies: (1) single-rank with increasing OpenMP threads (green solid line), which saturates around $2800$ seconds due to shared-memory bottlenecks, and (2) $P$ MPI ranks with single-threaded execution (red dashed line), which demonstrates effective strong scaling up to $P=32$, where the wall time drops significantly from $\sim 3000$\,s at $P=1$ to a minimum of $555$\,s at $P=32$. 
This confirms that the domain decomposition approach effectively distributes the assembly and linear solve burdens. 
However, at $P=64$, the pure MPI performance degrades (increasing to $781$\,s), a characteristic sign of the strong scaling limit where inter-rank communication overhead begins to dominate the diminishing local computational work. 
(3) The hybrid strategy (blue dotted line) tracks the Pure MPI performance in the linear scaling regime but offers superior robustness at the saturation point. 
At $P=64$, the optimal configuration was found to be a hybrid decomposition of $32$ MPI ranks with $2$ threads per rank ($r32/t2$). 
This configuration yielded a wall time of $560$\,s, mitigating the communication overhead observed in the pure MPI case ($r64/t1$) while maintaining the parallelism benefits. 
These results confirm that the solver scales effectively within a single multi-core node, with hybrid MPI+OpenMP decompositions extending the strong-scaling limit for fixed-size problems. 
Multi-node distributed-memory scaling and weak-scaling studies at constant work-per-rank are left for future work.

\subsection{Manufactured-solution verification (unsteady compressible duct flow)}
\label{subsec:mms_duct}

To provide an exact-reference verification of the fully compressible monolithic formulation, we employed the method of manufactured solutions (MMS) in a two-dimensional duct. 
In this test, smooth analytical fields $(\mathbf{u}^*, e^*, \rho^*, p^*)$ are prescribed and consistent source terms are added to the governing equations such that the prescribed fields satisfy the compressible system exactly. 
This approach verifies both the coupling of the primitive variables and the correctness of the weak-form implementation, independent of external solver comparisons.

\subsubsection{Definition of the unsteady MMS duct flow}
The computational domain is a rectangle $\Omega = [0,L_x]\times[0,L_y]$ with $(L_x,L_y)=(2.0,1.0)$. 
We prescribe a time-dependent velocity field composed of a duct-like streamwise profile with a weak spatial modulation and an oscillatory cross-stream component,
\begin{align}
u_x^*(x,y,t) &= U_{\mathrm{scale}} \, 4 \frac{y(L_y - y)}{L_y^2} \left[1 + 0.1 \sin\!\left(\frac{2\pi x}{L_x}\right) \cos(\omega t)\right], \\
u_y^*(x,y,t) &= U_{\mathrm{scale}} \, 0.05 \sin\!\left(\frac{\pi x}{L_x}\right) \sin\!\left(\frac{\pi y}{L_y}\right) \sin(\omega t).
\end{align}
where $U_{\mathrm{scale}}$ represents the scaling factor for the velocity amplitude. The scalar fields are specified as smooth, strictly positive functions:
\begin{align}
    e^*(x,y,t)=&1+0.2\cos\!\left(\frac{\pi x}{L_x}\right)\sin\!\left(\frac{\pi y}{L_y}\right)\cos(\omega t),\qquad \\
    \rho^*(x,y,t)=&1+0.1\sin\!\left(\frac{\pi x}{L_x}\right)\cos\!\left(\frac{\pi y}{L_y}\right)\sin(\omega t).
\end{align}
The pressure field is then defined through the nondimensional equation of state used in the solver,
\begin{align}
    p^*(x,y,t)=\frac{\rho^*(x,y,t)e^*(x,y,t)-1}{\gamma\,Ma^2},
\end{align}
ensuring exact consistency between $(\rho^*,e^*)$ and $p^*$.
To enforce the manufactured solution, we add source terms $(f_\rho,\mathbf{f}_u,f_e)$ to the right-hand sides. 
To verify the unsteady implementation without mixing in temporal truncation error, the forcing is constructed to satisfy the \emph{discrete} backward-Euler update by using the exact solution at consecutive time levels $t_n$ and $t_{n+1}$ (i.e., the manufactured residuals are formed using $(\cdot)^{*}_{n+1}$ and $(\cdot)^{*}_{n}$). 
With this construction, the exact manufactured fields satisfy the fully discrete system, so the measured error primarily reflects spatial discretization and nonlinear-solver accuracy.
Dirichlet conditions are imposed on the entire boundary $\partial\Omega$ using the analytical fields $(\mathbf{u}^*,e^*,\rho^*,p^*)$ evaluated at the current time, i.e., $\mathbf{u}=\mathbf{u}^*$, $e=e^*$, $\rho=\rho^*$, $p=p^*$ on $\partial\Omega$ for all $t$.
The initial condition is set by interpolating the manufactured fields at $t=0$ into the corresponding finite element spaces.

\subsubsection{Verification results}
\begin{figure}[!t]
    \centering
    \includegraphics[width=0.98\textwidth]{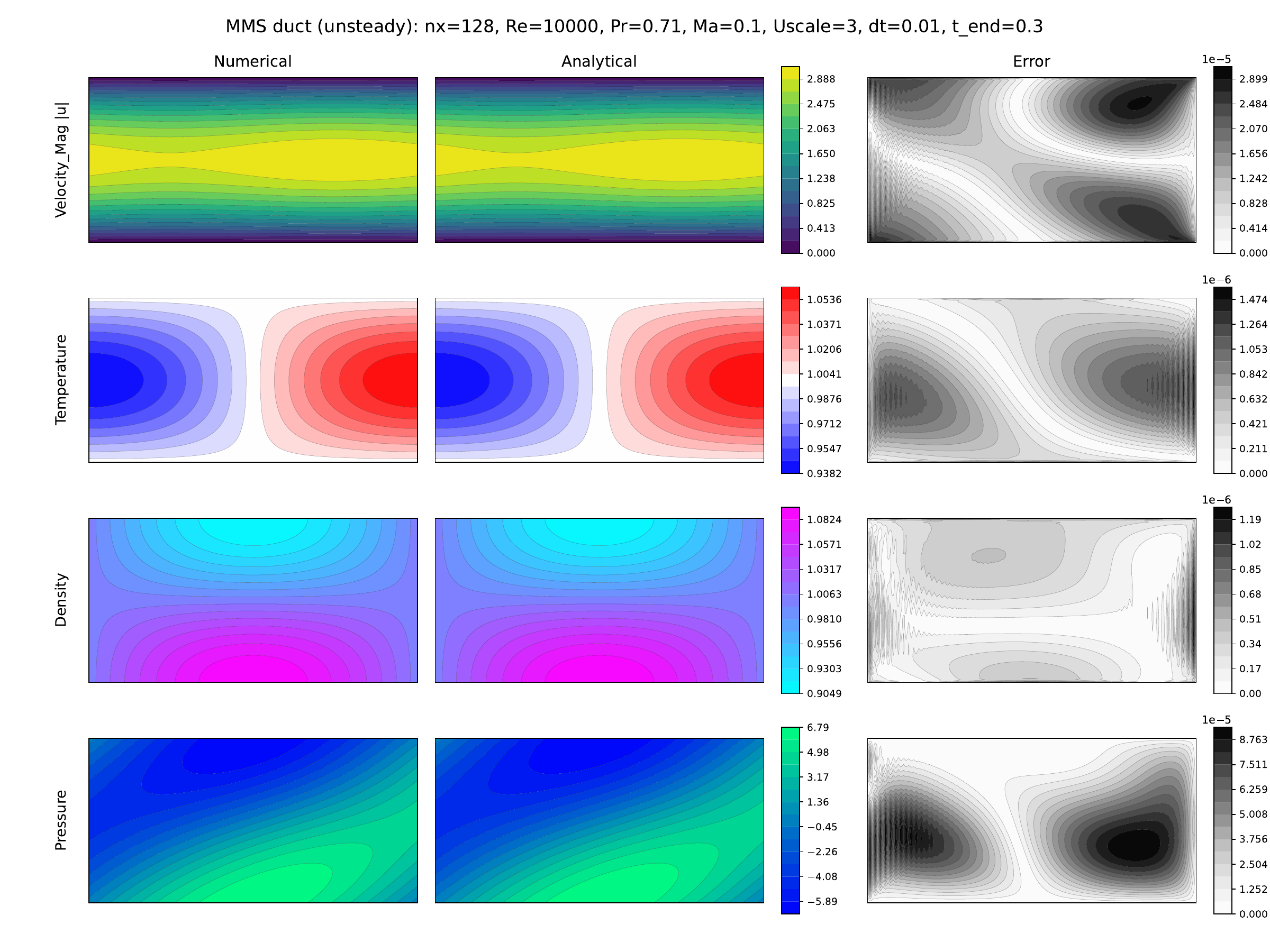}
    \caption{Unsteady MMS duct verification for the fully compressible formulation at $Re=10^{4}$, $Pr=0.71$, $Ma=0.1$, $\Delta t=0.01$, $t_{\mathrm{end}}=0.30$, $\omega=2\pi$, and $U_{\mathrm{scale}}=3$ on a $128\times128$ mesh. Numerical solution (left column), analytical manufactured solution (middle), and absolute error (right) for velocity magnitude, temperature/internal energy, density, and pressure at the final time.}
    \label{fig:mms_duct_fields}
\end{figure}

\begin{figure}[!t]
    \centering
    \includegraphics[width=0.6\textwidth]{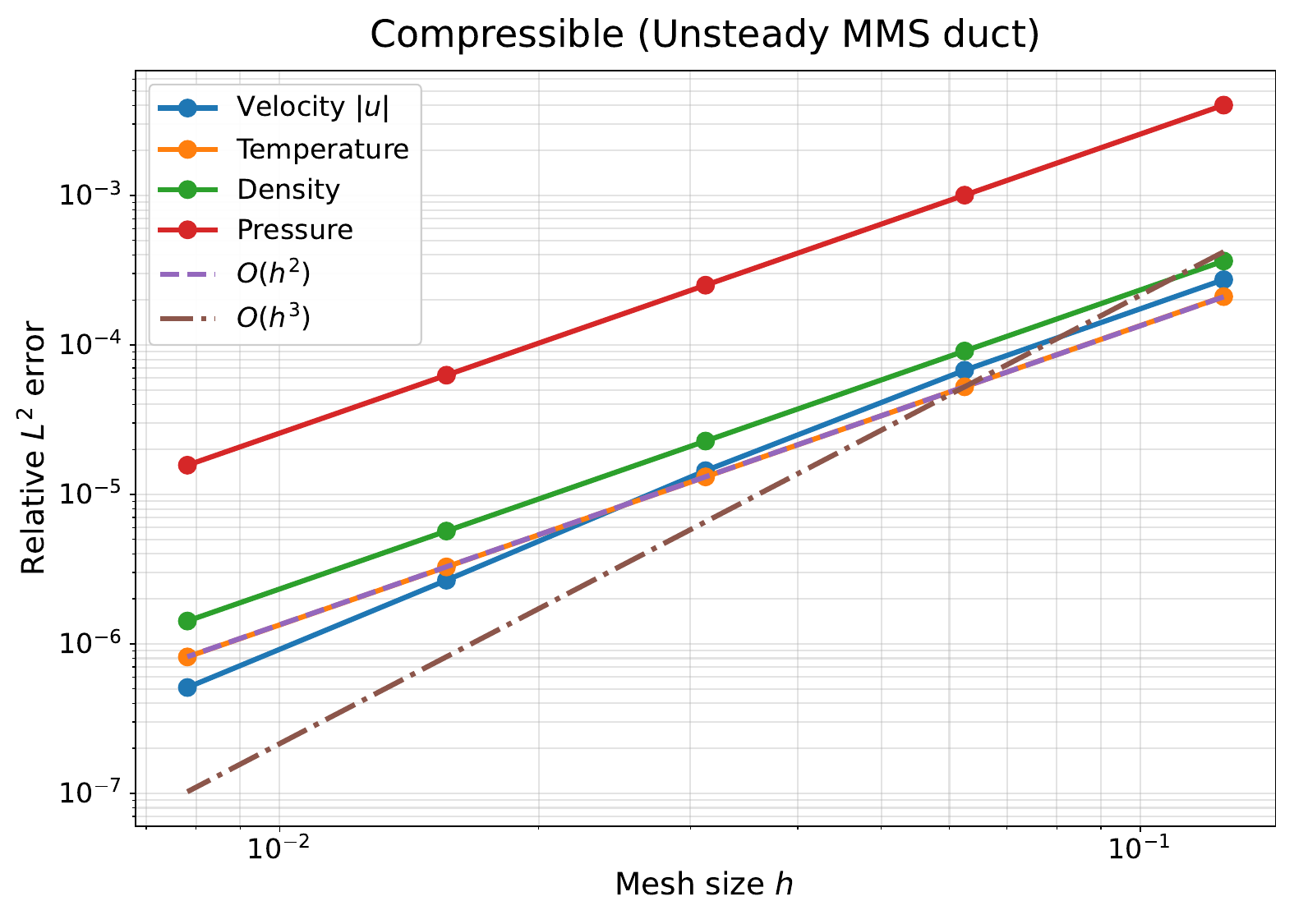}
    \caption{Mesh convergence for the unsteady MMS duct case (fully compressible formulation). Relative $L^2$ error versus mesh size $h$ at a fixed final time $t_{\mathrm{end}}=0.30$, compared with $O(h^2)$ and $O(h^3)$ reference slopes.}
    \label{fig:mms_duct_meshconv}
\end{figure}

Figure~\ref{fig:mms_duct_fields} presents a visual field-by-field comparison at the final time on a $128 \times 128$ mesh (approximately $16{,}400$ triangular elements).
The physical parameters are $\gamma = 1.4$, $Re = 10^4$, $Pr = 0.71$, $Ma = 0.1$, with backward Euler time integration ($\Delta t = 0.01$, $t_{\mathrm{end}} = 0.30$), temporal frequency $\omega = 2\pi$, and velocity scaling $U_{\mathrm{scale}} = 3$.
The left and center columns show the numerical and analytical solutions for velocity magnitude $|\mathbf{u}|$, temperature (internal energy $e$), density, and pressure.
The error fields (right column) are dominated by high-frequency noise and remain at the level of \(10^{-5}\) to \(10^{-6}\), indicating that no systematic spatial error remains.
Figure~\ref{fig:mms_duct_meshconv} quantifies the spatial convergence by computing relative $L^2$ errors on a sequence of refined meshes with $N =\{16,32,64,128,256\}$ at the same final time. 
The scalar fields exhibit near-second-order convergence consistent with their $P_1$ approximation, while the velocity magnitude shows a slope between $O(h^2)$ and $O(h^3)$, consistent with a $P_2$ velocity discretization applied to the smooth manufactured solution.
Notably, all fields track or exceed the theoretical $O(h^2)$ reference line, confirming that the spatial discretization achieves its designed accuracy order.
The excellent agreement between numerical and analytical solutions, combined with the convergence rates matching the finite element polynomial orders, provides rigorous code verification of the monolithic coupling, the weak-form assembly, the automatically-differentiated Jacobian, and the unsteady time-stepping implementation.

\section{Dataset Generation}
\label{sec:data}

Having validated the solver in Section~\ref{sec:results}, we now turn to the second pillar of this work: generating large, paired simulation datasets that are suitable for training a neural surrogate.
There is a growing need for open, large-scale simulation datasets that are statistically rich and broadly representative for machine learning and deep learning in fluid dynamics. Such datasets enable training and fair evaluation of surrogate models, facilitate generalization studies across regimes and boundary patterns, and support uncertainty quantification and robust model selection \cite{willard2020integrating,mangnike2024toward}.
In the case of our paper, we are interested in generating a dataset of paired simulations (same initial and boundary conditions) run using the Boussinesq and compressible flow models to use as training, validation, and test datasets for training a neural surrogate (Section \ref{sec:learning}) for correcting the model-form errors of Boussinesq flow.
While one could imagine simply randomizing initial fields, such a choice would be non-physical.
One could also imagine uniformly randomly sampling constant wall temperatures, as in \citet{mangnike2024toward}, but as we will show in this section, that choice does not provide good coverage of the space of possible thermally-driven cavity simulations.
Thus, in this section, we provide one way in 2D and one way in 3D to carefully specify the initial and boundary conditions such that resulting wall traces indicate a genuine stochastic process with controlled regularity and variance.
Section~\ref{sec:learning} reports both 2D and 3D results, including a controlled cross-dataset comparison that quantifies how the broader stochastic distribution introduced here improves surrogate generalization relative to the na\"ive dataset approach used in \citet{mangnike2024toward}.


\subsection{Stochastic temperature boundary conditions}
\label{Sto_data}

To construct these randomized thermal profiles, let $\Gamma\in\{\Gamma_L,\Gamma_R,\Gamma_B,\Gamma_T,\Gamma_F,\Gamma_{Bk}\}$ denote the boundary walls of the domain. Each active wall is assigned a random temperature field that fluctuates about the reference temperature $T_{\mathrm{init}}\in\mathbb{R}$ while remaining, by construction, within the physically admissible range $[T_{\mathrm{init}}-\tfrac12\Delta T,\ T_{\mathrm{init}}+\tfrac12\Delta T]$. Each wall additionally carries an integer complexity dial $n_\Gamma$ that governs the spatial richness of its profile and admits, as limiting cases, a spatially constant wall and a zero-flux (adiabatic) wall. 

\subsubsection{2D: Random Fourier series}

For 2D domains, the boundary walls are 1D line segments. Let $s\in[0,1]$ be the unit arclength along $\Gamma$ (for the square cavity, $s$ coincides with the tangential Cartesian coordinate). Each active wall carries an integer dial $n_\Gamma\in\{0,1,2,\dots\}$, which sets the number of Fourier modes $m_\Gamma=n_\Gamma-1$ on that wall: $n_\Gamma=0$ leaves the wall adiabatic (zero-flux Neumann), $n_\Gamma=1$ yields a spatially constant wall, and $n_\Gamma\ge2$ yields an $(n_\Gamma-1)$-mode fluctuating profile. The vertical walls $\Gamma_L,\Gamma_R$ are always Dirichlet ($n_\Gamma\ge1$); the horizontal walls $\Gamma_B,\Gamma_T$ may additionally be adiabatic ($n_\Gamma=0$).

For an active wall with $m_\Gamma=n_\Gamma-1\ge1$ modes, on a probability space $(\Omega,\mathcal{F},\mathbb{P})$ we fix a spectral decay exponent $\alpha>1/2$ and draw i.i.d.\ coefficients $\{a_k(\omega),b_k(\omega)\}$ with $a_k,b_k\sim\mathcal{N}(0,1)$, independent across $k$ and across walls. We form the truncated random Fourier series
\begin{equation}
S_\Gamma(s,\omega)=\sum_{k=1}^{m_\Gamma}\frac{a_k(\omega)\cos(2\pi k s)+b_k(\omega)\sin(2\pi k s)}{k^\alpha},\qquad s\in[0,1].
\label{eq:series}
\end{equation}
Letting $\bar S_\Gamma(\omega)=\int_0^1 S_\Gamma(r,\omega)\,dr$ denote the wall-mean, we center and normalize the series along the wall,
\begin{equation}
\widehat{S}_\Gamma(s,\omega)=\frac{S_\Gamma(s,\omega)-\bar S_\Gamma(\omega)}{\|S_\Gamma(\cdot,\omega)-\bar S_\Gamma(\omega)\|_{L^\infty(0,1)}+10^{-15}},
\end{equation}
so that $\int_0^1 \widehat{S}_\Gamma(r,\omega)\,dr=0$ and $\|\widehat{S}_\Gamma(\cdot,\omega)\|_{L^\infty}\le 1$, the constant $10^{-15}$ guards against division by zero in degenerate near-constant cases and is negligible relative to the field scales.

Rather than perturbing a fixed mean and projecting onto the admissible interval, we map the normalized shape into range by drawing a random mean and amplitude that keep the trace bounded \emph{by construction}. Writing $T_c=T_{\mathrm{init}}-\tfrac12\Delta T$ and $T_h=T_{\mathrm{init}}+\tfrac12\Delta T$, we draw
\begin{equation}
A_\Gamma(\omega)\sim\mathcal{U}\!\left(0,\tfrac12\Delta T\right),\qquad
\mu_\Gamma(\omega)\sim\mathcal{U}\!\left(T_c+A_\Gamma(\omega),\;T_h-A_\Gamma(\omega)\right),
\end{equation}
and set
\begin{equation}
T_\Gamma(s,\omega)=\mu_\Gamma(\omega)+A_\Gamma(\omega)\,\widehat{S}_\Gamma(s,\omega).
\label{eq:tmap}
\end{equation}
Since $\|\widehat{S}_\Gamma\|_{L^\infty}\le1$, every realization satisfies $T_\Gamma(s,\omega)\in[\mu_\Gamma-A_\Gamma,\ \mu_\Gamma+A_\Gamma]\subseteq[T_c,T_h]$ pointwise, so the thermodynamic bounds hold with no clamping projection. The degenerate dial $n_\Gamma=1$ ($m_\Gamma=0$) gives $\widehat{S}_\Gamma\equiv0$ and hence a constant wall $T_\Gamma\equiv\mu_\Gamma$ with $\mu_\Gamma\sim\mathcal{U}(T_c,T_h)$. At nodes shared by two Dirichlet walls, the corner value is inherited from the wall whose Dirichlet condition is assembled last (horizontal walls take precedence over vertical), since both incident traces lie in $[T_c,T_h]$, the assembled datum remains in range and the discrete Dirichlet condition is well defined.

The classical differentially-heated cavity is recovered deterministically as the $n_{\Gamma_L}=n_{\Gamma_R}=1$ member of this family: the left wall is held at the constant hot value $T_h$, the right wall at the constant cold value $T_c$, and the horizontal walls are adiabatic. The legacy benchmark configuration is therefore a strict subset of the stochastic training distribution, with the vertical-wall dial sampled uniformly on $\{1,2,3\}$.

Each realization $s\mapsto S_\Gamma(s,\omega)$ in Equation~\eqref{eq:series} is a trigonometric polynomial of degree $m_\Gamma$, hence $C^\infty$ in $s$, the centering, normalization, and random affine map $x\mapsto\mu_\Gamma+A_\Gamma x$ are continuous, so $s\mapsto T_\Gamma(s,\omega)$ is smooth and its trace lies in $H^{1/2}(\partial\Omega)$, an admissible Dirichlet datum for the variational problem. Joint Borel measurability on $[0,1]\times\Omega$ follows since $S_\Gamma$ is a finite sum of terms continuous in $s$ and measurable in $\omega$, preserved by the subsequent continuous maps, thus $\{T_\Gamma(s)\}_{s\in[0,1]}$ is a bona fide stochastic process. With $\alpha>1/2$ the pre-normalized series has finite, stationary covariance
\begin{equation}
\mathrm{Cov}\!\left(S_\Gamma(s),S_\Gamma(t)\right)=\sum_{k=1}^{m_\Gamma}\frac{\cos\!\big(2\pi k (s-t)\big)}{k^{2\alpha}},
\qquad
\mathrm{Var}\bigl[S_\Gamma(s)\bigr]=\sum_{k=1}^{m_\Gamma}\frac{1}{k^{2\alpha}}>0,
\end{equation}
so the decay exponent $\alpha$ controls spatial smoothness while the mode count $m_\Gamma$ controls spectral richness, and the non-degeneracy of $T_\Gamma$ follows since $A_\Gamma>0$ almost surely and $\widehat{S}_\Gamma$ is non-constant whenever $m_\Gamma\ge1$. In summary, the assembled wall temperature is a genuine stochastic boundary field with bounded, smooth ($C^\infty$) sample paths, in range by construction, with the constant ($n_\Gamma=1$) and adiabatic ($n_\Gamma=0$) walls arising as limiting members of the same family.

Each paired sample additionally draws a Rayleigh number $\mathrm{Ra}$ log-uniformly over $[10^2,10^6]$, with the Mach number tied to it by the calibrated monotone relation $\mathrm{Ma}(\mathrm{Ra})=8.27\times10^{-4}\,(\mathrm{Ra}/10^4)^{1/6}$ (clamped to $\mathrm{Ra}\in[10^2,10^6]$), which keeps the flow weakly compressible across the regime; the Reynolds number follows as $\mathrm{Re}=\sqrt{\mathrm{Ra}/\mathrm{Pr}}$. 

\subsubsection{3D: Smoothed-Voronoi stochastic fields}
\label{subsubsec:voronoi_3d}

For 3D domains, the boundary walls are 2D planar surfaces. In the differentially heated cavity, the natural-convection physics is driven by the two opposing thermal walls, and we therefore restrict the stochastic perturbation to $\Gamma\in\{\Gamma_{L},\Gamma_{R}\}$ (the $x{=}0$ and $x{=}L$ faces of the cube) while the remaining four walls carry adiabatic conditions consistent with Section~\ref{sec:results}. Let $\mathbf{x}\in\Gamma\subset\mathbb{R}^{2}$ be the spatial coordinate on the wall (e.g., parameterized by $(y,z)$ on the $x{=}0$ plane).

To produce spatially chaotic, non-isothermal boundary structures we sample a Voronoi-like seed configuration on $\Gamma$ and blend the resulting per-cell amplitudes into a smooth scalar field via Gaussian radial-basis-function (RBF) interpolation. On the probability space $(\Omega,\mathcal{F},\mathbb{P})$, we draw $n_{c}$ seed points $\mathbf{c}_{i}(\omega)\in[mL,(1{-}m)L]^{2}$ from a margin-padded uniform distribution ($m=0.15$) with a best-effort minimum-spacing rejection criterion (target separation $0.30L$ for $n_{c}\le 3$ and $0.15L$ otherwise); this prevents seeds from pinning to wall edges and yields well-separated centers reminiscent of a centroidal Voronoi tessellation \cite{doi:10.1137/S0036144599352836}. We then draw i.i.d.\ base amplitudes $\tilde v_{i}(\omega)\sim\mathcal{U}(T_{\min},T_{\max})$, where $T_{\min}=T_{\mathrm{init}}-\tfrac12\Delta T$ and $T_{\max}=T_{\mathrm{init}}+\tfrac12\Delta T$. Sampling directly inside the admissible range obviates the external clamping projection used in 2D.

To avoid uncorrelated jumps between adjacent cells, the base amplitudes are spatially correlated through a Gaussian kernel over seed positions,
\begin{equation}
v_{i}(\omega) \;=\; \sum_{j=1}^{n_{c}} A_{ij}(\omega)\,\tilde v_{j}(\omega),
\qquad
A_{ij}(\omega) \;=\; \frac{\exp\!\big(-\|\mathbf{c}_{i}-\mathbf{c}_{j}\|_{2}^{2}/(2\sigma^{2})\big)}{\sum_{k=1}^{n_{c}}\exp\!\big(-\|\mathbf{c}_{i}-\mathbf{c}_{k}\|_{2}^{2}/(2\sigma^{2})\big)},
\label{eq:voronoi_smoothing}
\end{equation}
where the row-stochastic matrix $A(\omega)$ acts as a discrete Gaussian smoother with correlation length $\sigma$ (default $\sigma=0.15L$). Because each row of $A$ sums to one, the smoothed amplitudes $v_{i}$ remain within $[T_{\min},T_{\max}]$.

The wall temperature at an arbitrary spatial point $\mathbf{x}\in\Gamma$ is then defined as a Gaussian-RBF-weighted convex combination of the smoothed amplitudes,
\begin{equation}
T_{\Gamma}(\mathbf{x},\omega) \;=\; \sum_{i=1}^{n_{c}} w_{i}(\mathbf{x},\omega)\,v_{i}(\omega),
\qquad
w_{i}(\mathbf{x},\omega) \;=\; \frac{\exp\!\big(-\|\mathbf{x}-\mathbf{c}_{i}(\omega)\|_{2}^{2}/(2(rL)^{2})\big)}{\sum_{k=1}^{n_{c}}\exp\!\big(-\|\mathbf{x}-\mathbf{c}_{k}(\omega)\|_{2}^{2}/(2(rL)^{2})\big)},
\label{eq:voronoi_blend}
\end{equation}
where $r$ is the blur radius expressed as a fraction of the wall side length $L$ (default $r=0.08$). The hard Voronoi partition is recovered in the limit $r\to 0^{+}$, in which $w_{i}(\mathbf{x},\omega)\to\mathbb{1}_{V_{i}(\omega)}(\mathbf{x})$ with $V_{i}(\omega)=\{\mathbf{x}\in\Gamma : \|\mathbf{x}-\mathbf{c}_{i}\|_{2}\le \|\mathbf{x}-\mathbf{c}_{j}\|_{2}\ \forall j\}$. For finite $r$, the wall trace is a convex combination of bounded amplitudes with smooth, strictly positive Gaussian weights, so $T_{\Gamma}(\cdot,\omega)\in C^{\infty}(\Gamma)$ almost surely and $T_{\Gamma}(\mathbf{x},\omega)\in[T_{\min},T_{\max}]$ pointwise. The map $\omega\mapsto T_{\Gamma}(\mathbf{x},\omega)$ is jointly Borel measurable as a finite composition of the measurable random variables $\{\mathbf{c}_{i},\tilde v_{i}\}$ with the continuous maps in Equations \ref{eq:voronoi_smoothing}--\ref{eq:voronoi_blend}. The resulting boundary trace is therefore a bona fide $L^{\infty}$-bounded, almost-surely smooth random field, and constitutes a valid Dirichlet trace for the Galerkin weak formulation.

A useful corner case is $n_{c}=1$: the generator then bypasses the Voronoi construction and applies the deterministic legacy protocol, holding the two active walls at the constant temperatures $T_{\max}=T_{\mathrm{init}}+\tfrac12\Delta T$ on $\Gamma_{L}$ ($x{=}0$) and $T_{\min}=T_{\mathrm{init}}-\tfrac12\Delta T$ on $\Gamma_{R}$ ($x{=}L$), exactly reproducing the hot-/cold-wall cavity of \citet{mangnike2024toward}. The legacy 3D dataset is thus realized as the $n_{c}=1$ special case of the present generator. The in-distribution training set draws $n_{c}\in\{1,\dots,3\}$, while the out-of-distribution test set introduced in Section~\ref{subsec:fno_3d} probes $n_{c}\in\{3,\dots,7\}$, extending the per-wall seed count up to seven, beyond the training maximum of three, so that the OOD walls carry markedly more spatial structure than any configuration seen during training.

\begin{figure}[!t]
    \centering
    \includegraphics[width=\textwidth]{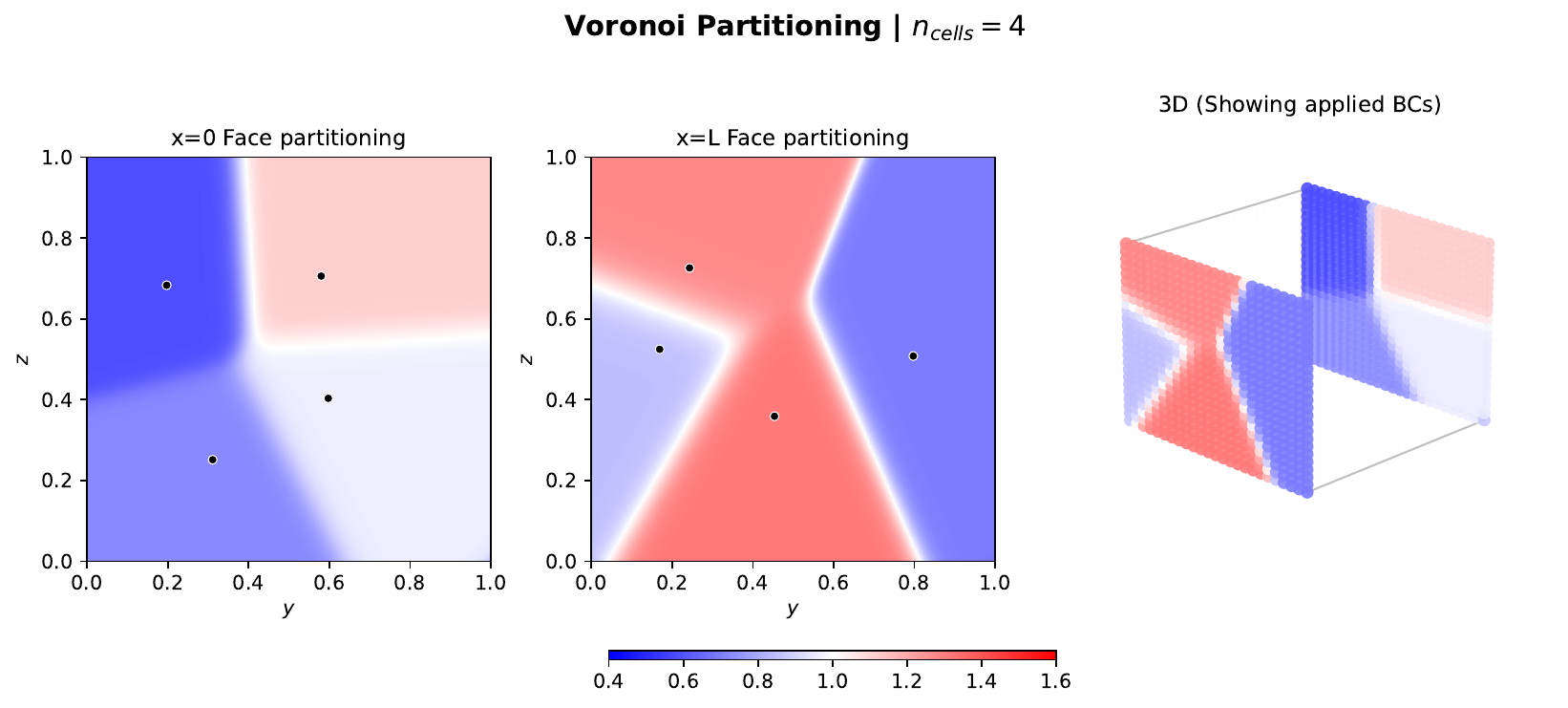}
    \caption{Representative realization of the stochastic Voronoi temperature boundary conditions applied to the 3D thermal cavity. (Left and Center) 2D heatmaps of the smooth-blended Voronoi partitions evaluated on the active left ($x=0$) and right ($x=L$) walls, with black markers indicating the randomly sampled cell seeds. (Right) The resulting 3D Dirichlet boundary condition trace correctly assigned to the active faces of the hexahedral domain.}
    \label{fig:voronoi_3d}
\end{figure}





\subsection{Dataset distribution and diversity analysis (2D)}
\label{sec:dataset_diversity}

To demonstrate that the generated dataset is statistically diverse, broadly representative, and suitable for training machine learning surrogate models, we perform quantitative and visual analysis of 500 stochastic simulations drawn at random from the larger pool used for surrogate training (Section~\ref{sec:learning}); this reduced set is purely for visualization, and the full pool is used downstream.

\subsubsection{Feature extraction pipeline}

\begin{figure}[!t]
    \centering
    \includegraphics[width = 0.85\textwidth]{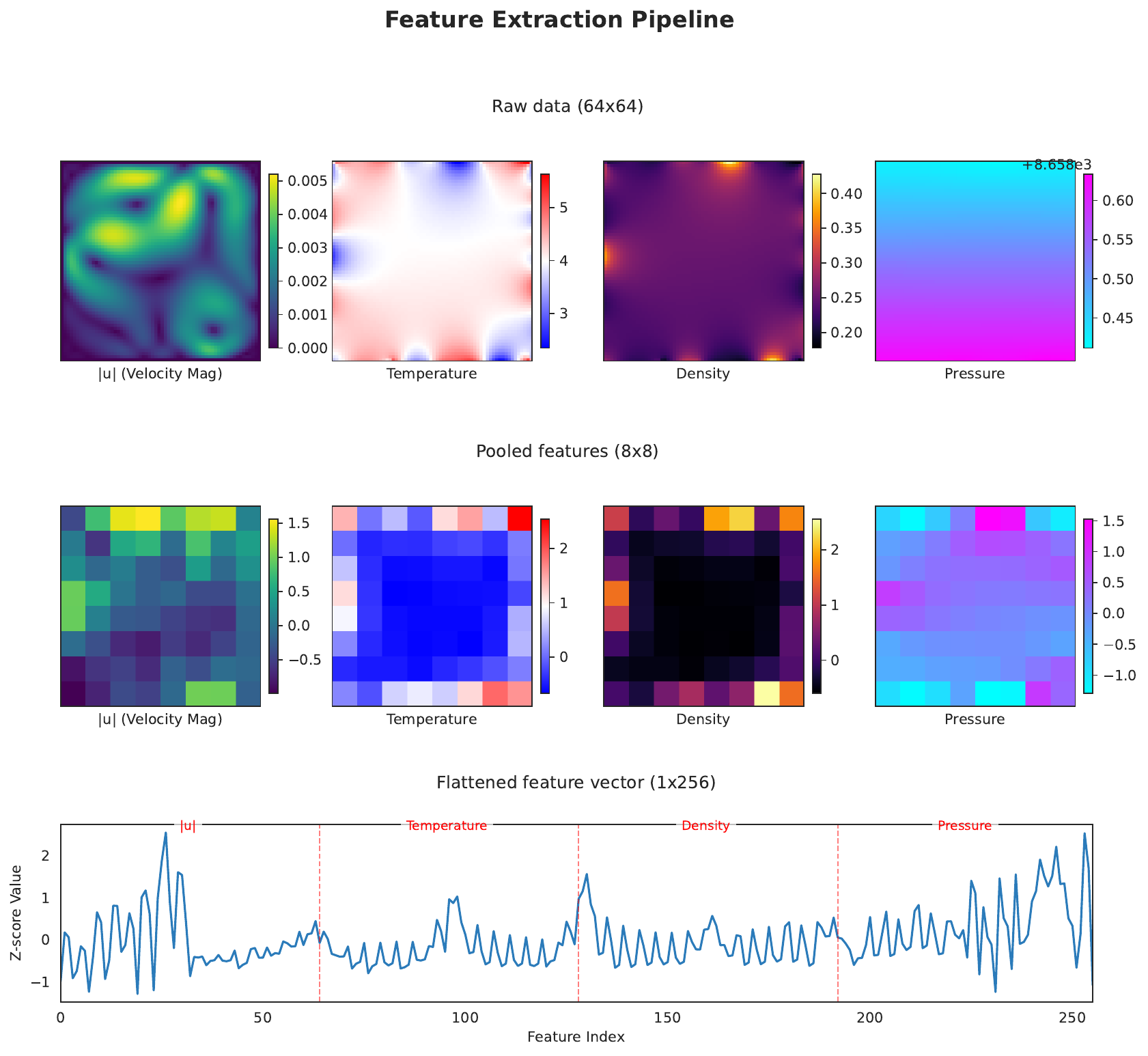}
    \caption{The feature extraction pipeline. (Top) Four raw physical fields on $64 \times 64$ grids. (Middle) Pooled features on $8 \times 8$ channels after Z-score normalization and edge detection. (Bottom) The final 256-dimensional flattened feature vector used for manifold analysis.} 
    \label{fig:PIPLINE}
\end{figure}

As shown in Figure~\ref{fig:PIPLINE}, each simulation is represented by a compact feature vector derived from the four primary solution fields: velocity magnitude $|\mathbf{u}|$, internal energy $e$ (temperature), density $\rho$, and pressure $p$, and interpolated onto uniform $64 \times 64$ grids.
These grids are normalized via Z-score transformation and processed using edge-aware Sobel filters to emphasize structural gradients.
The preprocessed fields are compressed through block-averaging ($8 \times 8$ pooling), reducing each field to an $8 \times 8$ feature map. 
These maps are concatenated into a 256-dimensional feature vector $\mathbf{x} \in \mathbb{R}^{256}$, effectively capturing the global morphology and local gradient information of the flow states.

\subsubsection{Dimensionality reduction and visualization}

\begin{figure}[!t]
    \centering
    \begin{minipage}{0.48\textwidth}
        \centering
        \includegraphics[width=\textwidth]{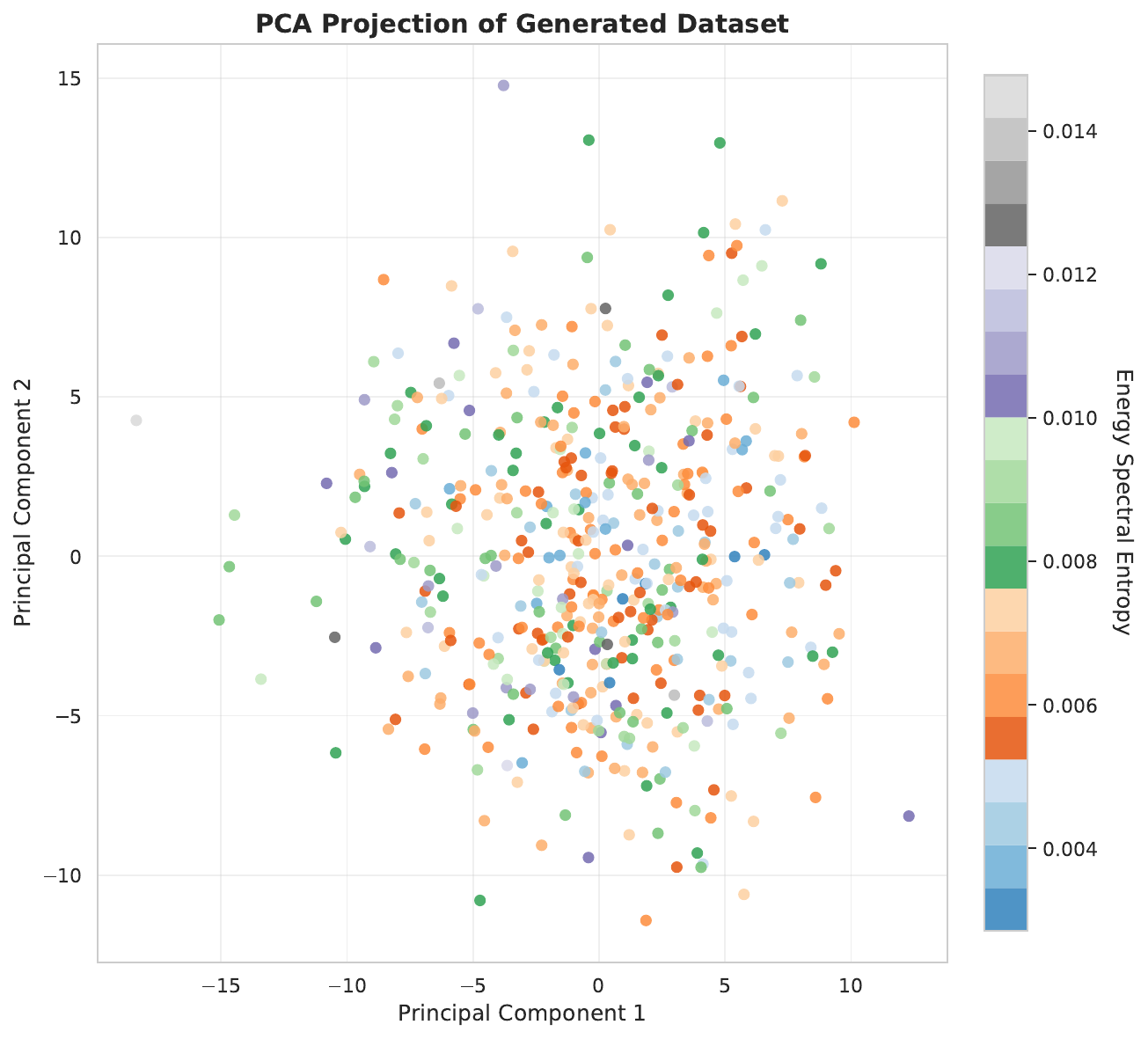}
        \caption{PCA projection of 500 shape feature vectors ($N=500$, $D=256$ dimensions). Points are colored by spectral entropy of the internal energy field. The full 2D scatter exhibits continuous, uniform distribution across PC space with no voids or clustering, confirming dataset diversity and homogeneity.}
        \label{fig:PCA}
    \end{minipage}
    \hfill
    \begin{minipage}{0.48\textwidth}
        \centering
        \includegraphics[width=\textwidth]{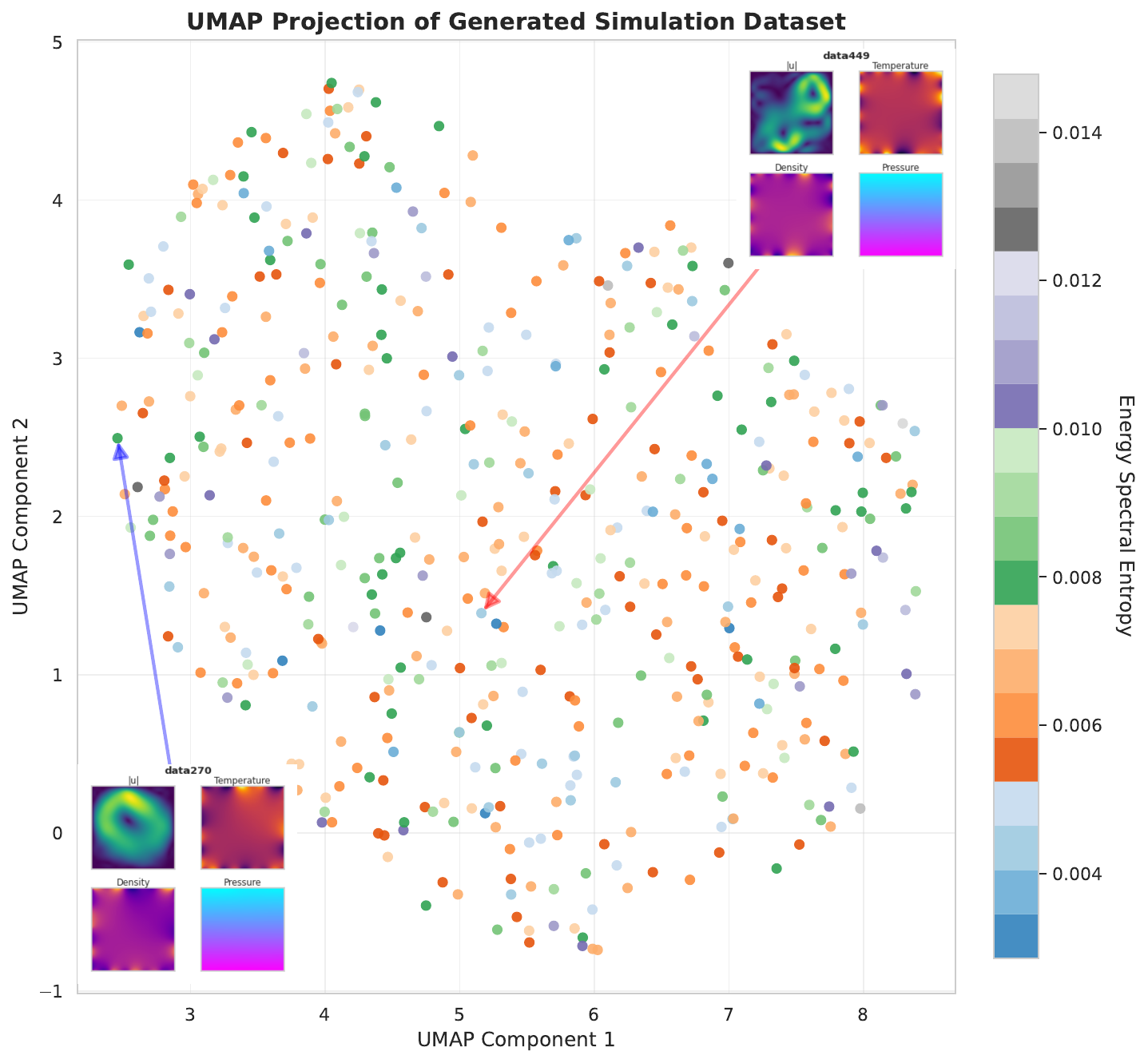}
        \caption{UMAP projection of 500 shape feature vectors ($N=500$, $D=256$ dimensions), colored by spectral entropy. Unlike PCA, UMAP preserves both local neighborhoods and global topology via fuzzy topological representation and cross-entropy optimization. Two random simulations are highlighted to show the difference between two distinct feature vectors in the UMAP 2D space.}
        \label{fig:UMAP}
    \end{minipage}
\end{figure}

To assess dataset coverage and homogeneity, we apply two complementary dimensionality-reduction techniques to the 500 feature vectors (each one a 256-dimensional summary of a single simulation, as constructed above):

Principal Component Analysis (PCA) (Figure~\ref{fig:PCA}) performs linear projection onto the two leading eigenvectors of the feature covariance matrix. The 2D scatter colored by spectral entropy $H_e$ (computed from the Fourier transform of the energy field) reveals a continuous distribution spanning the full PC space without clustering or voids, confirming the absence of degeneracies or preferred lower-dimensional subsets.

Uniform Manifold Approximation and Projection (UMAP) (Figure~\ref{fig:UMAP}) is a nonlinear dimensionality reduction technique that aims to preserve both local structure (nearest-neighbor relationships) and global topology of high-dimensional data. Unlike PCA's linear assumption, UMAP constructs a fuzzy topological representation in high dimensions and optimizes a low-dimensional ($d'=2$) embedding to preserve nearest-neighbor structure via cross-entropy minimization. The clear lack of isolated clusters indicates a well-sampled parameter space, confirming that the stochastic boundary conditions successfully produce a smooth and diverse spectrum of convection patterns. The two insets in Figure~\ref{fig:UMAP} demonstrate the qualitative differences between samples across the manifold, ranging from simple laminar patterns to high-entropy states with complex thermal structures. This diverse distribution confirms the dataset's suitability for robust machine learning applications in fluid dynamics.



\section{Neural Surrogate}
\label{sec:learning}
 
We now use the paired Boussinesq--compressible datasets generated in Section~\ref{sec:data} to train a Fourier neural operator (FNO) surrogate \citep{li2021fourier} $\mathcal{G}_\theta$ that maps a Boussinesq prediction onto its compressible counterpart in a single forward pass.
The surrogate is intended as a corrective post-processing step rather than as a replacement for the underlying solver: a practitioner runs the cheaper Boussinesq simulation, queries $\mathcal{G}_\theta$ once, and obtains an approximation to the corresponding fully compressible field.
 
\subsection{Architecture}
\label{subsec:fno_architecture}
 
The FNO is a particular instance of the broader neural operator framework \citep{kovachki2023neural}, in which a learned map between function spaces is represented as a stack of integral kernel operators. 
In the Fourier variant, the kernel integral is computed in Fourier space and parameterized by a finite set of complex spectral weights, augmented by a pointwise spatial residual to recover high-frequency content \citep{li2021fourier}. 
We use the standard FNO block design with three modifications that we found to matter on the cavity flow problem: (i) reflective input padding to mitigate the periodicity assumption inherent to the FFT for non-periodic boundary-value problems, (ii) a learned linear input--output skip that preserves the low-frequency content of the Boussinesq prediction, and (iii) a small $1{\times}1$ MLP readout head with optional depthwise smoothing to suppress speckle in the predicted fields.
 
\begin{figure*}[htbp]
    \centering
    \includegraphics[width=\textwidth]{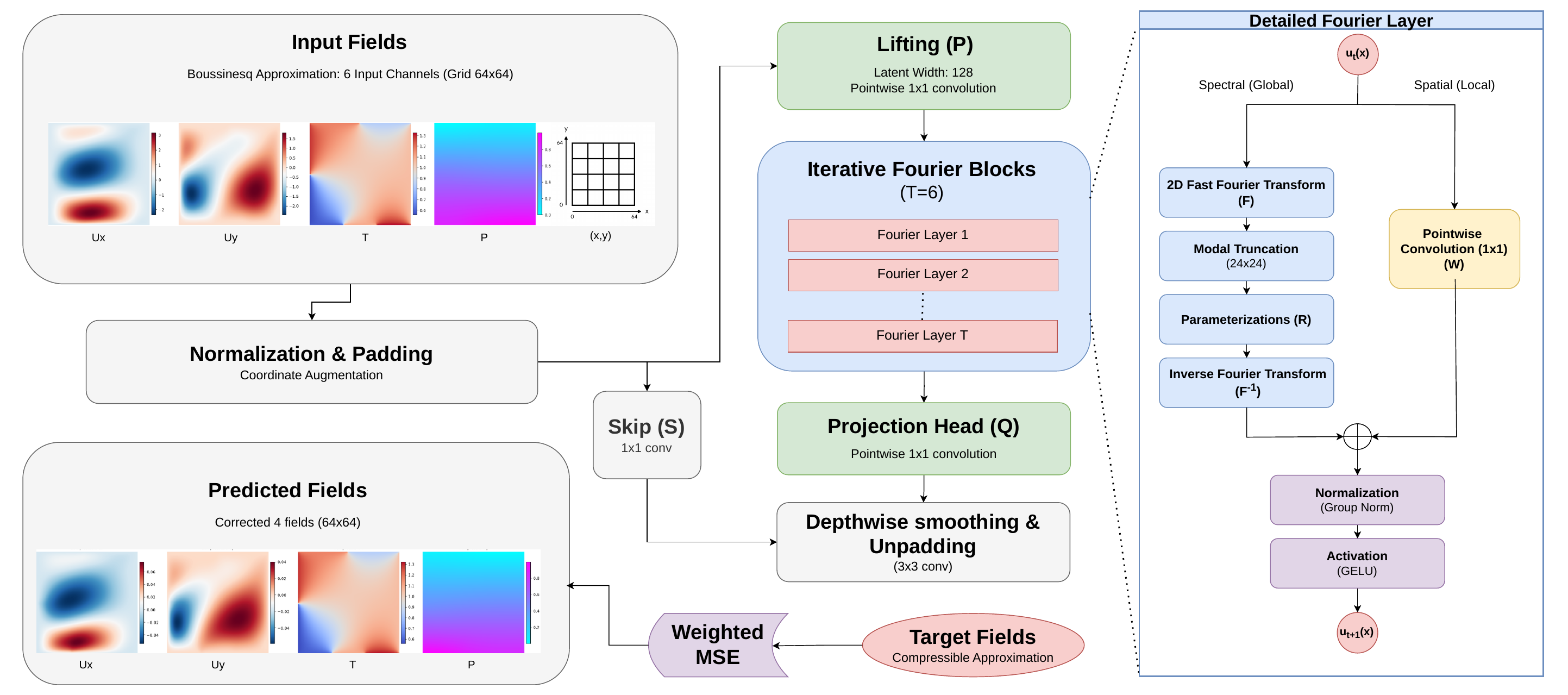}
    \caption{\textbf{2D Fourier Neural Operator (FNO2D) architecture.} The network maps a Boussinesq prediction to its fully compressible counterpart.
    \textbf{Left/Top (Macroscopic Pipeline):} The input consists of $6$ channels on a $64\times 64$ grid (four physical fields $u_{x}, u_{y}, T, p$ and two augmented coordinates $x,y$). The input is normalized and reflectively padded, then lifted to a latent width of $W=128$ via a pointwise $1{\times}1$ convolution $P$. A learned $1{\times}1$ skip $S$ bypasses the core blocks. The latent field passes through $L=6$ Fourier blocks; a readout head $Q$ ($1{\times}1$, hidden $2W$) followed by a depthwise $3{\times}3$ smoother and unpadding yields the four-channel compressible prediction. The network is optimized with a per-channel mean-squared error loss against the target.
    \textbf{Right/Bottom (Detailed Fourier Layer):} Each block splits its input into two parallel paths. The spectral (global) path applies a 2D FFT $\mathcal{F}$, truncates the spectrum to the lowest $24{\times}24$ modes, applies a learned complex weight $R$, and inverts via $\mathcal{F}^{-1}$. The spatial (local) path applies a pointwise $1{\times}1$ convolution $W$. The paths are summed, group-normalized, and activated by GELU to produce $v_{\ell+1}(x)$.}
    \label{fig:fno2d_architecture}
\end{figure*}

For the 2D model, the input is the four-channel Boussinesq field
\begin{equation}
a(x) \;=\; \big(u_x,\,u_y,\,T,\,p\big)(x), \qquad x=(x_1,x_2)\in\Omega\subset\mathbb{R}^{2},
\label{eq:fno_input2d}
\end{equation}
sampled on a $64{\times}64$ Cartesian grid. Two normalized coordinate channels $(\xi,\eta)\in[0,1]^{2}$ are concatenated to inject explicit positional information. The augmented input is reflectively padded by $p=8$ cells before entering the spectral layers and unpadded after the readout. A pointwise $1{\times}1$ convolution $P$ lifts the channel dimension to a width-$W$ latent field
\begin{equation}
v_{0}(x) \;=\; P\!\left(\tilde a^{(p)}(x)\right) \in \mathbb{R}^{W}, \qquad W=128,
\label{eq:fno_lift}
\end{equation}
which is then evolved by $L=6$ Fourier blocks of the form
\begin{equation}
v_{\ell+1}(x) \;=\; \sigma\!\Big(\mathrm{GN}\!\big(\mathcal{K}_{\ell}(v_{\ell})(x) + \mathcal{W}_{\ell}\,v_{\ell}(x)\big)\Big),
\qquad \ell=0,\dots,L-1,
\label{eq:fno_block}
\end{equation}
where $\sigma=\mathrm{GELU}$, $\mathrm{GN}$ denotes group normalization, and $\mathcal{W}_{\ell}$ is a $1{\times}1$ convolution that supplies a high-frequency local correction in physical space. The spectral operator $\mathcal{K}_{\ell}$ acts in Fourier space by truncating to the lowest $K_{x}{\times}K_{y}=24{\times}24$ modes and applying a learned, channel-mixing complex weight tensor $R_{\ell}$:
\begin{equation}
\widehat{\mathcal{K}_{\ell}(v_{\ell})}(k) \;=\;
\begin{cases}
R_{\ell}(k)\,\widehat{v_{\ell}}(k), & |k_{1}|\le K_{x},\ |k_{2}|\le K_{y}, \\[2pt]
0, & \text{otherwise,}
\end{cases}
\qquad
\mathcal{K}_{\ell}(v_{\ell}) \;=\; \mathcal{F}^{-1}\!\big(\widehat{\mathcal{K}_{\ell}(v_{\ell})}\big),
\label{eq:fno_spectral}
\end{equation}
where $\mathcal{F}$ is the 2D real-input FFT and the truncation sets all non-retained modes to zero. After $L$ blocks, a two-layer pointwise readout $Q:\mathbb{R}^{W}\to\mathbb{R}^{4}$ with hidden dimension $2W$, followed by an optional depthwise $3{\times}3$ smoother, projects back to the four output channels. A learned $1{\times}1$ skip $S$ connects the (padded) input directly to the output, so that the prediction is
\begin{equation}
\hat u(x) \;=\; \mathrm{Crop}\!\left(\,Q\!\big(v_{L}(x)\big) \;+\; S\!\big(\tilde a^{(p)}(x)\big)\,\right),
\label{eq:fno_skip}
\end{equation}
where $\mathrm{Crop}(\cdot)$ removes the reflective padding. The full 2D pipeline is summarized in Fig.~\ref{fig:fno2d_architecture}.
 
The 3D model uses the same block design, with three Cartesian coordinate channels concatenated to the input, a 3D real-input FFT, and four corner-mode weight tensors $R^{(1)}_{\ell},\dots,R^{(4)}_{\ell}$ to cover the $(\pm k_{1}, \pm k_{2}, k_{3}\!\geq\! 0)$ octants of the truncated spectrum. To keep the parameter count compatible with the smaller 3D training set (Section~\ref{subsec:fno_datasets}) we use the reduced hyperparameters $W=64$, $L=4$, $K_{x}=K_{y}=K_{z}=6$, replicate-padding $p=4$, and a non-zero feature dropout of $0.1$.

\subsection{Datasets}
\label{subsec:fno_datasets}

The surrogate is trained and evaluated using three data generators: the proposed stochastic-BC method of Section~\ref{Sto_data}, the constant-BC legacy protocol of \citet{mangnike2024toward}, and an out-of-distribution variant of the stochastic generator in which the Rayleigh number and boundary-condition complexity are pushed strictly beyond the training envelope. All samples are paired: each Boussinesq input is matched with the compressible solution computed under identical initial and boundary conditions, so that the surrogate's task is to reconstruct the gap between the two physical models rather than to redo the simulation from scratch. Table~\ref{tab:fno_datasets} summarises the resulting data layout.

\begin{table}[h]
\centering
\small
\caption{Dataset structure used to train and evaluate the FNO surrogate. Each sample is a paired (Boussinesq, compressible) tuple. The 2D test sets contain $200$ samples each; the 3D test sets are constrained to $100$ samples by the cost of the compressible reference solver at $24^{3}$ resolution.}
\label{tab:fno_datasets}
\begin{tabular}{llccc}
\toprule
Dim. & Role & Generator (Section~\ref{Sto_data}) & Tensor shape & $N$ \\
\midrule
\multirow{5}{*}{2D} & Train & Stochastic (Fourier-series BCs) & $(N, 4, 64, 64)$ & $4{,}000$ \\
                    & Train (legacy) & Constant left-hot/right-cold BCs \citep{mangnike2024toward} & $(N, 4, 64, 64)$ & $4{,}000$ \\
                    & Test \textsc{Old} ($\mathcal{D}_{\mathrm{old}}$)      & Constant BCs, in-distribution     & $(N, 4, 64, 64)$ & $200$ \\
                    & Test \textsc{Rand} ($\mathcal{D}_{\mathrm{rand}}$)     & Stochastic BCs, in-distribution   & $(N, 4, 64, 64)$ & $200$ \\
                    & Test \textsc{Rand-OOD} ($\mathcal{D}_{\mathrm{rand\text{-}ood}}$) & Stochastic BCs, OOD     & $(N, 4, 64, 64)$ & $200$ \\
\midrule
\multirow{4}{*}{3D} & Train      & Stochastic (Voronoi BCs)            & $(N, 5, 24, 24, 24)$ & $1{,}000$ \\
                    & Test 1 \textsc{Const}    & Constant left-hot/right-cold BCs \citep{mangnike2024toward} & $(N, 5, 24, 24, 24)$ & $100$ \\
                    & Test 2 \textsc{Rand}     & Stochastic BCs, in-distribution & $(N, 5, 24, 24, 24)$ & $100$ \\
                    & Test 3 \textsc{Rand-OOD} & Stochastic BCs, OOD             & $(N, 5, 24, 24, 24)$ & $100$ \\
\bottomrule
\end{tabular}
\end{table}

The four 2D channels are the velocity components $u_{x}, u_{y}$, temperature $T$, and pressure $p$, all sampled at the final simulation time. The 3D channels also include $u_{z}$.
The pressure is stored as a mean-removed perturbation and, like every channel, is min--max normalized to $[0,1]$ using dataset-wide (training-split) statistics rather than per-sample extrema, so that the normalization is consistent across samples (cf.\ Section~\ref{Sto_data}); the absolute pressure scale is not recoverable from the dataset, and the trained surrogate predicts only this normalized perturbation.
All pressure-channel errors reported in this paper, including in Tables~\ref{tab:cross_dataset_metrics} and \ref{tab:cross_dataset_metrics_3d}, refer to this normalized perturbation rather than the absolute pressure field.

Two FNO models are trained in 2D---one on the legacy training set and one on the stochastic training set---which enables the controlled cross-dataset study of Section~\ref{subsec:fno_xdataset}.
We denote these two surrogates $M_{\mathrm{old}}$ and $M_{\mathrm{rand}}$, and the three 2D test distributions $\mathcal{D}_{\mathrm{old}}$, $\mathcal{D}_{\mathrm{rand}}$, and $\mathcal{D}_{\mathrm{rand\text{-}ood}}$; this $(M,\mathcal{D})$ notation is used throughout Section~\ref{subsec:fno_xdataset}.
In 3D, only the stochastic generator is used at training time, and the three test distributions are evaluated against the same single model. The 3D constant-BC test set uses the same legacy left-hot/right-cold protocol as the 2D \textsc{Old} test set \citep{mangnike2024toward}, extended to the third spatial dimension by holding the remaining four walls at adiabatic boundary conditions; this allows the conclusions of the 2D cross-dataset experiment to be checked dimension-by-dimension under a comparable extrapolation regime.

The out-of-distribution test sets are constructed by widening the generator's parameter envelope strictly beyond what is seen during training. In 2D, the Rayleigh number is drawn from $Ra\in[10^{2},10^{8}]$ (versus the narrower training range used in Section~\ref{Sto_data}) and the random Fourier series is truncated at a mode count drawn from $\{1,\dots,5\}$, admitting both smoother and significantly more oscillatory wall-temperature profiles than appear during training. In 3D, the Rayleigh number is drawn from $Ra\in[10^{2},10^{7}]$ and the number of Voronoi seeds per active wall is drawn from $\{3,\dots,7\}$, both ranges strictly wider than at training time. The OOD comparison below is therefore between the surrogate prediction and the same-mesh compressible reference, and should be read as evidence that the surrogate tracks the reference solver across a wider parameter envelope rather than as a claim of physical correctness at these extremes.

\subsection{Training protocol}
\label{subsec:fno_training}

All inputs and targets are rescaled to $[0,1]$ using per-channel min--max statistics computed on the training split only, with the same statistics re-applied at inference time; all errors reported below are in this normalized space. 
The models are trained with a per-channel mean-squared error loss using AdamW, with an initial learning rate of $10^{-4}$ in 2D and $5\!\times\!10^{-4}$ in 3D, gradient clipping at $\|\nabla\|_{2}=1$. 
In 2D we use a plateau scheduler that halves the learning rate after $20$ epochs without validation improvement; in 3D we use a cosine annealing schedule with a $10$-epoch linear warmup and a minimum learning rate of $10^{-6}$.
We use a weight decay of $10^{-3}$ in 2D and $5\!\times\!10^{-3}$ in 3D, and a feature dropout rate of $0.05$ inside each Fourier block to provide mild regularization without introducing a measurable train and validation gap.
A random $20\%$ subset of each training set is held out as a validation split, the model with the lowest validation loss is retained, and we train for $200$ epochs at batch size $64$ in 2D and batch size $32$ in 3D.
All training and inference are performed on a single NVIDIA RTX A6000 GPU (48~GB VRAM), and the inference timings reported in Tables~\ref{tab:cross_dataset_metrics} and \ref{tab:cross_dataset_metrics_3d} are averaged over the corresponding test set.

\begin{figure}[!t]
  \centering
  \begin{subfigure}[t]{0.48\linewidth}
    \includegraphics[width=\linewidth]{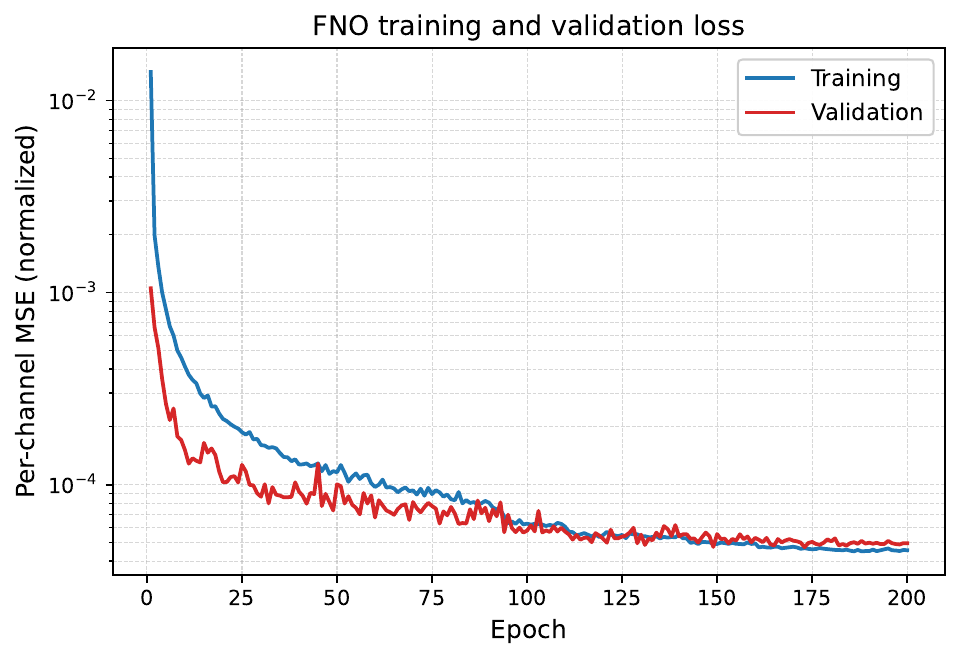}
    \caption{2D, stochastic dataset.}
    \label{fig:loss_curves_2d}
  \end{subfigure}
  \hfill
  \begin{subfigure}[t]{0.48\linewidth}
    \includegraphics[width=\linewidth]{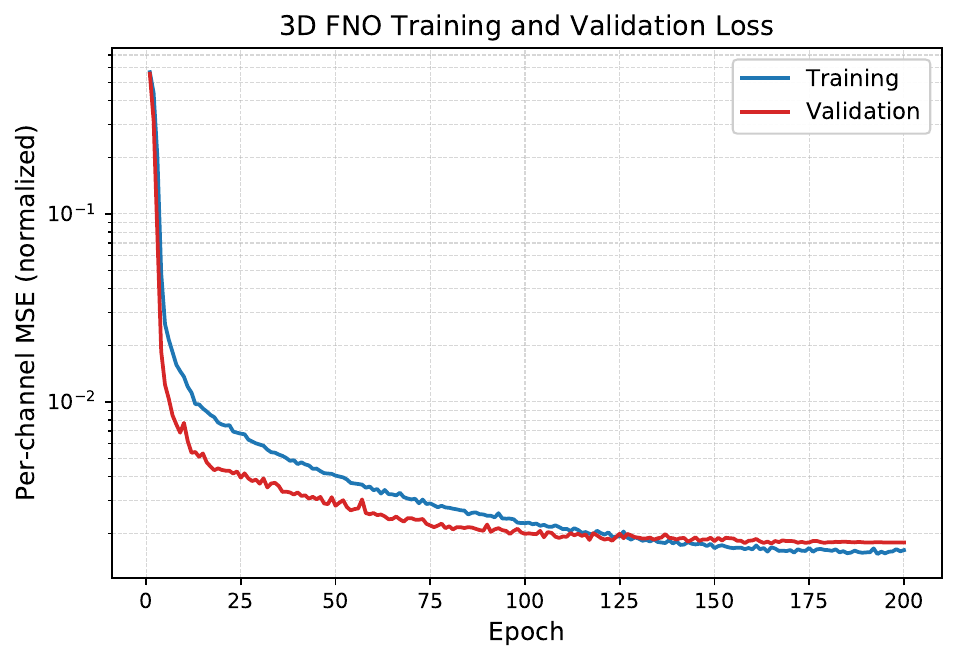}
    \caption{3D, stochastic dataset.}
    \label{fig:loss_curves_3d}
  \end{subfigure}
  \caption{Training and validation loss (per-channel MSE in normalized $[0,1]$ space) versus epoch for the 2D and 3D Fourier neural operator surrogates trained on the stochastic dataset. Both curves are plotted on a log $y$-axis. In both dimensions, training and validation loss decrease at similar rates and converge to the same plateau without divergence, indicating that the chosen weight decay and dropout are sufficient regularization for the dataset size. The validation curve lying slightly below the training curve during the early epochs is a measurement artifact: training loss is evaluated with dropout active (a noisier model) while validation runs in evaluation mode.}
  \label{fig:loss_curves}
\end{figure}

Figure~\ref{fig:loss_curves} shows the training and validation loss curves for the 2D and 3D stochastic models. 
In both dimensions, training and validation loss track each other closely as they decrease at similar rates and converge to the same plateau. Together with the consistent out-of-distribution behavior reported in the following subsections, this indicates that the chosen combination of weight decay and dropout is sufficient to prevent the surrogate from overfitting.

\subsection{Two-dimensional cross-dataset generalization and validation}
\label{subsec:fno_xdataset}

\begin{figure}[p]
  \centering
  \includegraphics[width=0.95\linewidth]{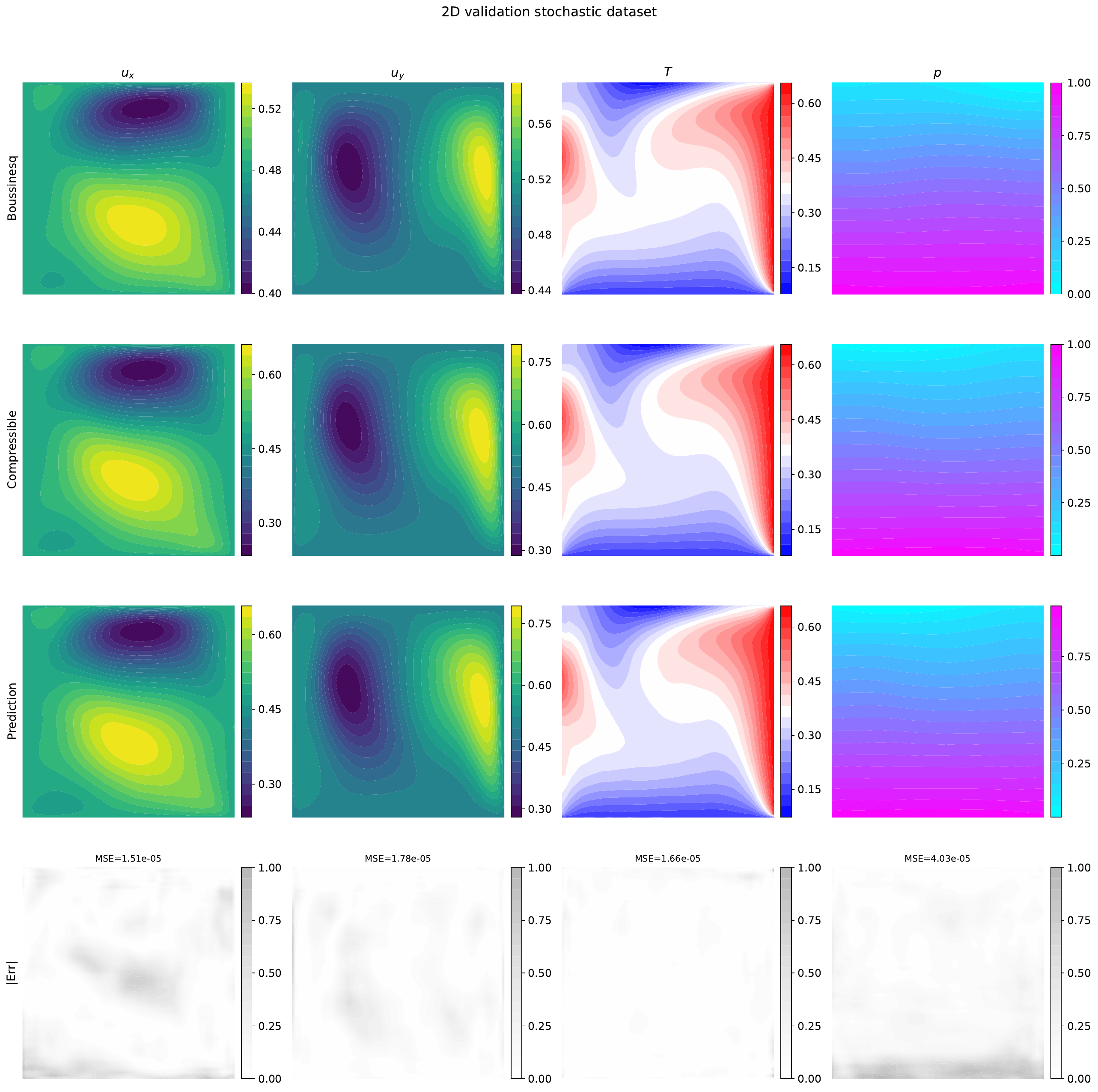}
  \caption{All four flow variables for a single representative in-distribution stochastic sample (the stochastic-trained surrogate evaluated on the stochastic test set, $M_{\mathrm{rand}}$ on $D_{\mathrm{rand}}$). Rows, from top to bottom: the normalized Boussinesq input, the compressible reference solution, the FNO prediction, and the per-field absolute error $|\hat{u}-u|$. Columns correspond to the four flow variables $u_{x}$, $u_{y}$, temperature $T$, and pressure $p$. All fields are shown in normalized $[0,1]$ space; the absolute-error row is mapped to a common normalized grey scale, with the per-field MSE annotated above each error panel. The prediction reproduces all four fields with no visible structure in the error, at per-field MSE of order $10^{-5}$ for this sample. Whereas Figure~\ref{fig:cross_dataset_fno_comparison} contrasts the four (model, data) settings for $u_{x}$ and $T$, this figure shows the complete field set for the in-distribution case.}
  \label{fig:fields_2d_rand}
\end{figure}

A central motivation for the stochastic dataset generator of Section~\ref{Sto_data} is to enlarge the support of the training distribution so that the learned surrogate generalizes beyond a narrow set of forcing patterns.
To quantify this effect, we perform a controlled cross-dataset study: we train the same FNO architecture, separately, on the legacy and stochastic 2D training sets, and evaluate each of the two trained models on the three 2D test sets summarised in Table~\ref{tab:fno_datasets}.
This $2{\times}3$ evaluation matrix isolates the effect of dataset coverage while holding the model class and optimization protocol fixed.

Let $\mathcal{D}_{\mathrm{old}}$, $\mathcal{D}_{\mathrm{rand}}$, and $\mathcal{D}_{\mathrm{rand\text{-}ood}}$ denote the three test distributions. For each $i\in\{\mathrm{old},\mathrm{rand}\}$ we train a surrogate $\mathcal{G}_{\theta_{i}}$ on the corresponding training set and evaluate it on each test distribution, defining the cross-dataset error
\begin{equation}
\mathcal{E}(i\rightarrow j)
\;=\;
\mathbb{E}_{(a,u)\sim \mathcal{D}_{j}}
\big[\,
\ell\big(\mathcal{G}_{\theta_{i}}(a),u\big)
\,\big],
\qquad i\in\{\mathrm{old},\mathrm{rand}\},\ j\in\{\mathrm{old},\mathrm{rand},\mathrm{rand\text{-}ood}\},
\label{eq:cross_eval}
\end{equation}
where $\ell$ is the per-channel mean-squared error (MSE) in normalized $[0,1]$ space; we additionally report the structural similarity index measure (SSIM) and the raw Boussinesq input baseline, which corresponds to setting $\mathcal{G}_{\theta_{i}}=\mathrm{Id}$, as a reference for the achievable error reduction.
In the qualitative figures we label each panel by its (model, data) pairing, writing $M_{i}$ for the surrogate trained on dataset $i$ and $D_{j}$ for the test distribution $j$; thus $M_{\mathrm{old}},\,D_{\mathrm{rand}}$ denotes the legacy-trained surrogate evaluated on stochastic data.

\begin{figure}[p]
  \centering
  \vspace*{\fill}
  \begin{minipage}{\textwidth}
    \centering
    \includegraphics[width=0.85\linewidth]{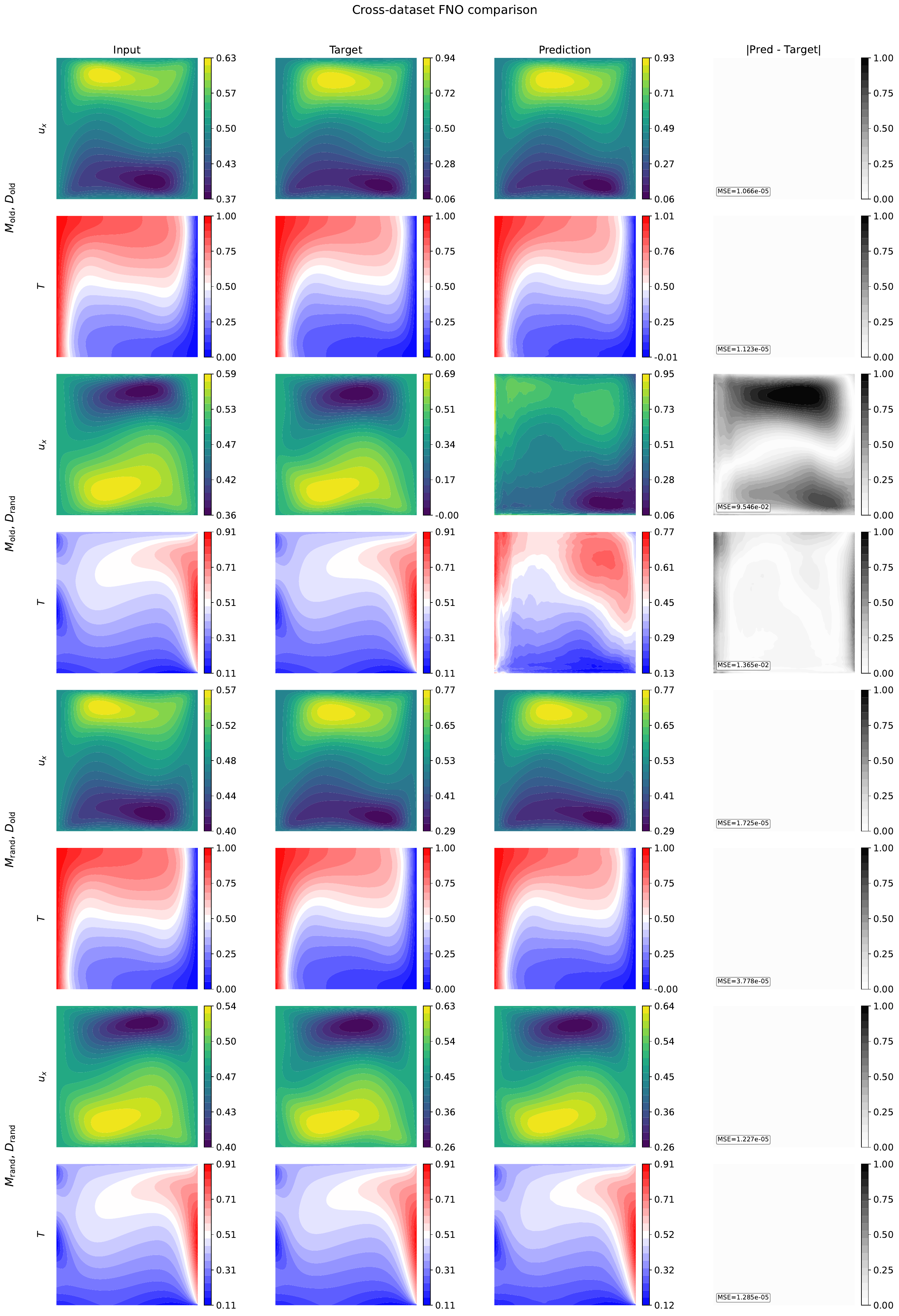}
    \caption{Cross-dataset FNO comparison on a single representative sample, shown for $u_{x}$ (viridis) and $T$ (red--blue). Columns are the normalized input, target, prediction, and absolute error $|\hat{u}-u|$. Each pair of rows is one (model, data) setting, labelled $M_{i},\,D_{j}$ at the far left, where $M_{i}$ is the surrogate trained on dataset $i$ and $D_{j}$ the test distribution: from top, $M_{\mathrm{old}},\,D_{\mathrm{old}}$; $M_{\mathrm{old}},\,D_{\mathrm{rand}}$; $M_{\mathrm{rand}},\,D_{\mathrm{old}}$; and $M_{\mathrm{rand}},\,D_{\mathrm{rand}}$. The error column is mapped to a common normalized $[0,1]$ grey scale and annotated with the per-panel MSE. Only $M_{\mathrm{old}},\,D_{\mathrm{rand}}$ -- the legacy-trained model on stochastic data -- produces a structured, high-magnitude error field; the stochastic-trained model generalizes to the legacy distribution ($M_{\mathrm{rand}},\,D_{\mathrm{old}}$) essentially as well as it performs in distribution.}
    \label{fig:cross_dataset_fno_comparison}
  \end{minipage}
  \vspace*{\fill}
\end{figure}

Three findings emerge from Table~\ref{tab:cross_dataset_metrics} and the qualitative comparisons in Figures~\ref{fig:fields_2d_rand} and \ref{fig:cross_dataset_fno_comparison}.
First, on the in-distribution test sets both surrogates collapse the Boussinesq error substantially---by up to three orders of magnitude (e.g.\ $u_{x}$ on $D_{\mathrm{old}}$ falls from $7.7\times10^{-3}$ to $1.3\times10^{-5}$)---reaching per-channel MSE of order $10^{-5}$ on their own training distribution, while SSIM rises from as low as $0.86$ on the raw input to $\geq 0.995$ on the prediction across all four channels.
Figure~\ref{fig:fields_2d_rand} shows the complete field set for a single in-distribution stochastic sample ($M_{\mathrm{rand}}$ on $D_{\mathrm{rand}}$): the prediction is visually indistinguishable from the compressible target for $u_{x}$, $u_{y}$, $T$, and $p$, and the absolute-error fields are structureless at the $10^{-5}$ level.
Second, the cross-dataset matrix is strongly asymmetric, as Figure~\ref{fig:cross_dataset_fno_comparison} makes immediately visible.
The stochastic-trained surrogate transfers to the legacy test set without meaningful degradation: $M_{\mathrm{rand}}$ on $D_{\mathrm{old}}$ gives per-channel MSE of $1.6$--$6.5\times10^{-5}$, comparable to its own in-distribution performance, with SSIM remaining $\geq 0.998$.
The legacy-trained surrogate, by contrast, not only fails to correct the Boussinesq input on stochastic data but actively amplifies it: $M_{\mathrm{old}}$ on $D_{\mathrm{rand}}$ rises to per-channel MSE of order $10^{-2}$ (e.g.\ $3.4\times10^{-2}$ for $u_{x}$), more than an order of magnitude above the raw Boussinesq baseline ($\sim 2\times10^{-3}$), with SSIM collapsing to $0.55$ for $u_{x}$.
The figure renders this asymmetry starkly: of the four (model, data) settings, only $M_{\mathrm{old}},\,D_{\mathrm{rand}}$ produces a dark, spatially structured error field, while the three remaining settings---including the cross transfer $M_{\mathrm{rand}},\,D_{\mathrm{old}}$---yield near-white, near-zero error maps.
The legacy generator's narrow constant-wall support does not populate the stochastic forcing manifold, and a model trained on it learns a correction that is actively harmful outside that support.
Third, the stochastic-trained surrogate degrades gracefully out of distribution: on $D_{\mathrm{rand\text{-}ood}}$ its per-channel MSE stays within roughly a factor of two of its in-distribution values (all $\leq 1.6\times10^{-4}$, SSIM $\geq 0.994$), whereas the legacy-trained surrogate continues to amplify the error at the same $\sim10^{-2}$ magnitude as on $D_{\mathrm{rand}}$, with no recovery.
Together, the asymmetric matrix and the graceful out-of-distribution behaviour of the stochastic-trained surrogate are direct empirical evidence that the stochastic generator expands the effective support of the training distribution beyond what the dataset's nominal parameter ranges suggest.

\begin{table*}[t!]
\centering
\small
\setlength{\tabcolsep}{4pt}
\caption{Cross-dataset generalization metrics for 2D FNO surrogates trained on the legacy (\textsc{Old}) and stochastic (\textsc{Rand}) datasets and evaluated on three 2D test distributions, averaged over the $N=200$ samples of each test set. Performance is measured using mean-squared error (MSE) and the structural similarity index measure (SSIM) across all flow variables in normalized $[0,1]$ space: the Boussinesq input and the compressible target are each min--max normalized using their own training statistics, so that all reported errors are bounded by unity and directly comparable across rows. The ``Time (s)'' column reports the average wall-clock time per sample for surrogate inference and full Boussinesq or compressible simulation. The per-sample values annotated in Figures~\ref{fig:fields_2d_rand} and \ref{fig:cross_dataset_fno_comparison} are single draws and differ from these averages.}
\label{tab:cross_dataset_metrics}
\resizebox{\textwidth}{!}{%
\begin{tabular}{lrrrrrrrrr}
\toprule
\multirow{3}{*}{Models} & \multicolumn{8}{c}{Flow Variables} & \multirow{3}{*}{Time (s)} \\
\cmidrule(lr){2-9}
 & \multicolumn{2}{c}{Velocity $x$-direction} & \multicolumn{2}{c}{Velocity $y$-direction} & \multicolumn{2}{c}{Temperature} & \multicolumn{2}{c}{Pressure} \\
\cmidrule(lr){2-3}\cmidrule(lr){4-5}\cmidrule(lr){6-7}\cmidrule(lr){8-9}
 & MSE & SSIM & MSE & SSIM & MSE & SSIM & MSE & SSIM \\
\midrule
Compressible (Reference Solution) & - & - & - & - & - & - & - & - & 158.87 \\
\midrule
\multicolumn{10}{l}{\textsc{Test Set 1: Constant BCs (Old)}} \\
Input Baseline (Boussinesq) & $7.69 \times 10^{-3}$ & 0.857 & $7.46 \times 10^{-3}$ & 0.855 & $1.47 \times 10^{-3}$ & 0.985 & $1.83 \times 10^{-2}$ & 0.860 & 11.22 \\
FNO (Trained on Old Data)   & $1.31 \times 10^{-5}$ & 0.999 & $1.19 \times 10^{-5}$ & 0.998 & $9.58 \times 10^{-6}$ & 0.999 & $2.30 \times 10^{-5}$ & 0.995 & $1.1 \times 10^{-3}$ \\
FNO (Trained on Rand Data)  & $3.89 \times 10^{-5}$ & 0.999 & $3.11 \times 10^{-5}$ & 0.999 & $6.53 \times 10^{-5}$ & 0.998 & $1.63 \times 10^{-5}$ & 0.999 & $1.1 \times 10^{-3}$ \\
\midrule
\multicolumn{10}{l}{\textsc{Test Set 2: Stochastic BCs (Rand)}} \\
Input Baseline (Boussinesq) & $2.03 \times 10^{-3}$ & 0.948 & $2.03 \times 10^{-3}$ & 0.946 & $1.32 \times 10^{-4}$ & 0.997 & $4.72 \times 10^{-3}$ & 0.945 & 11.22 \\
FNO (Trained on Old Data)   & $3.37 \times 10^{-2}$ & 0.551 & $1.44 \times 10^{-2}$ & 0.684 & $1.65 \times 10^{-2}$ & 0.769 & $7.10 \times 10^{-3}$ & 0.861 & $1.1 \times 10^{-3}$ \\
FNO (Trained on Rand Data)  & $8.11 \times 10^{-5}$ & 0.997 & $8.42 \times 10^{-5}$ & 0.997 & $4.74 \times 10^{-5}$ & 0.998 & $3.58 \times 10^{-5}$ & 0.999 & $1.1 \times 10^{-3}$ \\
\midrule
\multicolumn{10}{l}{\textsc{Test Set 3: Stochastic BCs OOD (Rand-OOD)}} \\
Input Baseline (Boussinesq) & $2.01 \times 10^{-3}$ & 0.949 & $2.00 \times 10^{-3}$ & 0.948 & $1.41 \times 10^{-4}$ & 0.997 & $3.96 \times 10^{-3}$ & 0.953 & 11.22 \\
FNO (Trained on Old Data)   & $3.60 \times 10^{-2}$ & 0.494 & $1.70 \times 10^{-2}$ & 0.628 & $1.70 \times 10^{-2}$ & 0.742 & $7.53 \times 10^{-3}$ & 0.844 & $1.1 \times 10^{-3}$ \\
FNO (Trained on Rand Data)  & $1.51 \times 10^{-4}$ & 0.994 & $1.42 \times 10^{-4}$ & 0.994 & $7.69 \times 10^{-5}$ & 0.996 & $4.82 \times 10^{-5}$ & 0.999 & $1.1 \times 10^{-3}$ \\
\bottomrule
\end{tabular}%
}
\end{table*}

\begin{table*}[t]
\centering
\small
\setlength{\tabcolsep}{4pt}
\caption{Generalization metrics for the 3D FNO surrogate (trained on stochastic BCs) evaluated on the three 3D test sets. Per-variable mean-squared error (MSE) and structural similarity index measure (SSIM) are reported in normalized $[0,1]$ space; the Boussinesq input and the compressible target are each min--max normalized by their own training statistics, so all reported errors are bounded by unity. SSIM in 3D is computed by averaging 2D SSIM over slices along each of the three spatial axes. The ``Time (s)'' column reports the average wall-clock time per sample for surrogate inference, the Boussinesq solver, and the full compressible simulation. Solver timings are measured at $24^{3}$ resolution on $16$ MPI ranks (average of $N{=}10$ for Boussinesq, $N{=}5$ converged runs for compressible); the FNO runs on a single GPU.}
\label{tab:cross_dataset_metrics_3d}
\resizebox{\textwidth}{!}{%
\begin{tabular}{lrrrrrrrrrrr}
\toprule
\multirow{3}{*}{Models} & \multicolumn{10}{c}{Flow Variables} & \multirow{3}{*}{Time (s)} \\
\cmidrule(lr){2-11}
 & \multicolumn{2}{c}{$u_x$} & \multicolumn{2}{c}{$u_y$} & \multicolumn{2}{c}{$u_z$} & \multicolumn{2}{c}{Temperature} & \multicolumn{2}{c}{Pressure} \\
\cmidrule(lr){2-3}\cmidrule(lr){4-5}\cmidrule(lr){6-7}\cmidrule(lr){8-9}\cmidrule(lr){10-11}
 & MSE & SSIM & MSE & SSIM & MSE & SSIM & MSE & SSIM & MSE & SSIM \\
\midrule
Compressible (Reference Solution) & - & - & - & - & - & - & - & - & - & - & 3245.58 \\
\midrule
\multicolumn{12}{l}{\textsc{Test Set 1: Constant BCs}} \\
Input Baseline (Boussinesq) & $6.69 \times 10^{-3}$ & 0.887 & $2.11 \times 10^{-3}$ & 0.905 & $4.35 \times 10^{-3}$ & 0.942 & $6.57 \times 10^{-3}$ & 0.900 & $2.31 \times 10^{-2}$ & 0.851 & 583.52 \\
FNO (Trained on Stochastic) & $1.04 \times 10^{-3}$ & 0.944 & $5.54 \times 10^{-4}$ & 0.948 & $5.46 \times 10^{-4}$ & 0.959 & $1.02 \times 10^{-3}$ & 0.945 & $6.46 \times 10^{-3}$ & 0.916 & 0.0016 \\
\midrule
\multicolumn{12}{l}{\textsc{Test Set 2: Stochastic BCs (In-Distribution)}} \\
Input Baseline (Boussinesq) & $1.31 \times 10^{-2}$ & 0.750 & $6.27 \times 10^{-3}$ & 0.802 & $3.43 \times 10^{-3}$ & 0.944 & $1.47 \times 10^{-2}$ & 0.757 & $1.40 \times 10^{-2}$ & 0.853 & 583.52 \\
FNO (Trained on Stochastic) & $2.02 \times 10^{-4}$ & 0.986 & $1.42 \times 10^{-4}$ & 0.983 & $9.43 \times 10^{-5}$ & 0.986 & $2.99 \times 10^{-4}$ & 0.983 & $1.55 \times 10^{-4}$ & 0.992 & 0.0016 \\
\midrule
\multicolumn{12}{l}{\textsc{Test Set 3: Stochastic BCs (Out-of-Distribution)}} \\
Input Baseline (Boussinesq) & $1.17 \times 10^{-2}$ & 0.823 & $2.93 \times 10^{-3}$ & 0.882 & $5.22 \times 10^{-3}$ & 0.904 & $4.87 \times 10^{-3}$ & 0.919 & $3.61 \times 10^{-2}$ & 0.767 & 583.52 \\
FNO (Trained on Stochastic) & $2.64 \times 10^{-3}$ & 0.910 & $1.04 \times 10^{-3}$ & 0.932 & $1.18 \times 10^{-3}$ & 0.937 & $9.30 \times 10^{-4}$ & 0.935 & $1.19 \times 10^{-2}$ & 0.876 & 0.0016 \\
\bottomrule
\end{tabular}%
}
\end{table*}
 
\subsection{Three-dimensional results}
\label{subsec:fno_3d}

\begin{figure}[p]
  \centering
  \vspace*{\fill}

  \begin{minipage}{\linewidth}
    \centering
    \includegraphics[width=0.98\linewidth]{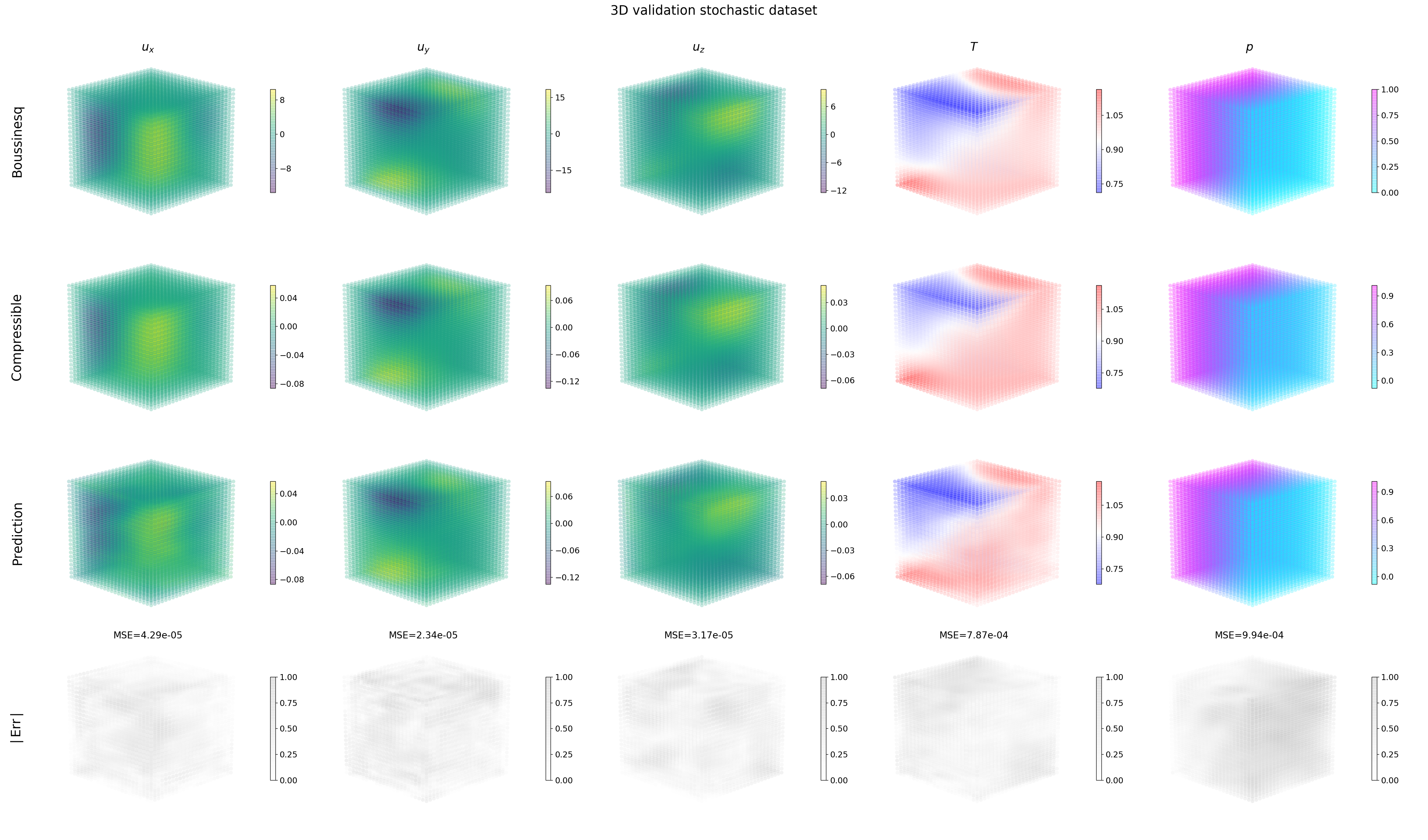}
    \caption{Per-channel qualitative comparison of the 3D stochastic-trained FNO surrogate on one in-distribution stochastic test sample. Rows show the Boussinesq input, the compressible target, the FNO prediction, and the pointwise absolute error $|\mathrm{Pred} - \mathrm{Comp}|$. Columns are the three velocity components $u_{x}$, $u_{y}$, $u_{z}$, the temperature $T$, and the pressure $p$. The error row reports the per-channel MSE (physical units); error values are normalized to $[0,1]$ relative to the per-channel maximum for visualization.}
    \label{fig:fno3d_validation_v2}
  \end{minipage}

  \vspace*{\fill}

  \begin{minipage}{\linewidth}
    \centering
    \includegraphics[width=0.95\linewidth]{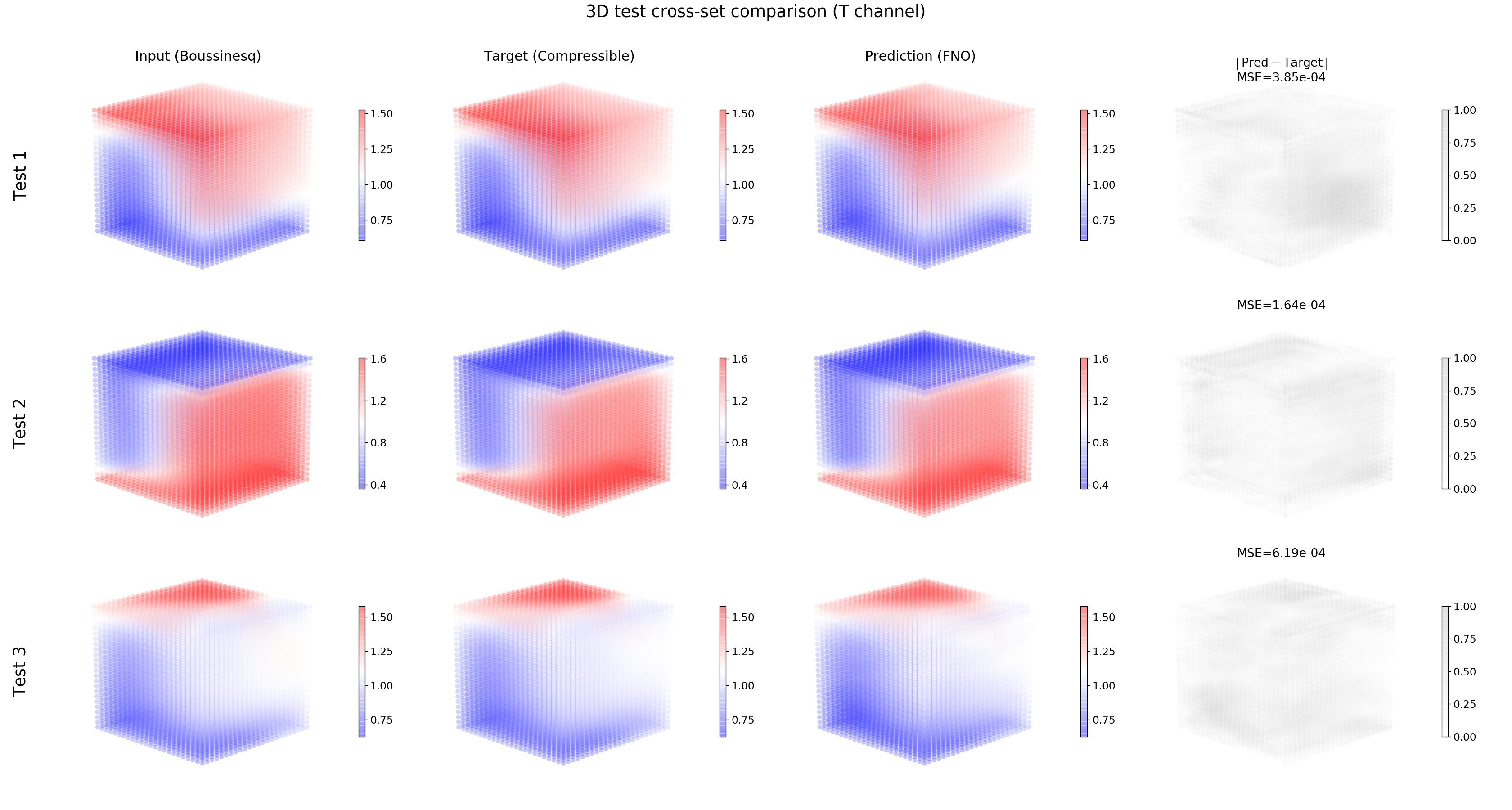}
    \caption{Cross test-set qualitative comparison of the 3D stochastic-trained FNO surrogate on the temperature field, for one random test sample (randomly picked from all three test sets). Rows correspond to the in-distribution stochastic test set (Test~1), the constant-BC test set (Test~2), and the out-of-distribution stochastic test set (Test~3). Columns show, from left to right, the Boussinesq input, the compressible target, the FNO prediction, and the pointwise absolute error. Per-row MSE is annotated on the error column.}
    \label{fig:fno3d_cross_T}
  \end{minipage}

  \vspace*{\fill}
\end{figure}

In three dimensions, for sake computational efficiency,
we train a single 3D FNO on the stochastic dataset and validate whether the stochastic-trained 3D surrogate generalizes to constant-BC inputs and to parameter regimes outside its training envelope.
Table~\ref{tab:cross_dataset_metrics_3d} reports the same metrics as Table~\ref{tab:cross_dataset_metrics}, evaluated on the three 3D test sets defined in Section~\ref{subsec:fno_datasets}.
The single trained model reduces the Boussinesq baseline MSE across all five flow variables on every test set: by factors of roughly $36$--$90$ (one to two orders of magnitude) in distribution, and by factors of roughly $3$--$8$ under the two extrapolation regimes, with in-distribution SSIM uniformly at or above $0.978$.
On the constant-BC test set, SSIM drops to $[0.916, 0.959]$ and the MSE gains are more modest (with SSIM on $u_z$ essentially unchanged from the baseline), but the surrogate still reduces the error on every channel.
On the OOD test set with wider $Ra$ range and richer wall patterns, performance remains qualitatively similar to the in-distribution stochastic case, with SSIM~$\in [0.876, 0.937]$ across all five channels and MSE within a factor of one order of magnitude from the in-distribution values for the velocity components and temperature.
A qualitative view of the in-distribution behavior is given in Figure~\ref{fig:fno3d_validation_v2}, which shows all five flow variables for one stochastic test sample. 
The FNO prediction visually matches the compressible target almost indistinguishably in all five channels, and the pointwise absolute error row quantifies this, showing residuals concentrated near the active walls where the Voronoi temperature pattern produces the strongest gradients, with per-channel MSE values consistent with Table~\ref{tab:cross_dataset_metrics_3d} (Test~2 rows).

Cross-regime behavior for the temperature channel is shown in Figure~\ref{fig:fno3d_cross_T} since this field is most directly tied to the boundary conditions and is therefore the most discriminating between regimes.
The figure displays the Boussinesq input, the compressible target, the FNO prediction, and the pointwise absolute error for one random sample on each of the three 3D test sets.
The qualitative agreement between target and prediction is essentially indistinguishable across the in-distribution and the two extrapolation regimes, despite the constant-BC and OOD samples having been produced by physical configurations the surrogate never saw during training.
The per-row MSE values confirm that the constant-BC and OOD cases sit within an order of magnitude of the in-distribution case, with no qualitative breakdown of the predicted temperature structure.

Taken together with the 2D cross-dataset matrix of Section~\ref{subsec:fno_xdataset}, the in-distribution, constant-BC, and OOD results show that our stochastic generator produces a rich and statistically diverse training set. 
The model trained on it still performs well on unseen distributions and holds up when pushed to Rayleigh numbers and boundary-condition complexities beyond the training range.

We close with a note on practical scope. The training cost of the surrogate is dominated by the compressible reference solver used to generate the paired dataset: $4{,}000$ runs at $\sim 159$~s each in 2D and $1{,}000$ runs at $\sim 3{,}246$~s each in 3D (Tables~\ref{tab:cross_dataset_metrics}--\ref{tab:cross_dataset_metrics_3d}). The 3D training set alone therefore requires approximately $900$ wall hours of compressible solves on the reference configuration. This cost is paid once. Each subsequent query of the trained model consists of one Boussinesq simulation ($\sim 11$~s in 2D, $\sim 584$~s in 3D) followed by a single FNO inference ($1.1$~ms in 2D, $1.6$~ms in 3D), replacing a full compressible solve. Equating cumulative cost---$(\text{Boussinesq} + \text{FNO} + \text{training})$ versus running the compressible solver directly---yields a break-even point near the training set size: roughly $4{,}300$ evaluations in 2D and $1{,}200$ in 3D under the timings reported above. The surrogate is therefore most useful in workflows that require many simulations on a fixed cavity geometry at fixed resolution: uncertainty quantification, parameter sweeps, design exploration, and ensemble runs.
At present, we make no claim that the trained surrogate transfers across geometries or mesh resolutions; both are fixed during training and inference, and applying the surrogate outside these conditions would require retraining.
Nonetheless, adding adaptors to make our surrogate mesh- and resolution-invariant is a natural direction for future work.

\subsection{Comparison against a convolutional U-Net baseline}
\label{sec:unet_comparison}
To isolate the role of the architecture from that of the training data, we compare the FNO against the task-specific convolutional U-Net introduced in~\citet{mangnike2024toward}, a modified U-Net developed for precisely this Boussinesq-to-compressible correction problem.
We retrain it in both 2D and 3D under a protocol identical to the FNO (same per-channel min--max normalization, AdamW optimizer, learning-rate schedule, batch size, and epoch budget), so that the surrogate architecture is the only difference.

The FNO and the U-Net sit at very different points on the size/speed axis, and the FNO's footprint is governed almost entirely by its spectral convolution weights.
Each spectral layer stores an independent complex weight for every channel pair and every retained Fourier mode, so the parameter count scales as $\mathcal{O}\!\left(\text{width}^{2}\cdot\textstyle\prod_{d}\text{modes}_{d}\cdot\text{depth}\right)$.
For our 2D configuration (width $128$, $24{\times}24$ modes, depth $6$) this term alone accounts for $99.9\%$ of the $113.4$\,M parameters, versus the U-Net's $1.9$\,M; the same factor sets the size of the per-layer Fourier-domain contractions and therefore drives inference cost ($\sim$$1.1\times10^{-3}$\,s/sample, about $10\times$ the U-Net's).
From this large budget we also train a compact FNO (width $64$, $16{\times}16$ modes, depth $4$); by the scaling above, this reduces parameter count to $8.4$\,M ($13.5\times$ fewer, now only $\sim$$4\times$ the U-Net) and lowers its inference cost to $\sim$$4\times10^{-4}$\,s/sample --- roughly $3\times$ faster than the full FNO, though still about $4\times$ the U-Net's.

Table~\ref{tab:unet_comparison} reports all surrogates on the in-distribution stochastic test set (Test~2); every surrogate is structurally faithful (SSIM $\ge 0.988$ in 2D, $\ge 0.85$ in 3D).
The full FNO is the most accurate, outperforming the U-Net by $62$--$78\%$ in 2D and by $80$--$98\%$ in 3D (where it reaches SSIM $\ge 0.98$ on every variable).
Crucially, this advantage survives compression: even at $\sim$$4\times$ the U-Net's parameter count, the compact FNO still reduces the U-Net's MSE by $32$--$65\%$ across all four 2D fields.
The spectral architecture therefore outperforms the convolutional baseline at similar model scale, indicating that the FNO's accuracy reflects the operator-learning formulation rather than merely its larger parameter budget.
In 3D the FNO is already compact relative to its 2D counterpart ($14.2$\,M, $\sim$$4\times$ the U-Net), so no reduction is required and the large margin persists.

All surrogates reduce the Boussinesq baseline by roughly one to two orders of magnitude in MSE for the velocity and pressure channels, at a tiny fraction of the solver cost; the temperature channel, already close to compressible in the Boussinesq input, improves more modestly.
This architecture/capacity comparison should ultimately be read against the dominant effect of the training distribution: the \textsc{Old}-trained FNO fails on the stochastic test sets (Table~\ref{tab:cross_dataset_metrics}) while the \textsc{Rand}-trained FNO succeeds, and we observe the identical pattern for the U-Net, confirming that cross-dataset generalization is governed primarily by the stochastic training data rather than by the choice of surrogate.

\begin{table*}[t]
\centering
\caption{FNO vs.\ the specialized convolutional U-Net of Mangnike and Hyde~\cite{mangnike2024toward} on the in-distribution stochastic test set (Test Set 2), in both 2D and 3D, with all surrogates trained on the stochastic (\textsc{Rand}) dataset. Per-variable MSE and SSIM are reported in normalized $[0,1]$ space (3D SSIM is the average of 2D SSIM over slices along the three spatial axes). The Boussinesq input and compressible target are each min--max normalized by their own training statistics so that all errors are bounded by unity. Parameter counts: 2D U-Net 1.9\,M, FNO 113.4\,M (full, width 128 / 24 modes / depth 6) and 8.4\,M (compact, width 64 / 16 modes / depth 4); 3D U-Net 3.2\,M, FNO 14.2\,M. The compact FNO probes whether the full model's accuracy stems from its spectral architecture or its parameter budget. The red sub-row gives the relative MSE change of each FNO versus the equally-trained U-Net per field ($\downarrow$/$\uparrow$ = lower/higher error). The ``Time (s)'' column reports the average wall-clock time per sample for surrogate inference and the Boussinesq solver.}
\label{tab:unet_comparison}
\resizebox{\textwidth}{!}{%
\begin{tabular}{l cc cc cc cc cc c}
\toprule
\multirow{2}{*}{Models} & \multicolumn{10}{c}{Flow Variables} & \multirow{2}{*}{Time (s)} \\
\cmidrule(lr){2-11}
 & \multicolumn{2}{c}{$u_x$} & \multicolumn{2}{c}{$u_y$} & \multicolumn{2}{c}{$u_z$} & \multicolumn{2}{c}{Temperature} & \multicolumn{2}{c}{Pressure} & \\
\cmidrule(lr){2-3}\cmidrule(lr){4-5}\cmidrule(lr){6-7}\cmidrule(lr){8-9}\cmidrule(lr){10-11}
 & MSE & SSIM & MSE & SSIM & MSE & SSIM & MSE & SSIM & MSE & SSIM & \\
\midrule
\multicolumn{12}{l}{\textsc{Two-Dimensional} (Test Set 2: Stochastic BCs)}\\
Input Baseline (Boussinesq) & $2.03\times10^{-3}$ & 0.948 & $2.03\times10^{-3}$ & 0.946 & -- & -- & $1.32\times10^{-4}$ & 0.997 & $4.72\times10^{-3}$ & 0.945 & 11.22 \\
U-Net (1.9\,M)              & $2.11\times10^{-4}$ & 0.989 & $2.41\times10^{-4}$ & 0.988 & -- & -- & $1.29\times10^{-4}$ & 0.995 & $1.63\times10^{-4}$ & 0.990 & 0.0001 \\
FNO, full (113.4\,M)        & $8.11\times10^{-5}$ & 0.997 & $8.42\times10^{-5}$ & 0.997 & -- & -- & $4.74\times10^{-5}$ & 0.998 & $3.58\times10^{-5}$ & 0.999 & 0.0011 \\
\quad vs.\ U-Net & \textcolor{red}{$\downarrow$62\%} & & \textcolor{red}{$\downarrow$65\%} & & & & \textcolor{red}{$\downarrow$63\%} & & \textcolor{red}{$\downarrow$78\%} & & \\
FNO, compact (8.4\,M)       & $1.44\times10^{-4}$ & 0.993 & $1.23\times10^{-4}$ & 0.994 & -- & -- & $5.75\times10^{-5}$ & 0.997 & $5.69\times10^{-5}$ & 0.996 & 0.0004 \\
\quad vs.\ U-Net & \textcolor{red}{$\downarrow$32\%} & & \textcolor{red}{$\downarrow$49\%} & & & & \textcolor{red}{$\downarrow$55\%} & & \textcolor{red}{$\downarrow$65\%} & & \\
\midrule
\multicolumn{12}{l}{\textsc{Three-Dimensional} (Test Set 2: Stochastic BCs)}\\
Input Baseline (Boussinesq) & $1.31\times10^{-2}$ & 0.750 & $6.27\times10^{-3}$ & 0.803 & $3.43\times10^{-3}$ & 0.944 & $1.47\times10^{-2}$ & 0.757 & $1.40\times10^{-2}$ & 0.853 & 583.52 \\
U-Net (3.2\,M)              & $4.70\times10^{-3}$ & 0.861 & $3.32\times10^{-3}$ & 0.858 & $7.04\times10^{-4}$ & 0.913 & $1.52\times10^{-3}$ & 0.934 & $7.58\times10^{-3}$ & 0.974 & 0.0007 \\
FNO (14.2\,M)               & $2.02\times10^{-4}$ & 0.986 & $1.42\times10^{-4}$ & 0.983 & $9.43\times10^{-5}$ & 0.986 & $2.99\times10^{-4}$ & 0.983 & $1.55\times10^{-4}$ & 0.992 & 0.0016 \\
\quad vs.\ U-Net & \textcolor{red}{$\downarrow$96\%} & & \textcolor{red}{$\downarrow$96\%} & & \textcolor{red}{$\downarrow$87\%} & & \textcolor{red}{$\downarrow$80\%} & & \textcolor{red}{$\downarrow$98\%} & & \\
\bottomrule
\end{tabular}%
}
\end{table*}

\section{Conclusions}
\label{sec:conc}

In this work, we developed and validated an implicit monolithic mixed finite element framework for simulating both Boussinesq and fully compressible natural convection flows.
By leveraging the flexibility of the FEniCSx platform, we closed an important gap in open-source tools capable of handling the stiffness associated with low-Mach number regimes without relying on segregating pressure-correction schemes or explicit time-stepping.
Our methodological approach focused on a globally coupled, fully implicit formulation using Taylor-Hood elements.
In contrast to traditional partitioned solvers that often struggle with the acoustic-convective timescale disparity in low-Mach flows, our monolithic approach maintains robustness and thermodynamic consistency by solving the momentum, continuity, and energy equations, along with the equation of state, simultaneously.

We then presented a novel stochastic data generation pipeline designed for scientific machine learning. 
We introduced Fourier- and Voronoi-based boundary condition perturbation methods that generate statistically independent, physically consistent training datasets.
This addresses a critical bottleneck in the field of data-driven CFD, where the lack of diverse, high-fidelity training data often hinders the generalization of physics-informed neural networks (PINNs) and surrogate models. 
By releasing this codebase as an open-source resource, we provide the community with a reproducible pipeline to generate ``ML-ready'' fluid dynamics data.
Finally, we constructed a novel FNO-based neural surrogate model that learns to correct Boussinesq flow results to more closely resemble the results of an equivalent compressible flow simulation.
This enables practitioners to run Boussinesq flow simulations yet obtain results that are much closer in quality to simulations run using compressible flow models, with in-distribution structural similarity close to unity and per-channel mean-squared error reduced by one to nearly three orders of magnitude, using only a single neural network evaluation.

Future work will focus on extending the framework to turbulent flow regimes by integrating Reynolds-Averaged Navier–-Stokes (RANS) or Large Eddy Simulation (LES) models within the monolithic variational form. 
Additionally, we plan to explore in-situ coupling between the FEniCSx solver and PyTorch, enabling active learning workflows where the solver dynamically generates data in regions of high varying uncertainty during model training. 

\section*{Acknowledgments}

D.H.\ was supported by NSF Grant No.\ 2324735.
The authors acknowledge Abhinav Gupta for help learning FEniCS at the outset of the project.
The authors also acknowledge Tianrun Gao for helpful proofreading of our manuscript and derivations.

\section*{Use of Generative AI}

The following generative AI models were used in preparation of this manuscript, including code and writing: GPT 5.1, GPT 5.2, Gemini 3, Claude Opus 4.5, and Claude Opus 4.8.

\newpage

\bibliography{bibliography}

@misc{BarattaEtal2023,
  title     = {{DOLFINx}: the next generation {FEniCS} problem solving environment},
  author    = {Baratta, Igor A. and Dean, Joseph P. and Dokken, J{\o}rgen S. and Habera, Michal and Hale, Jack S. and Richardson, Chris N. and Rognes, Marie E. and Scroggs, Matthew W. and Sime, Nathan and Wells, Garth N.},
  doi       = {10.5281/zenodo.10447666},
  year      = {2023},
  howpublished = {preprint},
  url={https://doi.org/10.5281/zenodo.10447666}
}

@book{boussinesq1903theorie,
  title={Th{\'e}orie analytique de la chaleur mise en harmonic avec la thermodynamique et avec la th{\'e}orie m{\'e}canique de la lumi{\`e}re: Refroidissement et {\'e}chauffement par rayonnement, conductibilit{\'e} des tiges, lames et masses cristallines, courants de convection, th{\'e}orie m{\'e}canique de la lumi{\`e}re},
  author={Boussinesq, Joseph},
  volume={2},
  year={1903},
  publisher={Gauthier-Villars}
}

@article{de1983natural,
  title={Natural convection of air in a square cavity: a bench mark numerical solution},
  author={de Vahl Davis, Graham},
  journal={International Journal for Numerical Methods in Fluids},
  volume={3},
  number={3},
  pages={249--264},
  year={1983},
  publisher={Wiley Online Library},
  doi={https://doi.org/10.1002/fld.1650030305},
  url={https://doi.org/10.1002/fld.1650030305}
}

@article{sharpdeep,
    author = {Gibou, Frederic and Hyde, David and Fedkiw, Ron},
    title = {Sharp Interface Approaches and Deep Learning Techniques for Multiphase Flows},
    journal = {Journal of Computational Physics},
    volume = {380},
    pages = {442--463},
    year = {2019},
    month = {},
    doi = {10.1016/j.jcp.2018.05.031},
    url = {https://www.sciencedirect.com/science/article/pii/S0021999118303371},
    publisher = {Elsevier Inc.}
}

@book{logg2012automated,
  title={Automated Solution of Differential Equations by the Finite Element Method: The FEniCS Book},
  author={Anders Logg and Kent‐Andre Mardal and Garth N. Wells},
  volume={84},
  year={2012},
  publisher={Springer Science \& Business Media},
  url={https://api.semanticscholar.org/CorpusID:59832704}
}

@inproceedings{mangnike2024toward,
author = {Mangnike, Nurshat and Hyde, David},
title = {Toward Improving {Boussinesq} Flow Simulations by Learning with Compressible Flow},
year = {2024},
isbn = {9798400706394},
publisher = {Association for Computing Machinery},
address = {New York, NY, USA},
url = {https://doi.org/10.1145/3659914.3659919},
doi = {10.1145/3659914.3659919},
booktitle = {Proceedings of the Platform for Advanced Scientific Computing Conference},
articleno = {5},
numpages = {12},
location = {Zurich, Switzerland},
series = {PASC '24}
}

@incollection{TU20181,
    title = {Chapter 1 - Introduction},
    editor = {Jiyuan Tu and Guan-Heng Yeoh and Chaoqun Liu},
    booktitle = {Computational Fluid Dynamics (Third Edition)},
    publisher = {Butterworth-Heinemann},
    edition = {Third Edition},
    pages = {1-31},
    year = {2018},
    isbn = {978-0-08-101127-0},
    doi = {https://doi.org/10.1016/B978-0-08-101127-0.00001-5},
    url = {https://www.sciencedirect.com/science/article/pii/B9780081011270000015},
    author = {Jiyuan Tu and Guan-Heng Yeoh and Chaoqun Liu}
}

@inproceedings{petsc-efficient,
  title={Efficient Management of Parallelism in Object-Oriented Numerical Software Libraries},
  author={Satish Balay and William Gropp and Lois Curfman McInnes and Barry F. Smith},
  booktitle={Modern Software Tools in Scientific Computing},
  year={1997},
  url={https://api.semanticscholar.org/CorpusID:56520908}
}

@article{brooks1982streamline,
title = {Streamline {upwind/Petrov-Galerkin} formulations for convection dominated flows with particular emphasis on the incompressible {Navier-Stokes} equations},
journal = {Computer Methods in Applied Mechanics and Engineering},
volume = {32},
number = {1},
pages = {199-259},
year = {1982},
issn = {0045-7825},
doi = {https://doi.org/10.1016/0045-7825(82)90071-8},
url = {https://www.sciencedirect.com/science/article/pii/0045782582900718},
author = {Alexander N. Brooks and Thomas J.R. Hughes}
}

@article{hughes1986new,
  title={A new finite element formulation for computational fluid dynamics: V. Circumventing the Babu{\v{s}}ka-Brezzi condition: A stable {Petrov-Galerkin} formulation of the {Stokes} problem accommodating equal-order interpolations},
  author={Thomas Joseph Robert Hughes and Leopoldo P. Franca and Marc Balestra},
  journal={Applied Mechanics and Engineering},
  year={1986},
  volume={59},
  pages={85-99},
  url={https://api.semanticscholar.org/CorpusID:120805215}
}

@article{lappa2022incompressible,
     author = {Marcello Lappa},
     title = {Incompressible flows and the {Boussinesq} approximation: 50 years of {CFD}},
     journal = {Comptes Rendus. M\'ecanique},
     pages = {75--96},
     year = {2022},
     publisher = {Acad\'emie des sciences, Paris},
     volume = {350},
     number = {S1},
     doi = {10.5802/crmeca.134},
     url={https://doi.org/10.5802/crmeca.134}
}

@article{TURKEL1987277,
title = {Preconditioned methods for solving the incompressible and low speed compressible equations},
journal = {Journal of Computational Physics},
volume = {72},
number = {2},
pages = {277-298},
year = {1987},
issn = {0021-9991},
doi = {https://doi.org/10.1016/0021-9991(87)90084-2},
url = {https://www.sciencedirect.com/science/article/pii/0021999187900842},
author = {Eli Turkel}
}

@article{turkel1999preconditioning,
  title={Preconditioning techniques in computational fluid dynamics},
  author={Turkel, Eli},
  journal={Annual Review of Fluid Mechanics},
  volume={31},
  number={1},
  pages={385--416},
  year={1999},
  publisher={Annual Reviews 4139 El Camino Way, PO Box 10139, Palo Alto, CA 94303-0139, USA},
  url={https://doi.org/10.1146/annurev.fluid.31.1.385},
  doi={10.1146/annurev.fluid.31.1.385}
}

@article{weiss1995preconditioning,
  title={Preconditioning Applied to Variable and Constant Density Flows},
  author={Jonathan M. Weiss and Wayne Smith},
  journal={AIAA Journal},
  year={1995},
  volume={33},
  pages={2050-2057},
  url={https://api.semanticscholar.org/CorpusID:119737915}
}

@article{guillard1999behaviour,
  title={On the behaviour of upwind schemes in the low Mach number limit},
  author={Guillard, Herv{\'e} and Viozat, C{\'e}cile},
  journal={Computers \& fluids},
  doi = {https://doi.org/10.1016/S0045-7930(98)00017-6},
  url = {https://www.sciencedirect.com/science/article/pii/S0045793098000176},
  volume={28},
  number={1},
  pages={63--86},
  year={1999},
  publisher={Elsevier}
}

@article{GRAY1976545,
title = {The validity of the {Boussinesq} approximation for liquids and gases},
journal = {International Journal of Heat and Mass Transfer},
volume = {19},
number = {5},
pages = {545-551},
year = {1976},
issn = {0017-9310},
doi = {https://doi.org/10.1016/0017-9310(76)90168-X},
url = {https://www.sciencedirect.com/science/article/pii/001793107690168X},
author = {Donald D. Gray and Aldo Giorgini}
}

@article{yanaoka2025numerical,
  title={A numerical method for low {Mach} number compressible flows by simultaneous relaxation of dependent variables},
  author={Yanaoka, Hideki and Sato, Yuji},
  journal={arXiv preprint arXiv:2502.08116},
  year={2025},
  doi={10.48550/arXiv.2502.08116},
  url={https://doi.org/10.48550/arXiv.2502.08116}
}

@article{rehm1978equations,
  title={The Equations of Motion for Thermally Driven, Buoyant Flows.},
  author={Ronald G. Rehm and Howard R. Baum},
  journal={Journal of Research of the National Bureau of Standards},
  year={1978},
  volume={83 3},
  pages={
          297-308
        },
  url={https://api.semanticscholar.org/CorpusID:13971757}
}

@article{le1992chebyshev,
title = {A {Chebyshev} collocation algorithm for {2D non-Boussinesq} convection},
journal = {Journal of Computational Physics},
volume = {103},
number = {2},
pages = {320-335},
year = {1992},
issn = {0021-9991},
doi = {https://doi.org/10.1016/0021-9991(92)90404-M},
url = {https://www.sciencedirect.com/science/article/pii/002199919290404M},
author = {Patrick L. {Le Quéré} and R Masson and Pierre Perrot}
}

@article{mortensen2011fenics,
title = {A {FEniCS}-based programming framework for modeling turbulent flow by the {Reynolds-averaged Navier–Stokes} equations},
journal = {Advances in Water Resources},
volume = {34},
number = {9},
pages = {1082-1101},
year = {2011},
issn = {0309-1708},
doi = {https://doi.org/10.1016/j.advwatres.2011.02.013},
url = {https://www.sciencedirect.com/science/article/pii/S030917081100039X},
author = {Mikael Mortensen and Hans Petter Langtangen and Garth N. Wells}
}

@article{vynnytska2013benchmarking,
title = {Benchmarking {FEniCS} for mantle convection simulations},
journal = {Computers \& Geosciences},
volume = {50},
pages = {95-105},
year = {2013},
issn = {0098-3004},
doi = {https://doi.org/10.1016/j.cageo.2012.05.012},
url = {https://www.sciencedirect.com/science/article/pii/S0098300412001689},
author = {Lyudmyla Vynnytska and Marie E. Rognes and Stuart R. Clark}
}

@article{zhang2016mixed,
title = {A mixed finite element solver for natural convection in porous media using automated solution techniques},
journal = {Computers \& Geosciences},
volume = {96},
pages = {181-192},
year = {2016},
issn = {0098-3004},
doi = {https://doi.org/10.1016/j.cageo.2016.08.012},
url = {https://www.sciencedirect.com/science/article/pii/S0098300416302746},
author = {Chao Zhang and Sadiq J. Zarrouk and Rosalind Archer}
}

@article{multiphysics1998introduction,
  title={Introduction to {COMSOL} multiphysics{\textregistered}},
  author={Multiphysics, COMSOL},
  journal={COMSOL Multiphysics, Burlington, MA},
  year={1998},
  url={https://www.comsol.com}
}

@article{armijo1966minimization,
  title={Minimization of functions having {Lipschitz} continuous first partial derivatives},
  author={Larry Armijo},
  journal={Pacific Journal of Mathematics},
  year={1966},
  volume={16},
  pages={1-3},
  doi={10.2140/PJM.1966.16.1},
  url={https://api.semanticscholar.org/CorpusID:54977700}
}

@inproceedings{amestoy2013mumps,
  title={{MUMPS} {MUltifrontal} {Massively} {Parallel} {Solver} Version 2.0},
  author={Patrick R. Amestoy and Iain S. Duff and Jean-Yves L’Excellent},
  year={1998},
  url={https://api.semanticscholar.org/CorpusID:60073183}
}

@article{saad1993flexible,
  title={A Flexible Inner-Outer Preconditioned {GMRES} Algorithm},
  author={Yousef Saad},
  journal={SIAM J. Sci. Comput.},
  year={1993},
  volume={14},
  pages={461-469},
  url={https://api.semanticscholar.org/CorpusID:12540446}
}

@article{yang2002boomeramg,
title = {{BoomerAMG}: A parallel algebraic multigrid solver and preconditioner},
journal = {Applied Numerical Mathematics},
volume = {41},
number = {1},
pages = {155-177},
year = {2002},
issn = {0168-9274},
doi = {https://doi.org/10.1016/S0168-9274(01)00115-5},
url = {https://www.sciencedirect.com/science/article/pii/S0168927401001155},
author = {Van Emden Henson and Ulrike Meier Yang}
}

@article{Spalart2016OnTR,
  title={On the role and challenges of {CFD} in the aerospace industry},
  author={Philippe R. Spalart and V. Venkatakrishnan},
  journal={The Aeronautical Journal},
  year={2016},
  volume={120},
  pages={209 - 232},
  url={https://api.semanticscholar.org/CorpusID:7731130}
}

@article{FUJII2005455,
title = {Progress and future prospects of {CFD} in aerospace—Wind tunnel and beyond},
journal = {Progress in Aerospace Sciences},
volume = {41},
number = {6},
pages = {455-470},
year = {2005},
issn = {0376-0421},
doi = {https://doi.org/10.1016/j.paerosci.2005.09.001},
url = {https://www.sciencedirect.com/science/article/pii/S0376042105001016},
author = {Kozo Fujii}
}

@article{CHINTALA2013709,
title = {A {CFD} (computational fluid dynamics) study for optimization of gas injector orientation for performance improvement of a dual-fuel diesel engine},
journal = {Energy},
volume = {57},
pages = {709-721},
year = {2013},
issn = {0360-5442},
doi = {https://doi.org/10.1016/j.energy.2013.06.009},
url = {https://www.sciencedirect.com/science/article/pii/S0360544213005070},
author = {Venkateswarlu Chintala and Kallipatti A. Subramanian}
}

@article{YE2024112639,
title = {Data-driven reduced-order modelling for blood flow simulations with geometry-informed snapshots},
journal = {Journal of Computational Physics},
volume = {497},
pages = {112639},
year = {2024},
issn = {0021-9991},
doi = {https://doi.org/10.1016/j.jcp.2023.112639},
url = {https://www.sciencedirect.com/science/article/pii/S0021999123007349},
author = {Dongwei Ye and Valeria Krzhizhanovskaya and Alfons G. Hoekstra}
}

@article{ESMAILYMOGHADAM201363,
title = {A modular numerical method for implicit {0D/3D} coupling in cardiovascular finite element simulations},
journal = {Journal of Computational Physics},
volume = {244},
pages = {63-79},
year = {2013},
note = {Multi-scale Modeling and Simulation of Biological Systems},
issn = {0021-9991},
doi = {https://doi.org/10.1016/j.jcp.2012.07.035},
url = {https://www.sciencedirect.com/science/article/pii/S0021999112004202},
author = {Mahdi {Esmaily Moghadam} and Irene E. Vignon-Clementel and Richard Figliola and Alison L. Marsden}
}

@Inbook{Reid2021,
author="Reid, Luke",
title="An Introduction to Biomedical Computational Fluid Dynamics",
bookTitle="Biomedical Visualisation: Volume 10",
year="2021",
publisher="Springer International Publishing",
address="Cham",
pages="205--222",
isbn="978-3-030-76951-2",
doi="10.1007/978-3-030-76951-2_10",
url="https://doi.org/10.1007/978-3-030-76951-2_10"
}

@article{babanezhad2020high,
  title={High-performance hybrid modeling chemical reactors using differential evolution based fuzzy inference system},
  author={Babanezhad, Meisam and Behroyan, Iman and Nakhjiri, Ali Taghvaie and Marjani, Azam and Rezakazemi, Mashallah and Shirazian, Saeed},
  journal={Scientific Reports},
  volume={10},
  number={1},
  pages={21304},
  year={2020},
  publisher={Nature Publishing Group UK London},
  doi={10.1038/s41598-020-78277-3},
  url={https://doi.org/10.1038/s41598-020-78277-3}
}

@inproceedings{zawawi2018review,
  title={A review: Fundamentals of computational fluid dynamics ({CFD})},
  author={Zawawi, Mohd Hafiz and Saleha, A. and Salwa, A. and Hassan, N.H. and Zahari, Nazirul Mubin and Ramli, Mohd Zakwan and Muda, Zakaria Che},
  booktitle={AIP conference proceedings},
  volume={2030},
  pages={020252},
  year={2018},
  organization={AIP Publishing LLC},
  doi={10.1063/1.5066893},
  url={https://doi.org/10.1063/1.5066893}
}

@article{KIM1999145,
title = {Application of {CFD} to environmental flows},
journal = {Journal of Wind Engineering and Industrial Aerodynamics},
volume = {81},
number = {1},
pages = {145-158},
year = {1999},
issn = {0167-6105},
doi = {https://doi.org/10.1016/S0167-6105(99)00013-6},
url = {https://www.sciencedirect.com/science/article/pii/S0167610599000136},
author = {Sung-Eun Kim and Ferit Boysan}
}

@article{TOMINAGA2024105741,
title = {{CFD} simulations of turbulent flow and dispersion in built environment: A perspective review},
journal = {Journal of Wind Engineering and Industrial Aerodynamics},
volume = {249},
pages = {105741},
year = {2024},
issn = {0167-6105},
doi = {https://doi.org/10.1016/j.jweia.2024.105741},
url = {https://www.sciencedirect.com/science/article/pii/S0167610524001041},
author = {Yoshihide Tominaga}
}

@article{LUO2012133,
title = {An implicit discontinuous {Galerkin} method for the unsteady compressible {Navier}–{Stokes} equations},
journal = {Computers \& Fluids},
volume = {53},
pages = {133-144},
year = {2012},
issn = {0045-7930},
doi = {https://doi.org/10.1016/j.compfluid.2011.10.009},
url = {https://www.sciencedirect.com/science/article/pii/S0045793011003082},
author = {Hong Luo and Hidehiro Segawa and Miguel R. Visbal}
}

@article{ISSA198640,
title = {Solution of the implicitly discretised fluid flow equations by operator-splitting},
journal = {Journal of Computational Physics},
volume = {62},
number = {1},
pages = {40-65},
year = {1986},
issn = {0021-9991},
doi = {https://doi.org/10.1016/0021-9991(86)90099-9},
url = {https://www.sciencedirect.com/science/article/pii/0021999186900999},
author = {Raad I. Issa}
}

@article{PATANKAR19721787,
title = {A calculation procedure for heat, mass and momentum transfer in three-dimensional parabolic flows},
journal = {International Journal of Heat and Mass Transfer},
volume = {15},
number = {10},
pages = {1787-1806},
year = {1972},
issn = {0017-9310},
doi = {https://doi.org/10.1016/0017-9310(72)90054-3},
url = {https://www.sciencedirect.com/science/article/pii/0017931072900543},
author = {Suhas V. Patankar and Dudley B. Spalding}
}

@article{rhie1989pressure,
  title={Pressure-based Navier-Stokes solver using the multigrid method},
  author={Rhie, Chae M.},
  journal={AIAA journal},
  volume={27},
  number={8},
  pages={1017--1018},
  year={1989},
  doi = {10.2514/6.1986-207},
  url={https://arc.aiaa.org/doi/pdf/10.2514/6.1986-207}
}

@article{karki1989pressure,
  title={Pressure based calculation procedure for viscous flows at all speedsin arbitrary configurations},
  author={Karki, Kailash C. and Patankar, Suhas V.},
  journal={AIAA journal},
  volume={27},
  number={9},
  pages={1167--1174},
  year={1989},
  url={https://doi.org/10.2514/3.10242}
}

@article{CHOI1993207,
title = {The Application of Preconditioning in Viscous Flows},
journal = {Journal of Computational Physics},
volume = {105},
number = {2},
pages = {207-223},
year = {1993},
issn = {0021-9991},
doi = {https://doi.org/10.1006/jcph.1993.1069},
url = {https://www.sciencedirect.com/science/article/pii/S0021999183710697},
author = {Yunho Choi and Charles L. Merkle}
}

@article{Doormaal01041984,
author={Jeffrey Peter van Doormaal and George D. Raithby},
title = {ENHANCEMENTS OF THE SIMPLE METHOD FOR PREDICTING INCOMPRESSIBLE FLUID FLOWS},
journal = {Numerical Heat Transfer},
volume = {7},
number = {2},
pages = {147--163},
year = {1984},
publisher = {Taylor \& Francis},
doi = {10.1080/01495728408961817},
URL = {https://doi.org/10.1080/01495728408961817}
}

@article{HENNINK2021109877,
title = {A pressure-based solver for low-{Mach} number flow using a discontinuous {Galerkin} method},
journal = {Journal of Computational Physics},
volume = {425},
pages = {109877},
year = {2021},
issn = {0021-9991},
doi = {https://doi.org/10.1016/j.jcp.2020.109877},
url = {https://www.sciencedirect.com/science/article/pii/S0021999120306513},
author = {Aldo Hennink and Marco Tiberga and Danny Lathouwers}
}

@article{WALL2002545,
title = {A Semi-implicit Method for Resolution of Acoustic Waves in Low Mach Number Flows},
journal = {Journal of Computational Physics},
volume = {181},
number = {2},
pages = {545-563},
year = {2002},
issn = {0021-9991},
doi = {https://doi.org/10.1006/jcph.2002.7141},
url = {https://www.sciencedirect.com/science/article/pii/S002199910297141X},
author = {Clifton Wall and Charles D. Pierce and Parviz Moin}
}

@article{TAYLOR197373,
title = {A numerical solution of the Navier-Stokes equations using the finite element technique},
journal = {Computers \& Fluids},
volume = {1},
number = {1},
pages = {73-100},
year = {1973},
issn = {0045-7930},
doi = {https://doi.org/10.1016/0045-7930(73)90027-3},
url = {https://www.sciencedirect.com/science/article/pii/0045793073900273},
author = {Cedric M. Taylor and P. Hood}
}

@article{Vierendeels2003BenchmarkSF,
  title={Benchmark solutions for the natural convective heat transfer problem in a square cavity with large horizontal temperature differences},
  author={Jan A. Vierendeels and Bart Merci and Erik Dick},
  journal={International Journal of Numerical Methods for Heat \& Fluid Flow},
  year={2003},
  volume={13},
  pages={1057-1078},
  url={https://api.semanticscholar.org/CorpusID:121701784}
}

@article{UECKERMANN2016390,
title = {Hybridizable discontinuous {Galerkin} projection methods for {Navier}–{Stokes} and {Boussinesq} equations},
journal = {Journal of Computational Physics},
volume = {306},
pages = {390-421},
year = {2016},
issn = {0021-9991},
doi = {https://doi.org/10.1016/j.jcp.2015.11.028},
url = {https://www.sciencedirect.com/science/article/pii/S0021999115007688},
author = {M.P. Ueckermann and P.F.J. Lermusiaux}
}

@article{SCHROEDER2017760,
title = {Stabilised {dG}-{FEM} for incompressible natural convection flows with boundary and moving interior layers on non-adapted meshes},
journal = {Journal of Computational Physics},
volume = {335},
pages = {760-779},
year = {2017},
issn = {0021-9991},
doi = {https://doi.org/10.1016/j.jcp.2017.01.055},
url = {https://www.sciencedirect.com/science/article/pii/S0021999117300712},
author = {Philipp W. Schroeder and Gert Lube}
}

@article{ELMAN20113900,
title = {Fast iterative solvers for buoyancy driven flow problems},
journal = {Journal of Computational Physics},
volume = {230},
number = {10},
pages = {3900-3914},
year = {2011},
issn = {0021-9991},
doi = {https://doi.org/10.1016/j.jcp.2011.02.014},
url = {https://www.sciencedirect.com/science/article/pii/S0021999111001033},
author = {Howard Elman and Milan Mihajlović and David Silvester}
}

@article{SONG2023112458,
title = {A numerical approach to the optimal control of thermally convective flows},
journal = {Journal of Computational Physics},
volume = {494},
pages = {112458},
year = {2023},
issn = {0021-9991},
doi = {https://doi.org/10.1016/j.jcp.2023.112458},
url = {https://www.sciencedirect.com/science/article/pii/S0021999123005533},
author = {Yongcun Song and Xiaoming Yuan and Hangrui Yue}
}

@article{YU2006424,
title = {A fictitious domain method for particulate flows with heat transfer},
journal = {Journal of Computational Physics},
volume = {217},
number = {2},
pages = {424-452},
year = {2006},
issn = {0021-9991},
doi = {https://doi.org/10.1016/j.jcp.2006.01.016},
url = {https://www.sciencedirect.com/science/article/pii/S0021999106000167},
author = {Zhaosheng Yu and Xueming Shao and Anthony Wachs}
}

@article{liu2003fourth,
  title={A fourth order scheme for incompressible {Boussinesq} equations},
  author={Liu, Jian-Guo and Wang, Cheng and Johnston, Hans},
  journal={Journal of Scientific Computing},
  doi={https://doi.org/10.1023/A:1021168924020},
  URL={https://link.springer.com/article/10.1023/A:1021168924020},
  volume={18},
  number={2},
  pages={253--285},
  year={2003},
  publisher={Springer}
}

@article{FUSEGI19911543,
title = {A numerical study of three-dimensional natural convection in a differentially heated cubical enclosure},
journal = {International Journal of Heat and Mass Transfer},
volume = {34},
number = {6},
pages = {1543-1557},
year = {1991},
issn = {0017-9310},
doi = {https://doi.org/10.1016/0017-9310(91)90295-P},
url = {https://www.sciencedirect.com/science/article/pii/001793109190295P},
author = {Toru Fusegi and Jae-min Hyun and Kunio Kuwahara and Bakhtier Farouk}
}

@article{mayeli2021buoyancy,
  title={Buoyancy-driven flows beyond the {Boussinesq} approximation: A brief review},
  author={Mayeli, Peyman and Sheard, Gregory J.},
  journal={International Communications in Heat and Mass Transfer},
  doi = {https://doi.org/10.1016/j.icheatmasstransfer.2021.105316},
  url = {https://www.sciencedirect.com/science/article/pii/S0735193321002098},
  volume={125},
  pages={105316},
  year={2021},
  publisher={Elsevier}
}

@article{zeytounian2003joseph,
  title={{Joseph} {Boussinesq} and his approximation: a contemporary view},
  author={Zeytounian, Radyadour Kh.},
  journal={Comptes Rendus Mecanique},
  doi = {https://doi.org/10.1016/S1631-0721(03)00120-7},
  url = {https://www.sciencedirect.com/science/article/pii/S1631072103001207},
  volume={331},
  number={8},
  pages={575--586},
  year={2003},
  publisher={Elsevier}
}

@article{cockburn2009unified,
  title={Unified hybridization of discontinuous Galerkin, mixed, and continuous Galerkin methods for second order elliptic problems},
  author={Cockburn, Bernardo and Gopalakrishnan, Jayadeep and Lazarov, Raytcho},
  journal={SIAM Journal on Numerical Analysis},
  doi = {10.1137/070706616},
  URL = {ttps://doi.org/10.1137/070706616},
  volume={47},
  number={2},
  pages={1319--1365},
  year={2009},
  publisher={SIAM}
}

@article{nguyen2009implicit,
  title={An implicit high-order hybridizable discontinuous {Galerkin} method for linear convection--diffusion equations},
  author={Nguyen, Ngoc Cuong and Peraire, Jaume and Cockburn, Bernardo},
  journal={Journal of Computational Physics},
  doi = {https://doi.org/10.1016/j.jcp.2009.01.030},
  url = {https://www.sciencedirect.com/science/article/pii/S0021999109000308},
  volume={228},
  number={9},
  pages={3232--3254},
  year={2009},
  publisher={Elsevier}
}

@article{DAMANIK20093869,
title = {A monolithic {FEM}-multigrid solver for non-isothermal incompressible flow on general meshes},
journal = {Journal of Computational Physics},
volume = {228},
number = {10},
pages = {3869-3881},
year = {2009},
issn = {0021-9991},
doi = {https://doi.org/10.1016/j.jcp.2009.02.024},
url = {https://www.sciencedirect.com/science/article/pii/S0021999109000941},
author = {H. Damanik and J. Hron and A. Ouazzi and S. Turek}
}

@article{Smethurst2013UnstructuredFE,
  title={Unstructured finite element method for the solution of the {Boussinesq} problem in three dimensions},
  author={Christopher A. Smethurst and David J. Silvester and Milan D. Mihajlovic},
  journal={International Journal for Numerical Methods in Fluids},
  year={2013},
  volume={73},
  url={https://api.semanticscholar.org/CorpusID:119789681}
}

@article{MILLER2022150,
title = {Versatile mixed methods for non-isothermal incompressible flows},
journal = {Computers \& Mathematics with Applications},
volume = {125},
pages = {150-175},
year = {2022},
issn = {0898-1221},
doi = {https://doi.org/10.1016/j.camwa.2022.08.044},
url = {https://www.sciencedirect.com/science/article/pii/S0898122122003704},
author = {Edward A. Miller and Xi Chen and David M. Williams}
}

@article{SADAKA2020107492,
title = {Parallel finite-element codes for the simulation of two-dimensional and three-dimensional solid–liquid phase-change systems with natural convection},
journal = {Computer Physics Communications},
volume = {257},
pages = {107492},
year = {2020},
issn = {0010-4655},
doi = {https://doi.org/10.1016/j.cpc.2020.107492},
url = {https://www.sciencedirect.com/science/article/pii/S0010465520302319},
author = {Georges Sadaka and Aina Rakotondrandisa and Pierre-Henri Tournier and Francky Luddens and Corentin Lothodé and Ionut Danaila}
}

@ARTICLE{11447375,
  author={Hyde, David and Kadapa, Chennakesava and Albustami, Abdallah Alalem and Taha, Ahmad F.},
  journal={Computing in Science \& Engineering}, 
  title={Progress and Perils of PINNs: Perspectives on Applying Physics-Informed Neural Networks to Science and Engineering Problems}, 
  url = {https://ieeexplore.ieee.org/document/11447375},
  year={2026},
  volume={},
  number={},
  pages={1-9},
  doi={10.1109/MCSE.2026.3675465}
}

@article{willard2020integrating,
  title={Integrating physics-based modeling with machine learning: A survey},
  author={Willard, Jared and Jia, Xiaowei and Xu, Shaoming and Steinbach, Michael and Kumar, Vipin},
  journal={arXiv preprint arXiv:2003.04919},
  volume={1},
  number={1},
  pages={1--34},
  year={2020},
  url={https://arxiv.org/pdf/2003.04919.pdf}
}

@article{karniadakis2021physics,
  title={Physics-informed machine learning},
  author={Karniadakis, George Em and Kevrekidis, Ioannis G. and Lu, Lu and Perdikaris, Paris and Wang, Sifan and Yang, Liu},
  journal={Nature Reviews Physics},
  doi={https://doi.org/10.1038/s42254-021-00314-5},
  URL={https://www.nature.com/articles/s42254-021-00314-5},
  volume={3},
  number={6},
  pages={422--440},
  year={2021},
  publisher={Nature Publishing Group UK London}
}

@ARTICLE{712178,
  author={Lagaris, I.E. and Likas, A. and Fotiadis, D.I.},
  journal={IEEE Transactions on Neural Networks}, 
  title={Artificial neural networks for solving ordinary and partial differential equations}, 
  doi={10.1109/72.712178},
  URL={https://ieeexplore.ieee.org/document/712178},
  year={1998},
  volume={9},
  number={5},
  pages={987-1000}
}

@article{raissi2019physics,
  title={Physics-informed neural networks: A deep learning framework for solving forward and inverse problems involving nonlinear partial differential equations},
  author={Raissi, Maziar and Perdikaris, Paris and Karniadakis, George E},
  journal={Journal of Computational physics},
  doi = {https://doi.org/10.1016/j.jcp.2018.10.045},
  url = {https://www.sciencedirect.com/science/article/pii/S0021999118307125},
  volume={378},
  pages={686--707},
  year={2019},
  publisher={Elsevier}
}

@article{dissanayake1994neural,
  title={Neural-network-based approximations for solving partial differential equations},
  author={Dissanayake, M.W.M. Gamini and Phan-Thien, Nhan},
  journal={Communications in Numerical Methods in Engineering},
  doi = {https://doi.org/10.1002/cnm.1640100303},
  url = {https://onlinelibrary.wiley.com/doi/abs/10.1002/cnm.1640100303},
  volume={10},
  number={3},
  pages={195--201},
  year={1994},
  publisher={Wiley Online Library}
}

@inproceedings{
    dgcm,
    author = {Kaneda, Ayano and Akar, Osman and Chen, Jingyu and Kala, Victoria Alicia Trevino and Hyde, David and Teran, Joseph},
    booktitle = {Proceedings of the 40th International Conference on Machine Learning},
    title = {A Deep Conjugate Direction Method for Iteratively Solving Linear Systems},
    year = {2023},
    pages = {15720--15736},
    editor = {Krause, Andreas and Brunskill, Emma and Cho, Kyunghyun and Engelhardt, Barbara and Sabato, Sivan and Scarlett, Jonathan},
    volume = {202},
    series = {Proceedings of Machine Learning Research},
    publisher = {PMLR},
    url = {https://proceedings.mlr.press/v202/kaneda23a.html}
}

@inproceedings{yang2023reinforcement,
  title={Reinforcement learning for adaptive mesh refinement},
  author={Yang, Jiachen and Dzanic, Tarik and Petersen, Brenden and Kudo, Jun and Mittal, Ketan and Tomov, Vladimir and Camier, Jean-Sylvain and Zhao, Tuo and Zha, Hongyuan and Kolev, Tzanio and others},
  url = 	 {https://proceedings.mlr.press/v206/yang23e.html},
  booktitle={International Conference on Artificial Intelligence and Statistics},
  pages={5997--6014},
  year={2023},
  organization={PMLR}
}

@article{viquerat2020supervised,
  title={A supervised neural network for drag prediction of arbitrary 2D shapes in laminar flows at low Reynolds number},
  author={Viquerat, Jonathan and Hachem, Elie},
  doi = {https://doi.org/10.1016/j.compfluid.2020.104645},
  url = {https://www.sciencedirect.com/science/article/pii/S0045793020302164},
  journal={Computers \& Fluids},
  volume={210},
  pages={104645},
  year={2020},
  publisher={Elsevier}
}

@article { SpatiallyLocalSurrogateModelingofSubgridScaleEffectsinIdealizedAtmosphericFlowsADeepLearnedApproachUsingHighResolutionSimulationData,
      author = "Muralikrishnan Gopalakrishnan Meena and Matthew R. Norman and David M. Hall and Michael S. Pritchard",
      title = "Spatially Local Surrogate Modeling of Subgrid-Scale Effects in Idealized Atmospheric Flows: A Deep Learned Approach Using High-Resolution Simulation Data",
      journal = "Artificial Intelligence for the Earth Systems",
      year = "2024",
      publisher = "American Meteorological Society",
      address = "Boston MA, USA",
      volume = "3",
      number = "4",
      doi = "10.1175/AIES-D-23-0043.1",
      pages=      "e230043",
      url = "https://journals.ametsoc.org/view/journals/aies/3/4/AIES-D-23-0043.1.xml"
}

@article{eiximeno2025deep,
  title={On deep-learning-based closures for algebraic surrogate models of turbulent flows},
  author={Eiximeno, Benet and Sanchis-Agudo, Marcial and Mir{\'o}, Arnau and Rodriguez, Ivette and Vinuesa, Ricardo and Lehmkuhl, Oriol},
  journal={Journal of Fluid Mechanics},
  DOI={10.1017/jfm.2025.10610},
  URL={https://www.cambridge.org/core/journals/journal-of-fluid-mechanics/article/on-deeplearningbased-closures-for-algebraic-surrogate-models-of-turbulent-flows/A1A32C52F9892284A7C290362CF501D2},
  volume={1020},
  pages={A36},
  year={2025},
  publisher={Cambridge University Press}
}

@inproceedings{bhattacharjee2026machine,
  title={A Machine Learning Model for the Prediction of Sub-Grid Interfacial Area in Two-Phase Turbulent Flows},
  author={Bhattacharjee, Anirban and Hatashita, Luis H. and Jain, Suhas S.},
  booktitle={AIAA SCITECH 2026 Forum},
  doi={https://doi.org/10.2514/6.2026-1731},
  URL={https://arc.aiaa.org/doi/10.2514/6.2026-1731},
  pages={1731},
  year={2026}
}

@article{stuckner2021optimal,
  title={Optimal experimental design with fast neural network surrogate models},
  author={Stuckner, Joshua and Piekenbrock, Matthew and Arnold, Steven M. and Ricks, Trenton M.},
  journal={Computational Materials Science},
  doi = {https://doi.org/10.1016/j.commatsci.2021.110747},
  url = {https://www.sciencedirect.com/science/article/pii/S0927025621004742},
  volume={200},
  pages={110747},
  year={2021},
  publisher={Elsevier}
}

@article{khorrami2023artificial,
  title={An artificial neural network for surrogate modeling of stress fields in viscoplastic polycrystalline materials},
  author={Khorrami, Mohammad S. and Mianroodi, Jaber R. and Siboni, Nima H. and Goyal, Pawan and Svendsen, Bob and Benner, Peter and Raabe, Dierk},
  journal={npj Computational Materials},
  doi={https://doi.org/10.1038/s41524-023-00991-z},
  URL={https://www.nature.com/articles/s41524-023-00991-z},
  volume={9},
  number={1},
  pages={37},
  year={2023},
  publisher={Nature Publishing Group UK London}
}

@article{zhou2026latent,
  title={Latent representation learning based model correction and uncertainty quantification for {PDEs}},
  author={Zhou, Wenwen and Feng, Xiaodong and Guo, Ling and Wu, Hao},
  journal={arXiv preprint arXiv:2603.24948},
  doi={http://dx.doi.org/10.2139/ssrn.6629548},
  url={https://api.semanticscholar.org/CorpusID:286790070},
  year={2026}
}

@article{vagnoli2025local,
  title={A local and explicit forcing correction for Lagrangian immersed boundary methods},
  author={Vagnoli, Giovanni and Scarpolini, Martino A and Verzicco, Roberto and Viola, Francesco},
  journal={Computer Physics Communications},
  doi = {https://doi.org/10.1016/j.cpc.2025.109741},
  url = {https://www.sciencedirect.com/science/article/pii/S0010465525002437},
  pages={109741},
  year={2025},
  publisher={Elsevier}
}

@article{duraisamy2019turbulence,
  title={Turbulence modeling in the age of data},
  author={Duraisamy, Karthik and Iaccarino, Gianluca and Xiao, Heng},
  journal={Annual review of fluid mechanics},
  doi={https://doi.org/10.1146/annurev-fluid-010518-040547},
  URL={https://www.annualreviews.org/content/journals/10.1146/annurev-fluid-010518-040547},
  volume={51},
  number={1},
  pages={357--377},
  year={2019},
  publisher={Annual Reviews}
}

@article{farchi2021using,
  title={Using machine learning to correct model error in data assimilation and forecast applications},
  author={Farchi, Alban and Laloyaux, Patrick and Bonavita, Massimo and Bocquet, Marc},
  journal={Quarterly Journal of the Royal Meteorological Society},
  doi = {https://doi.org/10.1002/qj.4116},
  url = {https://rmets.onlinelibrary.wiley.com/doi/abs/10.1002/qj.4116},
  volume={147},
  number={739},
  pages={3067--3084},
  year={2021},
  publisher={Wiley Online Library}
}

@book{hughes2000finite,
  author    = {Hughes, Thomas J. R.},
  title     = {The Finite Element Method: Linear Static and Dynamic
               Finite Element Analysis},
  publisher = {Dover Publications},
  address   = {Mineola, NY},
  year      = {2000},
  isbn      = {978-0-486-41181-1},
  note      = {Unabridged republication of the work originally published
               by Prentice-Hall, Englewood Cliffs, NJ, 1987}
}

@article{johnson2025software,
  title={Software-based automatic differentiation is flawed},
  author={Johnson, Daniel and Maxfield, Trevor and Jin, Yongxu and Fedkiw, Ronald},
  doi = {https://doi.org/10.1016/j.jcp.2025.114319},
  url = {https://www.sciencedirect.com/science/article/pii/S002199912500600X},
  journal={Journal of Computational Physics},
  pages={114319},
  year={2025},
  publisher={Elsevier}
}

@article{zhang2021petscsf,
  title={The {PetscSF} scalable communication layer},
  author={Zhang, Junchao and Brown, Jed and Balay, Satish and Faibussowitsch, Jacob and Knepley, Matthew and Marin, Oana and Mills, Richard Tran and Munson, Todd and Smith, Barry F and Zampini, Stefano},
  doi={10.1109/TPDS.2021.3084070},
  URL={https://www.computer.org/csdl/journal/td/2022/04/09442258/1tV59V9gKn6},
  journal={IEEE Transactions on Parallel and Distributed Systems},
  volume={33},
  number={4},
  pages={842--853},
  year={2021},
  publisher={IEEE}
}

@article{doi:10.1137/S0036144599352836,
author = {Du, Qiang and Faber, Vance and Gunzburger, Max},
title = {Centroidal {Voronoi} Tessellations: Applications and Algorithms},
journal = {SIAM Review},
volume = {41},
number = {4},
pages = {637-676},
year = {1999},
doi = {10.1137/S0036144599352836},
URL = {     
        https://doi.org/10.1137/S0036144599352836
},
eprint = {     
        https://doi.org/10.1137/S0036144599352836
}
}

@inproceedings{li2021fourier,
  title={Fourier Neural Operator for Parametric Partial Differential Equations},
  author={Li, Zongyi and Kovachki, Nikola and Azizzadenesheli, Kamyar and Liu, Burigede and Bhattacharya, Kaushik and Stuart, Andrew and Anandkumar, Anima},
  booktitle={International Conference on Learning Representations (ICLR)},
  year={2021},
  url={https://openreview.net/forum?id=c8P9NQVtmnO},
  eprint={2010.08895},
  archivePrefix={arXiv}
}

@article{kovachki2023neural,
  title={Neural Operator: Learning Maps Between Function Spaces with Applications to {PDE}s},
  author={Kovachki, Nikola and Li, Zongyi and Liu, Burigede and Azizzadenesheli, Kamyar and Bhattacharya, Kaushik and Stuart, Andrew and Anandkumar, Anima},
  doi={https://doi.org/10.48550/arXiv.2108.08481},
  URL={https://dl.acm.org/doi/10.5555/3648699.3648788},
  journal={Journal of Machine Learning Research},
  volume={24},
  number={89},
  pages={1--97},
  year={2023}
}

@article{rakotondrandisa2020finite,
title = {A finite-element toolbox for the simulation of solid--liquid phase-change systems with natural convection},
journal = {Computer Physics Communications},
volume = {253},
pages = {107188},
year = {2020},
issn = {0010-4655},
doi = {https://doi.org/10.1016/j.cpc.2020.107188},
url = {https://www.sciencedirect.com/science/article/pii/S0010465520300151},
author = {Aina Rakotondrandisa and Georges Sadaka and Ionut Danaila}
}

@article{wakitani2001numerical,
    author = {Wakitani, Shunichi },
    title = {Numerical Study of Three-Dimensional Oscillatory Natural Convection at Low Prandtl Number in Rectangular Enclosures },
    journal = {Journal of Heat Transfer},
    volume = {123},
    number = {1},
    pages = {77-83},
    year = {2000},
    month = {09},
    issn = {0022-1481},
    doi = {10.1115/1.1336508},
    url = {https://doi.org/10.1115/1.1336508},
    eprint = {https://asmedigitalcollection.asme.org/heattransfer/article-pdf/123/1/77/5485487/77_1.pdf},
}

\newpage
\appendix

\section{Derivation of Non-dimensionalized and Variational Forms}
\label{sec:app-derivations}

This appendix contains step-by-step derivations of the weak forms of the non-dimensionalized governing equations for Boussinesq and compressible flow (see Section \ref{sec:math_formulations}).
These derivations are included primarily for pedagogical purposes.

\subsection{Boussinesq flow equations}
\subsubsection{Non-dimensionalized formulation}
We begin from the dimensional Boussinesq model
\begin{align}
\nabla\cdot \boldsymbol{u} &= 0,
\label{eq:bo_dim_mass}\\
\rho_0\left(\frac{\partial \boldsymbol{u}}{\partial t} + \boldsymbol{u}\cdot\nabla \boldsymbol{u}\right)
&= -\nabla P + \mu\nabla^2\boldsymbol{u} - \rho_0\,g\,\beta\,(T-T_0) \boldsymbol{\hat{e}_g},
\label{eq:bo_dim_mom}\\
\frac{\partial T}{\partial t} + \boldsymbol{u}\cdot\nabla T &= \alpha\nabla^2 T,
\label{eq:bo_dim_energy}
\end{align}

Let $L$ be the characteristic length, and choose the thermal-diffusion velocity scale
\begin{align}
U=\frac{\alpha}{L},
\end{align}
which is common for natural convection non-dimensionalization. Define the non-dimensional variables
\begin{align}
\boldsymbol{x} = L\boldsymbol{x}^*, \qquad
t = \frac{L}{U}t^*, \qquad
\boldsymbol{u} = U\boldsymbol{u}^*, \qquad
T = T_0 + \Delta T\,T^*.
\end{align}
For pressure, a convenient scale consistent with the viscous term under the choice $U=\alpha/L$ is
\begin{align}
P = \rho_0 U^2\,P^*
\qquad\text{(equivalently, }P^*=\frac{P}{\rho_0U^2}=\frac{P\,L^2}{\rho_0\alpha^2}\text{)}.
\end{align}

The differential operators transform as
\begin{align}
\nabla = \frac{1}{L}\nabla^*, \qquad
\nabla^2=\frac{1}{L^2}\nabla^{*2}, \qquad
\frac{\partial}{\partial t}=\frac{U}{L}\frac{\partial}{\partial t^*}.
\end{align}
Note that the variables with asterisks are dimensionless.

Substituting $\boldsymbol{u}=U\boldsymbol{u}^*$ and $\nabla=(1/L)\nabla^*$ into Equation \ref{eq:bo_dim_mass} gives the non-dimensional continuity equation:
\begin{align}
\nabla\cdot\boldsymbol{u} = \frac{U}{L}\nabla^*\cdot\boldsymbol{u}^* = 0
\quad\Longrightarrow\quad
\nabla^*\cdot\boldsymbol{u}^* = 0.
\label{eq:bo_nd_mass}
\end{align}

We then rewrite each term in Equation \ref{eq:bo_dim_mom} under the above scalings.

Unsteady term:
\begin{align}
\rho_0\frac{\partial \boldsymbol{u}}{\partial t}
= \rho_0 \frac{\partial (U\boldsymbol{u}^*)}{\partial t}
= \rho_0 U\left(\frac{U}{L}\frac{\partial \boldsymbol{u}^*}{\partial t^*}\right)
= \rho_0\frac{U^2}{L}\frac{\partial \boldsymbol{u}^*}{\partial t^*}.
\end{align}

Convective term:
\begin{align}
\rho_0(\boldsymbol{u}\cdot\nabla)\boldsymbol{u}
= \rho_0 (U\boldsymbol{u}^*)\cdot\left(\frac{1}{L}\nabla^*\right)(U\boldsymbol{u}^*)
= \rho_0\frac{U^2}{L}\left(\boldsymbol{u}^*\cdot\nabla^*\right)\boldsymbol{u}^*.
\end{align}

Pressure gradient:
\begin{align}
\nabla P = \frac{1}{L}\nabla^*(\rho_0U^2P^*) = \rho_0\frac{U^2}{L}\nabla^*P^*.
\end{align}

Viscous diffusion:
\begin{align}
\mu\nabla^2\boldsymbol{u}
= \mu\left(\frac{1}{L^2}\nabla^{*2}\right)(U\boldsymbol{u}^*)
= \mu\frac{U}{L^2}\nabla^{*2}\boldsymbol{u}^*.
\end{align}

Buoyancy term:
\begin{align}
\rho_0\,\boldsymbol{g}\,\beta\,(T-T_0)
= \rho_0\,\boldsymbol{g}\,\beta\,(\Delta T\,T^*)
= \rho_0\,\boldsymbol{g}\,\beta\,\Delta T\,T^*.
\end{align}

Now divide the entire momentum equation by the common factor $\rho_0(U^2/L)$ to obtain
\begin{align}
\frac{\partial \boldsymbol{u}^*}{\partial t^*}
+\left(\boldsymbol{u}^*\cdot\nabla^*\right)\boldsymbol{u}^*
= -\nabla^*P^*
+ \underbrace{\frac{\mu}{\rho_0UL}}_{=\;\nu/(UL)}\,\nabla^{*2}\boldsymbol{u}^*
- \underbrace{\frac{g\beta\Delta T\,L}{U^2}}_{\text{buoyancy coefficient}}\,T^*\,\hat{\boldsymbol{e}}_g.
\label{eq:bo_nd_mom_raw}
\end{align}

With $U=\alpha/L$, the viscous prefactor becomes
\begin{align}
\frac{\nu}{UL}=\frac{\nu}{(\alpha/L)L}=\frac{\nu}{\alpha} = Pr,
\end{align}
and the buoyancy coefficient becomes
\begin{align}
\frac{g\beta\Delta T\,L}{U^2}
=\frac{g\beta\Delta T\,L}{(\alpha/L)^2}
=\frac{g\beta\Delta T\,L^3}{\alpha^2}
=\left(\frac{g\beta\Delta T\,L^3}{\nu\alpha}\right)\left(\frac{\nu}{\alpha}\right)
= Ra\,Pr.
\end{align}
Therefore, the non-dimensionalized momentum equation is
\begin{align}
\frac{\partial \boldsymbol{u}^*}{\partial t^*}
+\left(\boldsymbol{u}^*\cdot\nabla^*\right)\boldsymbol{u}^*
= -\nabla^*P^* + Pr\,\nabla^{*2}\boldsymbol{u}^* - Ra\,Pr\,T^*\,\hat{\boldsymbol{e}}_g.
\label{eq:bo_nd_mom}
\end{align}

For the heat equation component of the Boussinesq model, we substitute the appropriate scalings into Equation \ref{eq:bo_dim_energy}:
\begin{align}
\frac{\partial T}{\partial t}
&= \frac{\partial (T_0+\Delta T\,T^*)}{\partial t}
= \Delta T \frac{U}{L}\frac{\partial T^*}{\partial t^*},\\
\boldsymbol{u}\cdot\nabla T
&= (U\boldsymbol{u}^*)\cdot\left(\frac{1}{L}\nabla^*\right)(T_0+\Delta T\,T^*)
= \Delta T\,\frac{U}{L}\,\boldsymbol{u}^*\cdot\nabla^*T^*,\\
\alpha\nabla^2T
&= \alpha\left(\frac{1}{L^2}\nabla^{*2}\right)(T_0+\Delta T\,T^*)
= \Delta T\,\frac{\alpha}{L^2}\nabla^{*2}T^*.
\end{align}
Divide by $\Delta T(U/L)=\Delta T(\alpha/L^2)$ to obtain
\begin{align}
\frac{\partial T^*}{\partial t^*}+\boldsymbol{u}^*\cdot\nabla^*T^*=\nabla^{*2}T^*.
\label{eq:bo_nd_temp}
\end{align}

Dropping asterisks for clarity, the complete non-dimensionalized Boussinesq system is therefore
\begin{align}
\nabla\cdot\boldsymbol{u} &= 0,
\label{eq:bo_strong_nd_mass}\\
\frac{\partial \boldsymbol{u}}{\partial t}+\left(\boldsymbol{u}\cdot\nabla\right)\boldsymbol{u}
&= -\nabla P + Pr\,\nabla^2\boldsymbol{u} - Ra\,Pr\,T\,\hat{\boldsymbol{e}}_g,
\label{eq:bo_strong_nd_mom}\\
\frac{\partial T}{\partial t}+\boldsymbol{u}\cdot\nabla T &= \nabla^2T.
\label{eq:bo_strong_nd_temp}
\end{align}

\subsubsection{Variational (weak) formulation}

Let \(\Omega \subset \mathbb{R}^d\) (\(d = 2, 3\)) be the computational domain.
We employ the Taylor--Hood mixed finite element pairing:
\begin{align*}
\boldsymbol{u} \in \boldsymbol{V}_h = [\mathbb{P}_2]^d, \quad
T \in \Theta_h = \mathbb{P}_1, \quad
P \in Q_h = \mathbb{P}_1,
\end{align*}
where \(\mathbb{P}_2\) denotes piecewise quadratic Lagrange elements and \(\mathbb{P}_1\) denotes piecewise linear Lagrange elements.

Define the mixed space:
\begin{align*}
\mathcal{W}_h^\text{B}= \boldsymbol{V}_h \times \Theta_h \times Q_h,
\end{align*}
with trial and test functions:
\begin{align*}
w = 
\begin{bmatrix}
\boldsymbol{u} \\
T \\
P
\end{bmatrix} \in \mathcal{W}_h^\text{B}, \quad
\psi = 
\begin{bmatrix}
\boldsymbol{v} \\
\theta \\
q
\end{bmatrix} \in \mathcal{W}_h^\text{B}.
\end{align*}

Multiply Equation \ref{eq:bo_strong_nd_mass} by the test function \(q\) and integrate over \(\Omega\):
\begin{align}
\int_\Omega (\nabla \cdot \boldsymbol{u})\,q \, d\Omega = 0 \quad \forall q \in Q_h.
\label{eq:bouss_weak_mass_raw}
\end{align}

No integration by parts is needed for the continuity equation.
Multiply Equation~\ref{eq:bo_strong_nd_mom} by $\boldsymbol{v}$ and integrate over the domain:
\begin{align}
\int_\Omega \left( \frac{\partial \boldsymbol{u}}{\partial t} + \boldsymbol{u} \cdot \nabla \boldsymbol{u} \right) \cdot \boldsymbol{v} \, d\Omega
&= -\int_\Omega \nabla P \cdot \boldsymbol{v} \, d\Omega
+ Pr \int_\Omega \nabla^2 \boldsymbol{u} \cdot \boldsymbol{v} \, d\Omega
- Pr\,Ra \int_\Omega T\,(\boldsymbol{\hat{e}}_g \cdot \boldsymbol{v}) \, d\Omega .
\label{eq:bouss_weak_mom_raw}
\end{align}
The convective term is retained in advective form. The pressure gradient and the viscous term are integrated by parts,
\begin{align}
\int_\Omega \nabla P \cdot \boldsymbol{v} \, d\Omega
&= -\int_\Omega P\,(\nabla \cdot \boldsymbol{v}) \, d\Omega
+ \int_{\partial\Omega} P\,(\boldsymbol{v} \cdot \boldsymbol{n}) \, dS, \\
Pr \int_\Omega \nabla^2 \boldsymbol{u} \cdot \boldsymbol{v} \, d\Omega
&= -Pr \int_\Omega \nabla \boldsymbol{u} : \nabla \boldsymbol{v} \, d\Omega
+ Pr \int_{\partial\Omega} (\nabla \boldsymbol{u}\,\boldsymbol{n}) \cdot \boldsymbol{v} \, dS,
\end{align}
where both boundary integrals vanish since $\boldsymbol{v}=\boldsymbol{0}$ on $\partial\Omega$. Substituting and collecting all terms on one side gives the final weak form for the momentum conservation equation,
\begin{align}
\mathcal{F}_{\text{mom}} &= \int_\Omega \left( \frac{\partial \boldsymbol{u}}{\partial t} + \boldsymbol{u} \cdot \nabla \boldsymbol{u} \right) \cdot \boldsymbol{v} \, d\Omega
- \int_\Omega P\,(\nabla \cdot \boldsymbol{v}) \, d\Omega \nonumber \\
&\quad + Pr \int_\Omega \nabla \boldsymbol{u} : \nabla \boldsymbol{v} \, d\Omega
+ Pr\,Ra \int_\Omega T\,(\boldsymbol{\hat{e}}_g \cdot \boldsymbol{v}) \, d\Omega = 0 ,
\label{eq:bouss_weak_mom_final}
\end{align}
in agreement with Equation~\ref{eq:bo-weak}.

Next, for the energy equation, multiply Equation \ref{eq:bo_strong_nd_temp} by \(\theta\) and integrate:
\begin{align}
\int_\Omega \frac{\partial T}{\partial t}\,\theta \, d\Omega
+ \int_\Omega (\boldsymbol{u} \cdot \nabla T)\,\theta \, d\Omega
= \int_\Omega (\nabla^2 T)\,\theta \, d\Omega.
\label{eq:bouss_weak_energy_raw}
\end{align}

Using \(\nabla \cdot (\theta \nabla T) = (\nabla^2 T)\,\theta + \nabla T \cdot \nabla \theta\):
\begin{align}
\int_\Omega (\nabla^2 T)\,\theta \, d\Omega
= \int_{\partial \Omega} \theta\,(\nabla T \cdot \boldsymbol{n}) \, dS
- \int_\Omega \nabla T \cdot \nabla \theta \, d\Omega.
\end{align}

On adiabatic walls (top and bottom), \(\nabla T \cdot \boldsymbol{n} = 0\), on isothermal walls (left and right), \(\theta = 0\) since Dirichlet boundary conditions are applied \cite{hughes2000finite}.
Thus the boundary term vanishes:
\begin{align}
\int_\Omega (\nabla^2 T)\,\theta \, d\Omega = -\int_\Omega \nabla T \cdot \nabla \theta \, d\Omega.
\end{align}

Substituting this into the energy equation, Equation \ref{eq:bouss_weak_energy_raw}:
\begin{align}
\mathcal{F}_{\text{energy}} = \int_\Omega \frac{\partial T}{\partial t}\,\theta \, d\Omega
+ \int_\Omega (\boldsymbol{u} \cdot \nabla T)\,\theta \, d\Omega
+ \int_\Omega \nabla T \cdot \nabla \theta \, d\Omega = 0.
\label{eq:bouss_weak_energy_final}
\end{align}

Now we can state the complete variational problem: Find \(w = (\boldsymbol{u}, T, P) \in \mathcal{W}_h^\text{B}\) such that
\begin{align}
\mathcal{F}(w; \psi) = \mathcal{F}_{\text{mass}} + \mathcal{F}_{\text{mom}} + \mathcal{F}_{\text{energy}} = 0 \quad \forall \psi = (\boldsymbol{v}, \theta, q) \in \mathcal{W}_h^\text{B},
\end{align}
where:
\begin{align}
\mathcal{F}_{\text{mass}} &= \int_\Omega (\nabla \cdot \boldsymbol{u})\,q \, d\Omega, \\
\mathcal{F}_{\text{mom}} &= \int_\Omega \left( \frac{\partial \boldsymbol{u}}{\partial t} + \boldsymbol{u} \cdot \nabla \boldsymbol{u} \right) \cdot \boldsymbol{v} \, d\Omega
- \int_\Omega P\,(\nabla \cdot \boldsymbol{v}) \, d\Omega \nonumber \\
&\quad + Pr \int_\Omega \nabla \boldsymbol{u} : \nabla \boldsymbol{v} \, d\Omega
+ Pr\,Ra \int_\Omega T\,(\boldsymbol{\hat{e}}_g \cdot \boldsymbol{v}) \, d\Omega, \\
\mathcal{F}_{\text{energy}} &= \int_\Omega \frac{\partial T}{\partial t}\,\theta \, d\Omega
+ \int_\Omega (\boldsymbol{u} \cdot \nabla T)\,\theta \, d\Omega
+ \int_\Omega \nabla T \cdot \nabla \theta \, d\Omega.
\end{align}

This completes the derivation of the non-dimensional weak formulation for the Boussinesq approximation.

\subsection{Compressible flow equations}
The dimensional strong form of the compressible flow equations is:
\begin{align}
\frac{\partial \rho}{\partial t} + \nabla \cdot (\rho \boldsymbol{u}) &= 0, \label{eq:comp_dim_mass}\\
\frac{\partial (\rho \boldsymbol{u})}{\partial t} + \nabla \cdot (\rho \boldsymbol{u} \otimes \boldsymbol{u}) &= -\nabla p + \nabla \cdot \boldsymbol{\tau} + \rho g\,\boldsymbol{\hat{e}}_g, \label{eq:comp_dim_mom}\\
\frac{\partial (\rho E)}{\partial t} + \nabla \cdot (\rho E \boldsymbol{u}) &= \nabla \cdot (\kappa \nabla T) - \nabla \cdot (p\boldsymbol{u}) + \nabla \cdot (\boldsymbol{\tau} \cdot \boldsymbol{u}) + \rho g\,\boldsymbol{\hat{e}}_g \cdot \boldsymbol{u}, \label{eq:comp_dim_energy}
\end{align}
where $\rho$ is density, $\boldsymbol u$ velocity, $p$ pressure, $T$ temperature,
and $E$ is the total energy per unit mass
\begin{align}
E = e + \frac12 |\boldsymbol{u}|^2,
\label{eq:E_def_dim}
\end{align}
with internal energy $e=c_v T$ for an ideal gas and constant $c_v$.
The viscous stress tensor is
\begin{align}
\boldsymbol{\tau}
= \mu\left(\nabla\boldsymbol{u}+(\nabla\boldsymbol{u})^\top\right)
-\frac{2}{3}\mu(\nabla\cdot\boldsymbol{u})\boldsymbol{I},
\label{eq:tau_dim}
\end{align}
and the ideal gas equation of state reads
\begin{align}
p = (\gamma-1)\rho e,\qquad \gamma=\frac{c_p}{c_v}.
\label{eq:eos_dim}
\end{align}
\subsubsection{Non-dimensionalized formulation}
We introduce reference scales
$L$ (length), $U$ (velocity), and define nondimensional variables as:
\begin{subequations}
\label{eq:comp_nondim_vars}
\begin{align}
\boldsymbol{x}^* &= \frac{\boldsymbol{x}}{L}, \quad
t^* = \frac{tU}{L}, \quad
\boldsymbol{u}^* = \frac{\boldsymbol{u}}{U}, \\
\rho^* &= \frac{\rho}{\rho_0}, \quad
T^* = \frac{T}{T_0}, \quad
e^* = \frac{e}{e_0}, \\
p^* &= \frac{p - p_0}{\rho_0 U^2}.
\end{align}
\end{subequations}
Note that pressure is non-dimensionalized as a perturbation from the reference thermodynamic pressure \(p_0\).
We define non-dimensional parameters as:

\begin{align}
Ma &= \frac{U}{c} = \frac{U}{\sqrt{\gamma p_0 / \rho_0}}, \quad\text{(Mach number)} \\
Re &= \frac{\rho_0 U L}{\mu}, \quad\text{(Reynolds number)} \\
Pr &= \frac{\mu c_p}{\kappa}, \quad\text{(Prandtl number)} \\
Fr &= \frac{U}{\sqrt{gL}}. \quad\text{(Froude number)}
\end{align}

From the Mach number definition:
\begin{align}
Ma^2 = \frac{U^2}{\gamma p_0 / \rho_0} = \frac{\rho_0 U^2}{\gamma p_0}
\quad \Rightarrow \quad
p_0 = \frac{\rho_0 U^2}{\gamma Ma^2}.
\label{eq:p0_relation}
\end{align}

For an ideal gas, the specific internal energy and temperature are strictly coupled by the specific heat capacity at constant volume \(e = c_v T\). The reference scales satisfy this relationship:
\begin{align}
e_0 = c_v T_0.
\end{align}

Thus:
\begin{align}
e^* = \frac{e}{e_0} = \frac{c_v T}{c_v T_0} = \frac{T}{T_0} = T^*.
\end{align}

Therefore, the non-dimensionalized internal energy equals the non-dimensionalized temperature:
\begin{align}
e^* = T^*.
\label{eq:e_equals_T}
\end{align}

\textbf{Non-dimensionalizing the equation of state.}

Starting from Equation \ref{eq:eos_dim}, we make appropriate substitutions of dimensional variables with their non-dimensionalized counterparts, yielding
\begin{align}
p_0 + \rho_0 U^2 p^* = (\gamma - 1)\,(\rho_0 \rho^*)\,(e_0 e^*).
\end{align}

Dividing by \(\rho_0 U^2\),
\begin{align}
\frac{p_0}{\rho_0 U^2} + p^* = (\gamma - 1)\,\frac{e_0}{U^2}\,\rho^* e^*.
\end{align}

From Equation \ref{eq:p0_relation}, \(\frac{p_0}{\rho_0 U^2} = \frac{1}{\gamma Ma^2}\), and thus
\begin{align}
\frac{1}{\gamma Ma^2} + p^* = (\gamma - 1)\,\frac{e_0}{U^2}\,\rho^* e^*.
\label{eq:Ma_p}
\end{align}

Now, for an ideal gas,
\begin{align}
c_v = \frac{R}{\gamma - 1}, \quad
c_p = \frac{\gamma R}{\gamma - 1}, \quad
c^2 = \frac{\gamma p_0}{\rho_0} = \gamma R T_0.
\end{align}

Thus,
\begin{align}
e_0 = c_v T_0 = \frac{R T_0}{\gamma - 1} = \frac{c^2}{\gamma(\gamma - 1)}.
\end{align}

Since \(c = U / Ma\),
\begin{align}
e_0 = \frac{U^2}{\gamma(\gamma - 1) Ma^2}.
\end{align}

Therefore,
\begin{align}
\frac{e_0}{U^2} = \frac{1}{\gamma(\gamma - 1) Ma^2}.
\label{eq:e_Usquare}
\end{align}

Substituting this into Equation \ref{eq:Ma_p}:
\begin{align}
\frac{1}{\gamma Ma^2} + p^* = (\gamma - 1)\,\frac{1}{\gamma(\gamma - 1) Ma^2}\,\rho^* e^*
= \frac{1}{\gamma Ma^2}\,\rho^* e^*.
\end{align}

Multiply by \(\gamma Ma^2\):
\begin{align}
1 + \gamma Ma^2 p^* = \rho^* e^*.
\end{align}

Dropping asterisks, the non-dimensionalized equation of state is:
\begin{align}
\rho e = 1 + \gamma Ma^2 p.
\label{eq:eos_nondim_comp}
\end{align}

\textbf{Non-dimensionalizing the continuity equation.}

Starting from Equation \ref{eq:comp_dim_mass} and substituting in non-dimensionalized variables yields
\begin{align}
\frac{U}{L} \frac{\partial (\rho_0 \rho^*)}{\partial t^*} + \frac{1}{L} \nabla^* \cdot (\rho_0 \rho^* \cdot U \boldsymbol{u}^*) = 0.
\end{align}

Simplifying, we obtain
\begin{align}
\frac{\rho_0 U}{L} \frac{\partial \rho^*}{\partial t^*} + \frac{\rho_0 U}{L} \nabla^* \cdot (\rho^* \boldsymbol{u}^*) = 0.
\end{align}

Multiplying by \(\frac{L}{\rho_0 U}\), we then have
\begin{align}
\frac{\partial \rho^*}{\partial t^*} + \nabla^* \cdot (\rho^* \boldsymbol{u}^*) = 0,
\end{align}

and finally, dropping asterisks leaves
\begin{align}
\frac{\partial \rho}{\partial t} + \nabla \cdot (\rho \boldsymbol{u}) = 0.
\label{eq:comp_nondim_mass}
\end{align}

\textbf{Non-dimensionalizing the momentum equation.}

Similarly, starting from Equation \ref{eq:comp_dim_mom} and substituting non-dimensionalized variables gives:
\begin{align}
\frac{U}{L} \frac{\partial (\rho_0 \rho^* U \boldsymbol{u}^*)}{\partial t^*} &+ \frac{1}{L} \nabla^* \cdot (\rho_0 \rho^* U \boldsymbol{u}^* \otimes U \boldsymbol{u}^*) \nonumber \\
&= -\frac{1}{L} \nabla^* (p_0 + \rho_0 U^2 p^*) + \frac{1}{L} \nabla^* \cdot (\mu \boldsymbol{\tau}^*) + \rho_0 \rho^* g\,\boldsymbol{\hat{e}}_g,
\end{align}
where \(\boldsymbol{\tau}^*\) is the non-dimensionalized stress tensor.

After algebraic simplification, we obtain:
\begin{align}
\frac{\rho_0 U^2}{L} \frac{\partial (\rho^* \boldsymbol{u}^*)}{\partial t^*} + \frac{\rho_0 U^2}{L} \nabla^* \cdot (\rho^* \boldsymbol{u}^* \otimes \boldsymbol{u}^*)
&= -\frac{\rho_0 U^2}{L} \nabla^* p^* + \frac{\mu U}{L^2} \nabla^* \cdot \boldsymbol{\tau}^* + \rho_0 \rho^* g\,\boldsymbol{\hat{e}}_g.
\end{align}

Multiplying by \(\frac{L}{\rho_0 U^2}\) then gives
\begin{align}
\frac{\partial (\rho^* \boldsymbol{u}^*)}{\partial t^*} + \nabla^* \cdot (\rho^* \boldsymbol{u}^* \otimes \boldsymbol{u}^*)
= -\nabla^* p^* + \frac{\mu}{\rho_0 U L} \nabla^* \cdot \boldsymbol{\tau}^* + \frac{gL}{U^2}\,\rho^*\,\boldsymbol{\hat{e}}_g ,
\end{align}

and using \(Re = \frac{\rho_0 U L}{\mu}\) and \(Fr = \frac{U}{\sqrt{gL}}\), we obtain
\begin{align}
\frac{\partial (\rho^* \boldsymbol{u}^*)}{\partial t^*} + \nabla^* \cdot (\rho^* \boldsymbol{u}^* \otimes \boldsymbol{u}^*)
= -\nabla^* p^* + \frac{1}{Re} \nabla^* \cdot \boldsymbol{\tau}^* + \frac{\rho^*}{Fr^2}\,\boldsymbol{\hat{e}}_g.
\end{align}

Finally, dropping asterisks, we have:
\begin{align}
\frac{\partial (\rho \boldsymbol{u})}{\partial t} + \nabla \cdot (\rho \boldsymbol{u} \otimes \boldsymbol{u})
= -\nabla p + \frac{1}{Re} \nabla \cdot \boldsymbol{\tau} + \frac{\rho}{Fr^2}\,\boldsymbol{\hat{e}}_g.
\label{eq:comp_nondim_mom}
\end{align}

\textbf{Non-dimensionalizing the energy equation.}

First, scale the total energy:
\begin{align}
E=e+\frac12|\boldsymbol{u}|^2
=e_0 e^* + \frac12 U^2|\boldsymbol{u}^*|^2.
\end{align}
Referring to Equation \ref{eq:e_Usquare}, we have
\begin{align}
\frac{U^2}{e_0}
=\gamma(\gamma - 1) Ma^2.
\end{align}
Thus, it is natural to define the non-dimensionalized total energy as
\begin{align}
E^* = e^* + \frac{\gamma(\gamma - 1) Ma^2}{2}|\boldsymbol{u}^*|^2.
\label{eq:E_star_def}
\end{align}

Scaling on the left-hand side for Equation \ref{eq:comp_dim_energy}:
\begin{align}
\frac{\partial(\rho E)}{\partial t}
&=\frac{\partial(\rho_0\rho^*\,e_0E^*)}{\partial t}
=\rho_0e_0\frac{U}{L}\frac{\partial(\rho^*E^*)}{\partial t^*},\\
\nabla\cdot(\rho E\,\boldsymbol{u})
&=\frac{1}{L}\nabla^*\cdot(\rho_0\rho^*\,e_0E^*\,U\boldsymbol{u}^*)
=\rho_0e_0\frac{U}{L}\nabla^*\cdot(\rho^*E^*\boldsymbol{u}^*).
\end{align}
Hence the natural energy scale is $\rho_0e_0U/L$.
For the conduction term in the energy equation, Equation \ref{eq:comp_dim_energy}, we use $T=T_0T^*$ and $T^*=e^*$:
\begin{align}
\nabla\cdot(\kappa\nabla T)
=\nabla\cdot\!\left(\kappa\frac{T_0}{L}\nabla^*T^*\right)
=\kappa\frac{T_0}{L^2}\nabla^{*2}T^*.
\end{align}
Divide by $\rho_0e_0U/L$ and use $e_0=c_vT_0$, $\kappa=\mu c_p/Pr$:
\begin{align}
\frac{\kappa T_0/L^2}{\rho_0e_0U/L}
=\frac{\kappa}{\rho_0c_v}\frac{1}{UL}
=\frac{\mu c_p}{Pr\,\rho_0c_v}\frac{1}{UL}
=\frac{\mu\gamma}{Pr\,\rho_0UL}
=\frac{\gamma}{Re\,Pr}.
\end{align}
Therefore the non-dimensionalized conduction contribution is $\frac{\gamma}{RePr}\nabla^{*2}T^*$.
The same idea applies for the pressure-work flux:
\begin{align}
-\nabla\cdot(p\boldsymbol{u})
&=-\frac{1}{L}\nabla^*\cdot\!\left((p_0+\rho_0U^2p^*)\,U\boldsymbol{u}^*\right)
=-\frac{1}{L}\nabla^*\cdot\!\left(p_0U\boldsymbol{u}^*\right)
-\frac{1}{L}\nabla^*\cdot\!\left(\rho_0U^3p^*\boldsymbol{u}^*\right).
\end{align}
Divide by $\rho_0 e_0 U/L$ and use $p_0/(\rho_0 e_0) = \gamma-1$ and
$U^2/e_0 = \gamma(\gamma-1)Ma^2$:
\begin{equation}
-(\gamma-1)\,\nabla^*\!\cdot \boldsymbol{u}^*
-\gamma(\gamma-1)Ma^2\,\nabla^*\!\cdot(p^* \boldsymbol{u}^*)
= -(\gamma-1)\,\nabla^*\!\cdot\!\big[(1+\gamma Ma^2 p^*)\boldsymbol{u}^*\big].
\end{equation}

Similarly, viscous work flux:
\begin{align}
\nabla\cdot(\boldsymbol{\tau}\cdot\boldsymbol{u})
&=\frac{1}{L}\nabla^*\cdot\left(\mu\frac{U}{L}\boldsymbol{\tau}^*\cdot U\boldsymbol{u}^*\right)
=\mu\frac{U^2}{L^2}\nabla^*\cdot(\boldsymbol{\tau}^*\cdot\boldsymbol{u}^*).
\end{align}
Divide by $\rho_0e_0U/L$:
\begin{align}
\frac{\mu U^2/L^2}{\rho_0e_0U/L}
=\frac{\mu U}{\rho_0e_0 L}
=\frac{\mu}{\rho_0UL}\frac{U^2}{e_0} = \frac{1}{Re} \cdot \gamma(\gamma - 1) Ma^2.
\end{align}
Hence the non-dimensionalized viscous work term is $\frac{\gamma(\gamma - 1) Ma^2}{Re}\nabla^*\cdot(\boldsymbol{\tau}^*\cdot\boldsymbol{u}^*)$.

For the gravity work:
\begin{align}
\rho g\hat{\boldsymbol e}_g\cdot\boldsymbol{u}
=\rho_0\rho^* g\,U(\hat{\boldsymbol e}_g\cdot\boldsymbol{u}^*).
\end{align}
Divide by $\rho_0 e_0 U/L$; since $Fr^2 = U^2/(gL)$ and $U^2/e_0=\gamma(\gamma-1)Ma^2$,
the coefficient is $gL/e_0 = \gamma(\gamma-1)Ma^2/Fr^2$:
\begin{equation}
\frac{gL}{e_0}\,\rho^*(\boldsymbol{\hat{e}_g}\cdot \boldsymbol{u}^*)
= \frac{\gamma(\gamma-1)Ma^2}{Fr^2}\,\rho^*(\boldsymbol{\hat{e}_g}\cdot \boldsymbol{u}^*).
\end{equation}

Collecting terms in the energy equation, E \ref{eq:comp_dim_energy}, yields the non-dimensionalized total-energy equation:
\begin{align}
\frac{\partial (\rho E)}{\partial t} + \nabla \cdot (\rho E \boldsymbol{u}) &= \frac{\gamma}{Re\,Pr}\,\nabla^2 T - (\gamma - 1)\,\nabla \cdot [(\gamma Ma^2 p + 1)\boldsymbol{u}] \nonumber \\
&\quad + \frac{(\gamma - 1)\gamma Ma^2}{Re}\,\nabla \cdot (\boldsymbol{\tau} \cdot \boldsymbol{u}) + \frac{(\gamma - 1)\gamma Ma^2}{Fr^2}\,\rho\,(\boldsymbol{\hat{e}}_g \cdot \boldsymbol{u}).
\label{eq:comp_nondim_energy}
\end{align}

The complete non-dimensionalized system is:
\begin{align}
\frac{\partial \rho}{\partial t} + \nabla \cdot (\rho \boldsymbol{u}) &= 0, \\
\frac{\partial (\rho \boldsymbol{u})}{\partial t} + \nabla \cdot (\rho \boldsymbol{u} \otimes \boldsymbol{u}) &= -\nabla p + \frac{1}{Re}\,\nabla \cdot \boldsymbol{\tau} + \frac{\rho}{Fr^2}\,\boldsymbol{\hat{e}}_g, \\
\frac{\partial (\rho E)}{\partial t} + \nabla \cdot (\rho E \boldsymbol{u}) &= \frac{\gamma}{Re\,Pr}\,\nabla^2 T - (\gamma - 1)\,\nabla \cdot [(\gamma Ma^2 p + 1)\boldsymbol{u}] \nonumber \\
&\quad + \frac{(\gamma - 1)\gamma Ma^2}{Re}\,\nabla \cdot (\boldsymbol{\tau} \cdot \boldsymbol{u}) + \frac{(\gamma - 1)\gamma Ma^2}{Fr^2}\,\rho\,(\boldsymbol{\hat{e}}_g \cdot \boldsymbol{u}), \\
\rho e &= 1 + \gamma Ma^2 p.
\end{align}

\subsubsection{Variational (weak) formulation}

As with our Boussinesq formulation, we employ the Taylor--Hood mixed finite element pairing:
\begin{align*}
\boldsymbol{u} \in \boldsymbol{V}_h = [\mathbb{P}_2]^d, \quad
e \in \Theta_h = \mathbb{P}_1, 
\quad \rho \in R_h = \mathbb{P}_1, 
\quad P \in Q_h = \mathbb{P}_1,
\end{align*}
where \(\mathbb{P}_2\) denotes piecewise quadratic Lagrange elements and \(\mathbb{P}_1\) denotes piecewise linear Lagrange elements.

Define the mixed space:
\begin{align*}
\mathcal{W}_h^\text{C}= \boldsymbol{V}_h \times \Theta_h \times R_h \times Q_h,
\end{align*}
with trial and test functions:

\begin{align}
w=\begin{bmatrix}\boldsymbol{u}\\ e\\ \rho\\ p\end{bmatrix}\in \mathcal{W}_h^\text{C},
\qquad
\psi=\begin{bmatrix}\boldsymbol{v}\\ w_e\\ w_\rho\\ q\end{bmatrix}\in \mathcal{W}_h^\text{C}.
\end{align}

The strong form of the mass conservation equation is
\begin{align}
\frac{\partial \rho}{\partial t} + \nabla \cdot (\rho \boldsymbol{u}) = 0.
\end{align}
Multiplying by the test function $q$ and integrating over $\Omega$, with the convective flux retained in conservative (divergence) form, no integration by parts is required and the weak form for the mass conservation (continuity) equation is
\begin{align}
    \mathcal{F}_{\rho} = \int_{\Omega} \left( \frac{\partial \rho}{\partial t} + \nabla \cdot (\rho \boldsymbol{u}) \right) q \, d\Omega = 0,
\end{align}
in agreement with Equation~\ref{eq:weak_form_compressible}.

The strong form of the momentum conservation equation is:
\begin{align}
  \frac{\partial (\rho \boldsymbol{u})}{\partial t} +  \nabla \cdot (\rho \boldsymbol{u} \otimes \boldsymbol{u})   
    = -\nabla p + \frac{1}{Re} \nabla \cdot \boldsymbol{\tau} + \frac{\rho}{Fr^2} \boldsymbol{\hat{e}_g}
\end{align}
where $ \boldsymbol{\tau} $ is the viscous stress tensor.

Multiply the momentum equation by the test function \( \boldsymbol{v} \) and integrate over the domain:
\begin{align}
\int_\Omega \left( \frac{\partial ( \rho \boldsymbol{u})}{\partial t} + \nabla \cdot (\rho \boldsymbol{u} \otimes \boldsymbol{u} ) \right) \cdot \boldsymbol{v} \, d\Omega &= -\int_\Omega \nabla p \cdot \boldsymbol{v} \, d\Omega + \frac{1}{Re}\int_\Omega (\nabla \cdot \boldsymbol{\tau}) \cdot \boldsymbol{v} \, d\Omega \nonumber \\
&\quad + \int_\Omega \left ( \frac{\rho}{Fr^2} \boldsymbol{\hat{e_g}}\right ) \cdot \boldsymbol{v} \, d\Omega
\end{align}
The pressure term:
  \begin{align}
  \int_\Omega \nabla p \cdot \boldsymbol{v} \, d\Omega &= - \int_\Omega p \nabla \cdot \boldsymbol{v} \, d\Omega,
  \end{align}
The viscous term \( \nabla \cdot \boldsymbol{\tau} \) is integrated by parts to avoid second derivatives of velocity \( \boldsymbol{u} \):
  \begin{align}
  \frac{1}{Re}\int_\Omega (\nabla \cdot \boldsymbol{\tau}) \cdot \boldsymbol{v} \, d\Omega &= - \frac{1}{Re}\int_\Omega \boldsymbol{\tau} : \nabla \boldsymbol{v} \, d\Omega + \frac{1}{Re}\int_\Gamma (\boldsymbol{\tau}\mathbf n)\cdot\boldsymbol{v}\,\mathrm d\Gamma.
  \end{align}
The final weak form for the momentum conservation equation is:
\begin{align}
    \mathcal{F}_{\boldsymbol{u}} &= \int_{\Omega} \frac{\partial (\rho \boldsymbol{u})}{\partial t} \cdot \boldsymbol{v} \, d\Omega 
    + \int_{\Omega} \nabla \cdot (\rho \boldsymbol{u} \otimes \boldsymbol{u}) \cdot \boldsymbol{v} \, d\Omega \nonumber \\
    &- \int_{\Omega} p (\nabla \cdot \boldsymbol{v}) \, d\Omega 
    + \frac{1}{Re} \int_{\Omega} \boldsymbol{\tau} : \nabla \boldsymbol{v} \, d\Omega 
    - \int_{\Omega} \frac{\rho}{Fr^2} (\boldsymbol{\hat{e}_g} \cdot \boldsymbol{v}) \, d\Omega = 0,
\end{align}
where only the pressure and viscous terms are integrated by parts (the corresponding boundary integrals vanish under the no-slip condition).

The strong form of the energy equation is:
\begin{align}
\frac{\partial (\rho E)}{\partial t} + \nabla \cdot (\rho E\boldsymbol{u}) = \frac{\gamma}{RePr} \nabla \cdot (\nabla e) - (\gamma - 1)\nabla \cdot \big[(\gamma Ma^2 p +1)\boldsymbol{u}\big] \\
+ \frac{(\gamma -1)\gamma Ma^2}{Re}\nabla\cdot(\boldsymbol{\tau}\cdot\boldsymbol{u}) + \frac{(\gamma-1)\gamma Ma^2}{Fr^2}\rho\,(\boldsymbol{\hat e}_g\cdot\boldsymbol{u}),
\end{align}
where the deviatoric viscous stress is
\begin{equation}
\boldsymbol{\tau} \;=\; \nabla\boldsymbol{u}+(\nabla\boldsymbol{u})^\top
\;-\;\frac{2}{3}(\nabla\!\cdot\boldsymbol{u})\,\boldsymbol{I}.
\end{equation}
In the total-energy formulation the energy equation contains the viscous \emph{work} $\nabla\!\cdot(\boldsymbol{\tau}\cdot\boldsymbol{u})$, rather than the dissipation $\Phi=\boldsymbol{\tau}:\nabla\boldsymbol{u}$ alone; the two differ by $(\nabla\!\cdot\boldsymbol{\tau})\cdot\boldsymbol{u}$.

Multiply the energy equation by the test function \( w_e \) and integrate over the domain:
\begin{align}
\int_\Omega \left( \frac{\partial(\rho E)}{\partial t} + \nabla\!\cdot(\rho E\,\boldsymbol{u}) \right) w_e \, d\Omega
&= \frac{\gamma}{Re\,Pr} \int_\Omega \nabla\!\cdot(\nabla e)\, w_e \, d\Omega
- (\gamma-1)\int_\Omega \nabla\!\cdot\!\big[(\gamma Ma^2 p+1)\boldsymbol{u}\big]\, w_e \, d\Omega \nonumber\\
&\quad + \frac{(\gamma-1)\gamma Ma^2}{Re} \int_\Omega \nabla\!\cdot(\boldsymbol{\tau}\cdot\boldsymbol{u})\, w_e \, d\Omega
+ \frac{(\gamma-1)\gamma Ma^2}{Fr^2} \int_\Omega \rho\,(\boldsymbol{\hat{e}}_g\cdot\boldsymbol{u})\, w_e \, d\Omega .
\end{align}

For the diffusion term,
\begin{equation}
\int_\Omega \nabla\!\cdot(\nabla e)\, w_e\, d\Omega
= -\int_\Omega \nabla e \cdot \nabla w_e\, d\Omega
+ \int_{\partial\Omega} w_e\, \mathbf{n}\!\cdot\nabla e \, dS,
\end{equation}
and since $w_e=0$ on Dirichlet (isothermal) walls and $\mathbf{n}\!\cdot\nabla e=0$ on adiabatic walls,
the boundary contribution is zero. 

For the pressure-work term,
\begin{equation}
\int_\Omega \nabla\!\cdot\!\big[(\gamma Ma^2 p+1)\boldsymbol{u}\big]\, w_e\, d\Omega
= -\int_\Omega (\gamma Ma^2 p+1)\,\boldsymbol{u}\cdot\nabla w_e\, d\Omega
+ \int_{\partial\Omega} (\gamma Ma^2 p+1)\,(\boldsymbol{u}\cdot\mathbf{n})\, w_e \, dS,
\end{equation}
whose boundary term vanishes under the no-slip condition \(\boldsymbol{u}=\boldsymbol{0}\) on \(\partial\Omega\).

Similarly, integration by parts for the viscous work term:
\begin{align}
\int_\Omega \nabla \cdot (\boldsymbol{\tau} \cdot \boldsymbol{u})\,w_e \, d\Omega
= -\int_\Omega (\boldsymbol{\tau} \cdot \boldsymbol{u}) \cdot \nabla w_e \, d\Omega,
\end{align}
where the boundary term vanishes by the no-slip condition \(\boldsymbol{u}=\boldsymbol{0}\) on \(\partial\Omega\).
The final weak form for the energy equation is:
\begin{align}
    \mathcal{F}_{e} &= \int_\Omega \left( \frac{\partial (\rho E)}{\partial t} + \nabla \cdot (\rho E \boldsymbol{u}) \right) w_e \, d\Omega \\
    &+ \frac{\gamma}{Re Pr} \int_{\Omega} \nabla e \cdot \nabla w_e \, d\Omega \nonumber \\
    &- (\gamma - 1) \int_{\Omega} (1 + \gamma Ma^2 p) \boldsymbol{u} \cdot \nabla w_e \, d\Omega \nonumber \\
    &+ \frac{\gamma(\gamma - 1) Ma^2}{Re} \int_{\Omega} (\boldsymbol{\tau} \cdot \boldsymbol{u}) \cdot \nabla w_e \, d\Omega \nonumber \\
    &- \frac{\gamma(\gamma - 1) Ma^2}{Fr^2} \int_{\Omega} \rho (\boldsymbol{\hat{e}_g} \cdot \boldsymbol{u}) w_e \, d\Omega = 0.
\end{align}

\end{document}